\documentclass[apj,twocolumn,numberedappendix,twocolappendix]{openjournal}

\usepackage[breaklinks,colorlinks,citecolor=blue,urlcolor=blue]{hyperref}
\usepackage{amsmath}
\usepackage{graphicx}
\usepackage{multirow}
\usepackage{orcidlink}

\bibpunct[; ]{(}{)}{,}{a}{}{,}

\newcommand{\COone}{\ensuremath{\mathrm{CO}(1{\rightarrow}0)}}
\newcommand{\COtwo}{\ensuremath{\mathrm{CO}(2{\rightarrow}1)}}

\newcommand{\CO}{$^{12}$CO\,}

\newcommand{\msun}{\ensuremath{M_{\odot}}}
\newcommand{\mstar}{\ensuremath{M_{\star}}}

\newcommand{\bdm}{\begin{displaymath}}
\newcommand{\edm}{\end{displaymath}}
\newcommand{\beq}{\begin{equation}}
\newcommand{\eeq}{\end{equation}}
\newcommand{\bit}{\begin{itemize}}
\newcommand{\eit}{\end{itemize}}
\newcommand{\ben}{\begin{enumerate}}
\newcommand{\een}{\end{enumerate}}
\newcommand{\bfi}{\begin{figure}[htb]}
\newcommand{\bpfi}{\begin{figure}[p]}

\newcommand{\lir}{\ensuremath{L_{\rm IR}}}
\newcommand{\lco}{\ensuremath{L'_{\rm CO}}}
\newcommand{\aCO}{\ensuremath{\alpha_{\rm CO}}}

\newcommand{\mhtwo}{\ensuremath{M({\rm H_2})}}
\newcommand{\mHI}{\ensuremath{M({\rm HI})}}

\newcommand{\fgas}{\ensuremath{f_{\rm gas}}}
\newcommand{\tdep}{\ensuremath{\tau_{\rm dep.}}}
\newcommand{\sfr}{\rm SFR}

\newcommand{\ssfr}{\rm sSFR}
\newcommand{\ssfrMS}{\ensuremath{\langle\ssfr\rangle_{\rm MS}}}
\newcommand{\sfe}{\rm SFE}

\begin{document}

\title{The molecular gas content throughout the low-$z$ merger sequence$^\star$\\[-3em]}

\author{M.~T. Sargent$^{1,2}$\,\orcidlink{0000-0003-1033-9684}}
\author{S.~L. Ellison$^3$\,\orcidlink{0000-0002-1768-1899}}
\author{J.~T. Mendel$^{4,5,6}$\,\orcidlink{0000-0002-6327-9147}}
\author{A. Saintonge$^{7,8}$\,\orcidlink{0000-0003-4357-3450}}
\author{D.~Cs. Moln\'ar$^2$\,\orcidlink{0000-0002-8971-9462}}
\author{T. Schwandt$^2$}
\author{J.~M. Scudder$^{9,2}$\,\orcidlink{0000-0002-8798-3972}}
\author{G. Violino$^{10}$}
\affiliation{$^1$~International Space Science Institute (ISSI), Hallerstrasse 6, CH-3012 Bern, Switzerland; \href{mailto:mark.sargent@issibern.ch}{mark.sargent@issibern.ch}}
\affiliation{$^2$~Astronomy Centre, Department of Physics and Astronomy, University of Sussex, Brighton, BN1 9QH, UK}
\affiliation{$^3$~Department of Physics and Astronomy, University of Victoria, Finnerty Road, Victoria, British Columbia, V8P 1A1, Canada}
\affiliation{$^4$~Research School of Astronomy and Astrophysics, Australian National University, Canberra, ACT 2611, Australia}
\affiliation{$^5$~ARC Center of Excellence for All Sky Astrophysics in 3 Dimensions (ASTRO 3D), Australia}
\affiliation{$^6$~Max-Planck-Institut f{\"u}r Extraterrestrische Physik, Giessenbachstrasse, D-85748 Garching, Germany}
\affiliation{$^7$~Department of Physics \& Astronomy, University College London, Gower Place, London WC1E 6BT, UK}
\affiliation{$^8$~Max-Planck-Institut für Radioastronomie (MPIfR), Auf dem H\"ugel 69, D-53121 Bonn, Germany }
\affiliation{$^9$~Department of Physics and Astronomy, Oberlin College, Oberlin, OH 44074, USA}
\affiliation{$^{10}$~Centre for Astrophysics Research, University of Hertfordshire, College Lane, Hatfield AL10 9AB, UK}

\thanks{$^\star$Based on observations carried out under project number 196-14 with the IRAM 30\,m Telescope. IRAM is supported by INSU/CNRS (France), MPG (Germany) and IGN (Spain).} 

\shorttitle{The molecular gas content throughout the low-$z$ merger sequence}
\shortauthors{Sargent et al.}

\begin{abstract}
Exploiting IRAM 30\,m CO spectroscopy, we find that SDSS post-merger galaxies display gas fractions and depletion times enhanced by 25-50\%, a mildly increased CO excitation, and standard molecular-to-atomic gas ratios, compared to non-interacting galaxies with similar redshift, stellar mass ($M_{\star}$) and star-formation rate (SFR). To place these results in context, we compile further samples of interacting or starbursting galaxies, from pre-coalescence kinematic pairs to post-starbursts, carefully homogenising gas mass, $M_{\star}$ and SFR measurements in the process. We explore systematics by duplicating our analysis for different SFR and $M_{\star}$ estimators, finding good qualitative agreement in general. Molecular gas fractions and depletion times are enhanced in interacting pairs, albeit by less than for post-mergers. Among all samples studied, gas fraction and depletion time enhancements appear largest in young (a few 100 Myr) post-starbursts. While there is only partial overlap between post-mergers and post-starbursts, this suggests that molecular gas reservoirs are boosted throughout most stages of galaxy interactions, plausibly due to torque-driven inflows of halo gas and gas compression. The gas fraction and depletion time offsets of mergers and post-starbursts anti-correlate with their distance from the galaxy main sequence $\Delta({\rm MS})$, evidencing the role of star-formation efficiency (SFE) in driving the high SFRs of the strongest starbursts. Post-starbursts display the steepest dependency of gas fraction and SFE-offsets on $\Delta({\rm MS})$, with an evolving normalisation that reflects gas reservoir depletion over time. Our multi-sample analysis paints a coherent picture of the starburst-merger connection throughout the low-$z$ merger sequence. It reconciles contradictory literature findings by highlighting that gas fraction enhancements and SFE variations both play their part in merger-driven star formation.
\end{abstract}



\section{Introduction}
\label{sect:intro}

The characterisation of galaxies into two broad populations can be traced back almost a century to Hubble's tuning fork classification scheme. Although Hubble focused primarily on morphologies, we now know that many galactic properties exhibit a bimodal distribution. Broadly speaking, galaxies tend to be either blue, disk-dominated and actively star-forming, or red ellipticals with very low rates of star formation \citep[e.g.][]{strateva01, baldry04, balogh04, schawinski14}. Large spectroscopic galaxy surveys have traced this bimodal behaviour in the distribution of star-formation rates (SFRs) with stellar mass \citep[e.g.][]{brinchmann04, salim07}: SFR distributions at fixed stellar mass display a tight (0.2-0.3\,dex dispersion) `main sequence' of star-forming galaxies, complemented by the so-called `quenched' galaxy population in a long, broad tail of the SFR distribution that extends to SFRs lower by at least an order of magnitude compared to the galaxy main sequence \citep[e.g.][]{renzinipeng15, bluck16, feldmann17}.\medskip

Understanding what causes galaxies to transition from the main sequence of star-forming galaxies to the quenched population has emerged as one of extragalactic astronomy's linchpin endeavours of the last decade. One long-standing idea is that feedback from active galactic nucleus (AGN) accretion can effectively remove gas \citep[e.g.][]{dimatteo05, hopkins06}; bereft of its fuel supply, the galaxy's star formation must promptly cease.  However, several problems exist with this rapid gas expulsion-driven AGN quenching scenario. First, it is well documented that most AGN host galaxies are still actively forming stars \citep{mullaney12, rosario13, ellison16, aird18, birchall23}. Relatedly, there is no anti-correlation between AGN accretion rate and current star-formation rate \citep[e.g.][]{azadi15, stanley15}. Finally, AGN host galaxies generally still possess plentiful gas reservoirs, as measured in both the atomic \citep{fabello11, gereb15, ellison19} and molecular \citep{jarvis20, shangguan20, koss21, molina23} phases. The impact of the AGN on gas content has only been glimpsed at either very low masses \citep{bradford18, ellison19} or on nuclear (rather than global) scales \citep{izumi20, ellison21, garcia-burillo21, garcia-burillo24}.\medskip

However, the observation that AGN are both gas-rich and star-forming actually only indicates that accretion-driven feedback fails to \textit{promptly} shut down star formation.  There has been a growing body of observational evidence that the impact of AGN feedback may actually be a slow process that can be deduced from the link between quiescence and nuclear properties, such as the central concentration of stellar matter \citep[e.g.][]{bluck14, lang14, omand14, woo15}, the velocity dispersion of this central stellar component \citep{bluck16, bluck20, teimoorinia16, brownson22} or black hole mass, either measured directly \citep{terrazas16} or inferred from scaling relations \citep{piotrowska22}.  These observations suggest that it is the cumulative effect of accretion over the lifetime of the galaxy (that is captured by the total black hole mass and its concomitant relation with central velocity dispersion and bulge mass) that eventually leads to quenching, an effect observed to be in place over a range of redshifts \citep{bluck22, bluck24}.\medskip

Cosmological simulations agree: even though AGN are the mechanism by which quenching is achieved in many contemporary models, the host galaxies of AGN in these simulations are gas-rich and star-forming \citep{ward22}. And yet these same simulations show a clear connection between black hole mass and quiescence \citep{piotrowska22, bluck23}, again supporting the idea that integrated, rather than instantaneous accretion is the key to quenching. This idea of a gradual, rather than instantaneous, impact of the AGN has become known as `preventative' feedback, as it leads to an aggregate effect on the larger gaseous halo, eventually cutting off replenishment of the gas supply to the inner star-forming regions.\medskip

Although AGN may represent the primary route to slow quenching (with this argued to be the dominant timescale for the shut down of star formation, e.g. \citealp{peng15, trussler20}), it is very likely that it is not the \textit{only} route to quenching. In an influential paper, \citet{schawinski14} suggested that both slow (on Gyr time scales) and fast (time scale $\sim$10$^{-1}$\,Gyr) quenching pathways are needed to explain observations.  Modern simulations also exhibit a diversity of quenching timescales \citep[e.g.][]{rodriguez-montero19, wright19, walters22}. Whilst the slow pathway is associated with a gradual strangulation of the gas supply (i.e. consistent with preventative feedback), fast quenching is linked to a prompt removal of the galactic interstellar medium (ISM). In an AGN-related pathway that is likely less prevalent than preventative feedback, `ejective' supermassive black hole-driven processes can lead galaxy quenching \citep[e.g.][]{deugenio23, belli24} on short time scales. The existence of `post-starburst' (PSB) galaxies are further proof that a rapid quenching mode must exist. PSBs (previously also referred to as E+A or k+A) galaxies are characterised by spectra exhibiting strong Balmer absorption features from A and F type stars, and yet no emission lines, indicating recent active star formation, followed by a rapid quenching event \citep[e.g.][]{goto05, pawlik18, french21}. Although rare in the nearby universe, PSBs become increasingly common at high redshift \citep[e.g.][]{wild16, belli19, park23, setton23}. Indeed, at early times fast quenching is the only plausible quenching pathway \citep{walters22, tacchella22}, given the brevity of time since the Big Bang. The abundance of quiescent galaxies at high redshift recently uncovered by the James Webb Space Telescope \citep{carnall23} underscores the importance of understanding the fast quenching mechanism.\medskip

Studying the properties of PSB galaxies provides clues to the mechanism behind the fast quenching mechanism. It is well documented that many low-$z$ PSBs show signs of recent interactions \citep{goto05, yang08, pawlik18}. In the largest study of its kind to date, and benefitting from both high quality imaging and a carefully selected control sample, \citet{wilkinson22} have shown that $\sim$20--40 per cent of $z\,{\sim}$\,0.1 PSBs are mergers, depending on PSB definition and merger identification method\footnote{Given the imperfect performance of all merger identification methods, these fractions are likely to be lower limits, e.g. \citet{wilkinson24}.}.  These PSB merger fractions exceed the expected (control) fraction by factors of between three and 50, robustly demonstrating that PSBs are preferentially linked to mergers.  Further evidence of the link between mergers and rapid quenching comes from the excess of PSBs observed in late stage mergers \citep{ellison22, li23, ellison24}.\medskip

Having established that mergers can lead to rapid quenching (which has been a long-standing idea in the literature, e.g. \citealp{sandersmirabel96, hopkins08, schawinski14}) we must now understand why.  Is it a return to the classic `prompt gas removal' scenario linked to the enhanced star formation \citep[e.g.][]{ellison08, scudder12, bickley22} and AGN \citep{ellison11, ellison13b, bickley23, bickley24} activity known to ensue as a result of the interaction? Or is the gas reservoir prevented from continuing to form stars due to elevated turbulence \citep[e.g.][]{smercina22}? The first step towards understanding why mergers can lead to rapid quenching is thus to investigate whether interactions deplete the host galaxy's gas content, and then subsequently assess any changes in star formation efficiency (SFE).\medskip

Studies of the atomic gas content of interacting galaxies are numerous, having achieved large samples and encompassed both the pre-merger (pair) and post-merger (remnant) regimes \citep[e.g.][]{ellison15, ellison18, dutta18, dutta19, lisenfeld19, bok20, bok22, diaz-garcia20, yu22}.  Results are mixed, with evidence for both enhanced HI \citep[e.g.][]{casasola04, ellison18, dutta18, dutta19} and modest suppression \citep[e.g.][]{kaneko17, lisenfeld19, yu22}.  However, measuring the molecular gas content is perhaps more germane for the issue of understanding the quenching of star formation. In contrast to the body of work on the atomic gas content, studies of molecular gas in mergers are heavily biased to the pre-coalescence regime \citep[e.g.][]{combes94, zhu99, pan18, violino18, lisenfeld19, garay-solis23}. These works unanimously find elevated molecular gas fractions.\medskip

In the work presented here, we aim to extend these previous studies by conducting a statistical study of the molecular gas content of interacting galaxies into the post-merger regime. Moreover, by combining our sample with others in the literature, all of which have been carefully re-assessed within our own measurement methodology, we investigate the time evolution of molecular gas fractions from the pre-coalescence phase, through recent merging and into late post-coalescence. In this way, we can establish not only if, but also when, the molecular gas reservoir is impacted by the interaction, both in terms of its global gas fraction, and also its SFE.\medskip

We present our sample of post-merger galaxies and describe our CO line observations and flux measurements in Sect. \ref{sect:PMintro}. Further galaxy samples spanning a variety of interaction or starburst stages/remnants are then introduced in Sect. \ref{sect:compsampintro}. Our approach for determining shifts in the molecular gas properties of interacting and post-starburst galaxies, relative to control-matched, normal galaxies is explained in Sect. \ref{sect:controlintro}.  In Sect. \ref{sect:physquant} we derive the physical galaxy properties relevant to our analysis -- star-formation rates, stellar masses and molecular gas masses -- and compare these with other frequently used estimates in the literature. We present our results in Sect. \ref{sect:results}, before discussing their implications in Sect. \ref{sect:discussion} and summarizing our findings in Sect. \ref{sect:summary}. A $\Lambda$CDM cosmology with $\Omega_m$\,=\,0.315, $\Omega_{\Lambda}$\,+\,$\Omega_m$\,=\,1 and $H_0$\,=\,67.4 km\,s$^{-1}$\,Mpc$^{-1}$ \citep{2020A&A...641A...6P} and a \citet{chabrier03} initial mass function (IMF) are assumed throughout.

\begin{table*}
\centering
\caption{Galaxy samples used in this paper.}
\label{tab:samples}
\begin{tabular}{l @{\quad\vline\quad} ccccc}
\hline\hline \\[-2ex]
Sample & Acronym & \#Objects & Selection & Lit. source & Described in\\[0.5ex]
(1) & \multicolumn{1}{c}{(2)} & (3) & (4) & (5) & (6)\\[1ex]
\hline \hline
\multicolumn{6}{c}{\it \raisebox{-1ex}{Interacting and post-starburst galaxies}}\\[2ex]
\hline \\[-2ex]
Post-mergers & PMs & 39 & optical morphology & this paper & Sect. \ref{sect:PMintro}\\
Interacting pair galaxies & IPGs & 39 & $r_p$ \& $\Delta v$ thresholds & \citet{violino18}, & Sect. \ref{sect:IPGintro}\\
& & & & \citet{pan18} & \\
Young post-starbursts & yPSBs & 11 & optical spectral properties & \citet{rowlands15} & Sect. \ref{sect:PSBintro}\\
Mature post-starbursts & mPSBs & 31 & `E+A' spectral features & \citet{french15} & Sect. \ref{sect:PSBintro}\\
Dust lane early-type galaxies & DETGs & 17 & ETG structure, & \citet{davis15} & Sect. \ref{sect:DETGintro}\\
& & & dust lane features & & \\[1ex]
\hline
\multicolumn{6}{c}{\it \raisebox{-1ex}{Pool of control galaxies}}\\[2ex]
\hline \\[-2ex]
Extended CO Legacy Database & xCG & 488 & mass-limited (${>}10^9$\,\msun), & \citet{saintonge17} & Sect. \ref{sect:controlintro}\\
for GASS (xCOLD GASS) & & & representative sampling & & \\[1ex]
\hline\hline
\end{tabular}
\end{table*}

\section{Galaxy samples and data}

In the following sections we introduce the different galaxy samples used in the present paper. For a high-level overview, Table \ref{tab:samples} lists sample sizes, selection methodology and data sources.

\subsection{The SDSS post-merger sample}
\label{sect:PMintro}

\subsubsection{Selection of post-merger galaxies}
\label{sect:PMsel}

We select a sample of post-merger (PM) galaxies via two routes detailed in the following. In both cases, the identification of PMs is based on the presence of morphological disturbances in SDSS colour-composite images. Following \citep{ellison13b} the identification of galaxies as PMs relies on Galaxy Zoo morphological information from \citet{darg10}, and involves further visual inspection steps in SDSS $gri$ image cut-outs, plus the additional requirement that galaxies have a stellar mass estimate from the \citet{mendel14} catalog. For all further details on PM selection we refer to \citet{ellison13b, ellison15}.\medskip

The majority of PMs selected with this approach are isolated systems with distinctive shells and tidal features. (SDSS cut-out stamps for all PMs are included in Appendix \ref{appsect:PMintro_part2}.) A small fraction of our PMs are overlapping galaxies, where coalescence is at an advanced stage (e.g. PMs \#9, 34, 36 \& 39). Our fiducial PM sample consists of galaxies that have compelling evidence in their SDSS images for gravitational disturbances. Whereas this leads to a sample that is certainly incomplete, it is pure with high confidence. PMs selected via our approach are unambiguously at a later merger stage than visibly distinct pair galaxies, and are likely the remnants of gas-rich major mergers. Although it is not possible to know the exact masses of the colliding progenitors, simulations favour mass ratios ranging from of order 4:1 to approaching 1:1 equality \citep{lotz10, ji14}.\medskip

To begin, we select 24 objects from a larger sample of 93 PMs from \citet{ellison15}. Specifically, among the HI-detected galaxies in the \cite{ellison15} sample (43 objects), we identify those PMs with at least one `control galaxy' (see Sect. \ref{sect:matching} for a detailed description of our final control/reference matching strategy) that is detected in both HI and CO in the GASS \citep{catinella13} and COLD GASS \citep{saintonge11} samples, respectively. This (a) prevents a spurious overestimate of the gas content of the PM phase -- brought about by a simple restriction to HI-detected PMs without the ability to assess their gas content {\it in comparison} to non-interacting, normal galaxies with similar physical properties -- and (b) ensures a well-defined constraint on the total (atomic and molecular) gas content of PMs.\medskip

We additionally performed an {\it a posteriori} visual inspection of galaxies in the Extended CO Legacy Database for GASS (xCOLD GASS; `xCG' for short in the following) DR1 catalog \citep[see also Sect. \ref{sect:controlintro}]{saintonge17}. This is the currently largest homogeneous data base of single-dish molecular gas measurements for low-redshift galaxies. We identified a further 15 PM candidates in the xCG sample (see Fig. \ref{appfig:addedPMstamps} in Appendix \ref{appsect:PMintro_part2}). 13 of these were not included in the \citet{ellison13b} sample because they did not receive high merger vote fractions in Galaxy Zoo, despite the presence of structural features suggesting a PM state when applying identical criteria as in \citet{ellison13b}. Two additional systems received high merger vote fractions, but were placed into a binary merger catalog by \citet{darg10}, rather than the Darg et al. PM list, which served as the starting point for sample selection in \citet{ellison13b}.\medskip

Our final sample comprises 39 PMs with redshifts in the range 0.014\,${<}\,z\,{<}$\,0.053. Their physical properties are summarised in Table \ref{tab:prop}).

\begin{table*}
\centering
\caption{Summary of physical properties of post-merger galaxies.}
\label{tab:prop}
\begin{tabular}{r @{\quad\vline\quad} lccccccc}
\hline\hline \\[-2ex]
\# & SDSS-ID & R.A. & Dec. & $z_{\rm SDSS}$ & log(\mstar) & log(SFR) & 12+log[O/H] & \texttt{bptclass}\\
& & [J2000] & [J2000] & & [\msun] & [\msun/yr] &\\[0.5ex]
(1) & \multicolumn{1}{c}{(2)} & (3) & (4) & (5) & (6) & (7) & (8) & (9)\\[1ex]
\hline \hline
\multicolumn{9}{c}{\it \raisebox{-1ex}{Newly observed post-merger galaxies (IRAM program 196-14)}}\\[2ex]
\hline \\[-2ex]
 1 & 588010880366477449 & 172.628998 &  5.891780 & 0.034933 &  10.28$_{-0.18}^{+0.11}$ &   0.06$_{-0.13}^{+0.16}$ &  --- &  2\\
 2 & 587739132428157191 & 237.067993 & 25.527000 & 0.041730 &  10.14$\pm$0.17 &   0.35$_{-0.15}^{+0.20}$ &  --- &  1\\
 3 & 587736543096799321 & 226.324997 &  8.153550 & 0.039118 &  10.11$_{-0.15}^{+0.22}$ &   1.20$_{-0.14}^{+0.17}$ & 8.85 &  3\\
 4 & 587739382067822837 & 234.005997 & 25.551001 & 0.036157 &  10.55$_{-0.25}^{+0.12}$ &   0.75$\pm$0.13 &  --- &  1\\
 5 & 587736808845082795 & 223.057007 & 12.060700 & 0.052500 &  10.49$_{-0.19}^{+0.13}$ &   0.80$_{-0.16}^{+0.20}$ & 8.77 &  1\\
 6 & 588010880378404942 & 199.990997 &  5.807880 & 0.021339 &  10.52$_{-0.21}^{+0.08}$ &   0.29$\pm$0.08 &  --- &  3\\
 7 & 588016840167653827 & 117.375999 & 18.957800 & 0.047237 &  10.31$_{-0.17}^{+0.13}$ &   0.26$_{-0.15}^{+0.18}$ &  --- &  1\\
 8 & 587739159266590725 & 153.544006 & 34.342999 & 0.037583 &  10.07$_{-0.24}^{+0.10}$ &   1.09$_{-0.14}^{+0.17}$ & 8.89 &  1\\
 9 & 587739305286303750 & 172.322006 & 35.577301 & 0.034566 &   10.01$_{-0.12}^{+0.03}$ &   0.46$_{-0.12}^{+0.13}$ &  --- &  1\\
10 & 587741533323526200 & 173.781006 & 29.891001 & 0.046238 &  10.72$\pm$0.18 &   1.09$_{-0.14}^{+0.17}$ & 8.97 &  1\\
11 & 587739507154288785 & 185.065002 & 33.660801 & 0.021574 &  10.40$_{-0.21}^{+0.13}$ &   0.04$_{-0.13}^{+0.15}$ &  --- &  1\\
12 & 587724232100282512 &   8.777040 & 14.354800 & 0.038013 &  10.29$\pm$0.09 &   0.54$_{-0.14}^{+0.17}$ & 9.01 &  1\\
13 & 587726101750546619 & 210.550995 &  4.585730 & 0.040307 &  10.93$_{-0.18}^{+0.07}$ &   0.12$_{-0.10}^{+0.11}$ &  --- & -1\\
14 & 587728881418502205 & 164.380005 &  5.698730 & 0.053201 &  10.93$_{-0.07}^{+0.08}$ &   0.49$_{-0.13}^{+0.15}$ & 9.12 &  3\\
15 & 587729158966345769 & 194.587997 &  4.885650 & 0.036125 &  10.44$_{-0.08}^{+0.10}$ &   1.12$_{-0.14}^{+0.16}$ & 9.33 &  3\\
16 & 587732577774469219 & 163.518005 &  6.725390 & 0.026585 &  10.45$_{-0.15}^{+0.07}$ &   0.12$_{-0.12}^{+0.14}$ &  --- &  3\\
17 & 587734622163763258 & 120.866997 & 25.102699 & 0.027584 &  10.20$_{-0.19}^{+0.15}$ &   0.98$\pm$0.11 & 8.99 &  3\\
18 & 587734891683119143 & 188.681000 &  9.004710 & 0.043057 &  10.77$_{-0.24}^{+0.08}$ &   1.16$_{-0.14}^{+0.17}$ & 9.08 &  1\\
19 & 587735347489341442 & 156.546997 & 12.573800 & 0.030963 &  10.19$_{-0.16}^{+0.06}$ &   0.12$_{-0.13}^{+0.15}$ & 8.90 &  1\\
20 & 587735349086912675 & 126.597000 &  8.423770 & 0.045945 &  11.00$\pm$0.01 &   0.12$\pm$0.14 &  --- &  5\\
21 & 587738409251635215 & 150.264008 & 11.461400 & 0.036294 &  10.38$_{-0.12}^{+0.19}$ &   1.18$\pm$0.08 & 9.04 &  1\\
22 & 587738410860806225 & 146.826996 & 12.158500 & 0.047507 &  10.85$_{-0.25}^{+0.10}$ &   0.74$_{-0.14}^{+0.18}$ &  --- &  1\\
23 & 588010359073931336 & 151.835999 &  4.079310 & 0.028591 &  10.48$_{-0.28}^{+0.13}$ &   0.44$_{-0.15}^{+0.19}$ & 9.13 &  1\\
24 & 588010360148656136 & 154.000000 &  4.954760 & 0.032029 &  10.12$\pm$0.21 &   0.94$_{-0.14}^{+0.17}$ & 8.92 &  1\\[1ex]
\hline
\multicolumn{9}{c}{\it \raisebox{-1ex}{Post-merger galaxies from xCOLD GASS}}\\[2ex]
\hline \\[-2ex]
25 & 587724199359021099 &  32.889801 & 13.917100 & 0.026510 &  10.80$_{-0.06}^{+0.01}$ &   0.78$_{-0.15}^{+0.19}$ & 9.16 &  1\\
26 & 587744638561484901 & 119.332001 & 11.206100 & 0.046400 &  10.72$_{-0.12}^{+0.01}$ &   0.21$_{-0.13}^{+0.14}$ &  --- &  3\\
27 & 588016891172618376 & 147.936996 & 35.622101 & 0.026990 &  10.64$_{-0.11}^{+0.07}$ &   0.61$_{-0.15}^{+0.20}$ & 8.97 &  1\\
28 & 587738617018122300 & 149.714005 & 32.073101 & 0.027030 &  10.74$_{-0.16}^{+0.14}$ &   0.16$_{-0.13}^{+0.16}$ &  --- &  1\\
29 & 587732701792174316 & 160.337997 &  6.278970 & 0.033920 &   9.92$_{-0.16}^{+0.20}$ &   0.12$_{-0.12}^{+0.13}$ & 8.97 &  1\\
30 & 587726032242016326 & 165.134995 &  2.116060 & 0.039390 &  11.21$\pm$0.02 &   0.77$\pm$0.01 &  --- &  4\\
31 & 588017992295972989 & 194.268997 & 10.620700 & 0.046280 &  11.18$\pm$0.02 &   0.35$\pm$0.11 &  --- &  4\\
32 & 587726016150110325 & 196.962006 &  3.194670 & 0.038550 &  11.03$\pm$0.03 &   0.81$\pm$0.15 &  --- &  4\\
33 & 587741722826244233 & 201.345001 & 27.249100 & 0.034480 &  10.31$_{-0.17}^{+0.11}$ &   0.30$_{-0.13}^{+0.16}$ &  --- &  1\\
34 & 588017569779744827 & 203.100998 & 11.106400 & 0.031440 &  10.09$_{-0.18}^{+0.16}$ &   1.16$_{-0.13}^{+0.16}$ & 8.95 &  1\\
35 & 587722984440135708 & 215.811005 &  0.978339 & 0.040070 &  10.21$_{-0.12}^{+0.16}$ &   0.64$_{-0.12}^{+0.14}$ & 9.16 &  1\\
36 & 587736585506062428 & 227.020004 & 34.387199 & 0.045490 &  10.93$_{-0.19}^{+0.11}$ &   1.40$_{-0.14}^{+0.17}$ &  --- &  2\\
37 & 587727221945466971 & 348.381027 & 14.327429 & 0.039220 &  10.52$_{-0.16}^{+0.12}$ &  -1.58$_{-0.32}^{+0.89}$ &  --- & -1\\
38 & 587739406784790531 & 192.170929 & 34.477608 & 0.014178 &   9.06$_{-0.12}^{+0.20}$ &  -0.03$_{-0.11}^{+0.13}$ & 8.71 &  1\\
39 & 587735662082064461 & 139.977260 & 32.933280 & 0.049160 &  11.10$_{-0.21}^{+0.11}$ &   1.40$_{-0.14}^{+0.17}$ & 9.20 &  3\\[1ex]
\hline
\multicolumn{9}{c}{\it \raisebox{-1ex}{Interacting pair galaxy observed during IRAM program 196-14}}\\[2ex]
\hline \\[-2ex]
40 & 587724232641937419 &  20.011000 & 14.361800 & 0.031044 &  10.69$_{-0.08}^{+0.14}$ &   0.85$_{-0.14}^{+0.18}$ &  --- &  3\\[1ex]
\hline\hline
\end{tabular}
\tablecomments{ID numbers in col. 2 are from SDSS DR7. Oxygen abundances assume the \citet{kewleydopita02} metallicity calibration, stellar masses (\mstar) and star-formation rates (SFR) a \citet{chabrier03} IMF. For galaxies with \texttt{bptclass} flag values of 1--3 (galaxies dominated by star formation (\texttt{bptclass}\,=\, 1/2), or composite objects) stellar mass measurements are taken from \citet{mendel14} and star-formation rates are `hybrid' SFRs combining GALEX and WISE photometry. For AGN hosts (\texttt{bptclass}\,=\, 4/5) and unclassified spectral types (\texttt{bptclass}\,=\,-1) SFRs and stellar masses are taken from \citet{salim16}, after correction for small systematic offsets as described in Sect. \ref{sect:physquant}.}
\end{table*}

\subsubsection{IRAM 30\,m observations and data reduction}
\label{sect:redu}
For the 15 PMs identified in xCG DR1, the $^{12}$\COone\ transition (simply denoted ``\COone" henceforth) is detected for all but one galaxy, with line signal-to-noise ratios ranging from $S/N$\,=\,4.3 to 104 (median $S/N\,{\sim}$\,13). For the object with SDSS ObjID 587727221945466971 (PM \#37) we set a 3\,$\sigma$ upper limit on the line flux by scaling the channel rms noise by the square root of the ratio between the channel width (20\,km/s) and the default 300\,km/s velocity width we adopt for non-detections in all galaxy samples studied in this paper (see further justification and discussion in Appendix \ref{appsect:meashomog}). The \COtwo\ transition is detected toward 7 of the PMs from xCG, with the remaining 8 galaxies lacking \COtwo\ follow-up (such that no line flux upper limit can be inferred for them). $S/N$ values for the 7 \COtwo\ detections vary between 4.3 and 160 (median $S/N\,{\sim}$\,12).\medskip

We observed the 24 additional PMs with the IRAM 30\,m telescope through program 196-14 (PI: Sargent) for a total project time (including calibration overheads) of 49.6 hours from January 6 to March 18, 2015. Integration times for individual PMs ranged from 0.4 to 3.2 hours, with an average time per target of 1.42 hours. In keeping with our control-matching approach that utilizes non-interacting galaxies in the xCG sample as a pool of normal reference galaxies (Sect. \ref{sect:matching}), our observing and data reduction strategy closely follows that of the (x)CG survey \citep{saintonge11, saintonge17}. All observations were executed in wobbler-switching mode with a 120$''$ throw between the ON and OFF positions. CO spectroscopy was carried out with the Eight Mixer Receiver \citep[EMIR;][]{carter12} and the Fourier Transform Spectrometer (FTS) as backend. This combination allows for simultaneous observations in two receiver bands and provides, within a given receiver band, 7.78\,GHz of instantaneous bandwidth per linear polarisation in two separate sidebands. Given the narrow redshift interval (0.021\,${<}\,z\,{<}$\,0.053) from which these 24 PMs are drawn, a single tuning centred at 109.742\,GHz in the 3\,mm band (E090) was sufficient to cover the $^{12}$\COone\ transition for all, and the $^{13}$\COone\ transition for three quarters of our targets. We used parallel observations in the 1.3\,mm band (E230) to additionally follow up the $^{12}$\COtwo\ line for all targets. Within the allocated observing time we detected the $^{12}$\COone\ transition for all PM galaxies, and $^{12}$\COtwo\ for 23 of the 24 targets.\medskip

We reduced the IRAM 30\,m spectra with the \texttt{CLASS} package following the same steps as adopted in \citet{saintonge11, saintonge17} for (x)CG galaxies. The final spectra with 20\,km/s velocity binning are shown for each PM galaxy in Fig. \ref{appfig:spectra}. Brightness temperatures on the $T_a^*$ scale and in units of Kelvin were converted to main-beam temperatures in Jansky units using the Kelvin-to-Jansky conversion factor 3.906\,$F_{\rm eff}/A_{\rm eff}$, where the frequency-dependent exact values of the forward efficiency $F_{\rm eff}$ and the aperture efficiency $A_{\rm eff}$ were derived by linear interpolation of the data tabulated for the upper and lower end of EMIR band E090 and E230 on the IRAM 30\,m website\footnote{\url{http://www.iram.es/IRAMES/mainWiki/Iram30mEfficiencies}}. To extract velocity-integrated \COone\ and \COtwo\ line fluxes from the EMIR spectra we follow the approach adopted by \citet{saintonge17} for the xCG control galaxies. Further details on our line flux measurements for PMs, and on the aperture correction methodology we adopt to convert observed fluxes to galaxy-integrated line fluxes, are provided in Appendix \ref{appsect:PMintro_part2}. The \COone\ and \COtwo\ line fluxes for all PMs can be found in the associated Table \ref{tab:fluxes}.

\subsection{Comparison samples -- interacting SDSS galaxies with IRAM 30m CO follow-up}
\label{sect:compsampintro}

The samples of comparison galaxies included here are in a specific evolutionary phase, often also potentially linked to a galaxy interaction, such that a comparison of their gas content to that of PMs can provide insight into the time evolution of interstellar medium (ISM) conditions throughout the interaction process. As for the PM sample, we will also reference the comparison galaxies to appropriately chosen control galaxies (Sect. \ref{sect:controlintro}) to quantify how they differ from the `normal', non-interacting population.\medskip

All comparison galaxies are covered by the same SDSS imaging and spectroscopic surveys as the PMs and xCG control galaxies. Similarly, all comparison galaxies also have CO spectra from the IRAM 30\,m single-dish telescope. In Appendix \ref{appsect:meashomog}, we summarize how the CO line flux measurements for galaxies in all comparison samples are homogenised. The aperture corrections derived as outlined and tested in Sect. \ref{appsect:apcorr} are applied to the galaxies in all comparison samples. Together with the control-matching approach, which uses the same pool of xCG control galaxies for all comparison samples, the homogeneous nature of the multi-wavelength ancillary data and of the molecular gas measurements -- underpinning the physical properties inferred in Sect. \ref{sect:physquant} -- minimises the potential for bias or spurious evolutionary trends.\newpage

\subsubsection{Interacting Pair Galaxies -- IPGs}
\label{sect:IPGintro}
We complement our post-merger data with a sample of interacting galaxies in the pre-coalescence phase, drawn from the analyses of \citet{violino18} and \citet{pan18}. The selection criteria adopted in these studies single out members of close kinematic pair systems, with a projected separation of $r_p\,{\leq}$\,30\,kpc\,$h^{-1}$ plus a velocity offset -- measured in SDSS spectra -- of $\Delta v\,{\leq}$\,300\,km/s in \citet{violino18}, and somewhat larger projected and velocity space separations in \citet[][$r_p\,{\leq}$\,50\,kpc\,$h^{-1}$ and $\Delta v\,{\leq}$\,500\,km/s, respectively]{pan18}. As the majority of the selected interacting pair galaxies (IPGs) don't have strongly perturbed morphologies \citep[see, e.g., Figs. 2 \& 3 in][]{violino18}, and with a median projected separation $r_p$ of $\sim$40\,kpc, it is likely that most are caught either prior to the first encounter, or soon after the first pericentric passage.\medskip

The IPGs in \citet{pan18} have \COone\ spectra from a mixture of JCMT and IRAM 30\,m observations, with the latter coming from the xCG data base. For our analysis we take all pair galaxies with IRAM 30\,m data in the \citet{pan18} sample, with the exception of one object (J130750.80+031140.7; SDSS ObjID 587726016150110325). The high mass contrast of 107:1 between this galaxy and its nearest neighbour at $r_p$\,=\,43.5\,kpc points to its tidal distortion features originating from a preceding interaction with a larger object, and we therefore include it in our PM sample as one of the 15 objects identified in the xCG catalog (\#32; see Sect. \ref{sect:PMsel}). The IPGs in \citet{violino18} were specifically selected for a targeted follow-up campaign at the IRAM 30\,m telescope, with an observing strategy similar to that used for the xCG project. To the 26 (11) IPGs from \citet{pan18} \citep{violino18} we add the pair galaxy with SDSS ObjID 587724232641937419 from our own IRAM 30\,m program (see Appendix \ref{appsect:ourpair}), as well as SDSS ObjID 587739114697195849 from the xCG data base, which is part of a system of two galaxies during final coalescence and was not included in the pair study of \citet{pan18}. In total, our IPG sample consists of 39 galaxies, and at a threshold of $S/N$\,=\,3, the \COone\ transition is detected toward all of these.

\subsubsection{Young/Mature Post-Starbursts -- yPSBs \& mPSBs}
\label{sect:PSBintro}
\citet{french15} and \citet{rowlands15} presented two samples of SDSS-selected post-starburst galaxies with CO spectra from the IRAM 30\,m telescope. Post-{\it starbursts} and post-{\it mergers} are not mutually inclusive populations, nor is either of the populations simply a subset of the other. There is, however, substantial overlap between PMs and PSBs, as evidenced by the substantial fraction of morphologically identified mergers in PSB samples (several 10s of percent at least or even approaching 100\%, see e.g. \citealp{sazonova21, wilkinson22}, and references therein), and by kinematical studies \citep[e.g.][]{otter22}. At the same time, recent work has revealed a higher incidence by up to a factor 50 of PSBs among PMs \citep{ellison22, li23} compared to normal control galaxies, with PMs in the first 0.5\,Gyr since coalescence showing the highest excess of PSBs \citep{ellison24}. By  including both post-{\it starburst} and post-{\it merger} galaxies in our analysis we can thus gain insight into the starburst-merger connection. At low redshift, galaxy mergers may be particularly conducive to starburst activity and gas consumption due the lower initial, pre-interaction ISM turbulence in these systems compared to higher redshift galaxies \citep[e.g.,][]{hopkins09, fensch17, patton20}. In this context, measurements of molecular gas fractions and star-formation efficiencies for both PMs and PSBs directly address the role of the gas reservoir in merger-related star-formation quenching.\medskip

\citet{french15} select PSBs using SDSS galaxy-integrated optical fibre spectra; the presence of strong stellar Balmer absorption lines (characterised via the Lick H$\delta$ index) serves as an indicator of recent starburst activity, and low nebular emission line strength (as constrained by the equivalent width of the H$\alpha$ line) as evidence of low on-going star formation. The galaxies selected for CO spectroscopy from an initially larger pool of `E+A' PSBs are bright in the near- and mid-IR ({\it Spitzer} or {\it WISE}), and were subsequently targeted for photometric and spectroscopic observations with {\it Herschel} (see \citealp{french15} and \citealp{smercina18} for details). From the 32 galaxies with \COone\ data in \citet{french15} we exclude object EAH01, for which resolved \COtwo\ imaging with ALMA in \citet{french18b} showed that the IRAM 30\,m molecular line detection is in fact associated with a companion galaxy. The 31 remaining PSBs from \citet{french15} span a broad range of ages since burst onset, varying from 0.14 to 1.84\,Gyr as per the multi-component stellar population synthesis in \citet{french18a} who model the PSB spectra with an old population plus up to two superimposed burst episodes. 17/31 PSBs in the sample have \COone\ lines detections with $S/N\,{\geq}$\,3 (16 detections were already reported by \citealp{french15}, with one further galaxy exceeding the detection threshold upon homogenisation of the $S/N$ estimates across all samples -- see Appendix \ref{appsect:meashomog}).\medskip

As in \citet{french15}, the selection of PSB galaxies in \citet{rowlands15} relies on two characteristic spectral features; the Balmer absorption line strength plus the 4000\,\AA\ break strength, which are combined and re-expressed in form of a new set of spectroscopic indices via the principle component (PC) analysis described in \citet{wild07}. From the locus in PC-space which contains the evolutionary tracks of ageing starbursts, \citet{rowlands15} picked 11 PSBs spanning a representative range of burst ages up to $\sim$0.6\,Gyr, and with typical H$\alpha$ luminosities for their starburst age. At a threshold of $S/N$\,=\,3, the \COone\ transition is detected toward all 11 PSBs, one of these having been upgraded to a detection as a consequence of our homogenisation of $S/N$ estimates described in Appendix \ref{appsect:meashomog}.\medskip

\citet{french18a} re-calculated ages for the \citet{rowlands15} PSBs using the same approach as adopted for their `E+A' PSBs above. This allows for a self-consistent age comparison between the two PSB samples (see Tables 1 \& 3 in \citealp{french18a}). The median age since the start of the burst episode is 0.225\,Gyr for the \citet{rowlands15} PSBs ($\sim$40\% lower than the 0.394\,Gyr median age based on the original age estimates in \citealp{rowlands15}). The median PSB age of the \citet{french15} sample -- estimated on the 21 galaxies with age measurements -- is 1.375\,Gyr. In the following we will thus refer to these two samples as `young' and `mature' post-starbursts (yPSBs \& mPSBs, respectively). There is little overlap in burst ages between the yPSB and mPSB samples, with 75\% of the PSBs in the \citet{french15} sample being older than all but just one of the post-starbursts from \citet{rowlands15}. yPBSs and mPSBs are also noticeably different in terms of their SDSS spectral types, with the vast majority of the latter being classified as classified as AGN/LINER, while 64\% (36\%) of yPSBs fall in the star-forming (AGN/LINER or composite) category.

\subsubsection{Dust Lane Early-Type Galaxies -- DETGs}
\label{sect:DETGintro}
As a final comparison sample we include the Dust lane Early-Type Galaxies (DETGs) from \citet{davis15}, which span Hubble types from elliptical to bulge-dominated Sa-type spirals. Joint analysis of these objects in \citet{kaviraj13} and \citet{davis15} -- based on gas and dust measurements and multi-wavelength detections from the far-ultraviolet to the far-IR -- points to minor mergers\footnote{Based on our visual assessment of SDSS images of DETGs -- analogous to the selection approach for PMs outlined in Sect. \ref{sect:PMsel} -- we would have classified $\sim$50\% as post-mergers, i.e. as convincing cases of merger-induced features. This fraction is likely a lower bound as the shallow SDSS imaging is ill-suited for the selection of complete post-merger samples \citep[e.g.,][]{wilkinson24}.} with gas-rich, low-mass satellites (median mass ratio of 1:40) being the origin and driver of their ISM content and star-formation activity. These galaxies were drawn from a statistically complete sample of $\sim$350 massive, early-type SDSS galaxies, with dust lane features originally identified via the Galaxy Zoo project \citep{kaviraj12}. Full multi-wavelength coverage was available for a sub-sample of 23 DETGs compiled by \citet{kaviraj13}. The 17 DETGs with the largest dust masses were subsequently observed with the IRAM 30\,m telescope, and 15 detected with $S/N\,{\geq}\,3$.

\subsection{Pool of control galaxies and control-matching procedure}
\label{sect:controlintro}

{\it Control} galaxies enable us to study the impact of the merging process, or of historical starburst activity, by referencing the gas content of PMs and the other comparison samples to that of `normal', non-interacting galaxies with similar global properties (the details of the control matching procedures are described in Sect. \ref{sect:matching}). This approach minimizes systematics that could otherwise arise in such comparisons due to different sample selection strategies.

\subsubsection{Non-interacting control galaxies from xCOLD GASS}

Through its DR1 catalog\footnote{\url{http://www.star.ucl.ac.uk/xCOLDGASS/}} the xCOLD GASS project has provided \COone\ spectra for 532 SDSS galaxies \citep{saintonge11, saintonge17}. A subset of 336 (34) xCG galaxies additionally have \COtwo\ spectra from the IRAM 30\,m (APEX) telescope, with 29 objects having been targeted by both the IRAM 30\,m telescope and APEX. In this paper we focus mostly on \COone\ line flux measurements, with the exception of Appendix \ref{appsect:apcorr}, where we used \COtwo\ data from APEX and the IRAM 30\,m telescope to test the validity of our aperture correction factors, and Sect. \ref{sect:PM_r21} where we will compare the relative excitation of the \COtwo\ transition of PMs with the excitation level of control-matched xCG galaxies.\medskip

\citet{saintonge11} and \citet{saintonge17} detail the xCG observing strategy and sample selection. The xCG sample spans redshifts 0.01\,${<}\,z\,{<}$\,0.05 and includes galaxies with stellar mass \mstar\,${\gtrsim}\,10^{9}\,\msun$. It covers a broad range of galaxy star-formation activity, from starburst galaxies with high specific SFR that lie above the main sequence of star-forming galaxies to quiescent galaxies with low SFR. Given the xCG observing strategy, whereby on-source integration is stopped as soon as a gas fraction threshold of 2.5\% (assuming a Milky-Way-like CO-to-H$_2$ conversion factor) is reached, CO non-detections are more prevalent among xCG galaxies with low SFRs. At a detection threshold of $S/N$\,=\,3, 199 xCG galaxies have a \COone\ line flux upper limit, rather than a line detection.\medskip

The xCG sample contains a small number of interacting galaxies: 15 visually identified PMs (see Sect. \ref{sect:PMsel}) and 27 galaxies identified as being part of an interacting galaxy pair (see Sect. \ref{sect:IPGintro}). We exclude these when defining our pool of non-interacting control galaxies. We also exclude one galaxy (SDSS ObjID 588010359070458037), where the available imaging and spectroscopy does not allow us to unambiguously classify this object as a pair or an isolated galaxy associated with the nearby object due to projection effects. After accounting for the fact that PM \#3 (SDSS ObjID 587736543096799321; see Sect. \ref{sect:PMintro}) has been observed by both xCG and our IRAM 30\,m program, the pool of control galaxies thus consists of 488 objects or $\sim$92\% of the xCG DR1 sample.

\subsubsection{Control matching and calculation of offset distributions}
\label{sect:matching}
Selection of suitable control galaxies is either done in the 2-D space of redshift and stellar mass, or in 3-D with star-formation rate as an additional parameter. We require a minimum of $N_{\rm min}$\,=\,5 control galaxies to be matched to each PM (or comparison) galaxy and adopt the following initial search intervals: $\Delta{z}$\,=\,0.01, $\Delta{{\rm log}(\mstar)}$\,=\,0.1\,dex and $\Delta{{\rm log(SFR)}}$\,=\,0.1\,dex. If fewer than $N_{\rm min}$ galaxies lie within in these default search intervals, we iteratively grow the search space by 5\% in each iteration until at least $N_{\rm min}$\,=\,5 control galaxies are found. In the narrow redshift range spanned by all PM (or comparison) galaxies, evolutionary effects for the galaxy properties for which we wish to compute offset distributions are much less important than variations with \mstar\ and SFR. Consequently, if the xCG sample contains a sufficient number of galaxies with matching \mstar\ and SFR, but if at first $N\,{<}\,N_{\rm min}$ due to a lack of control galaxies within the initial $\Delta{z}$-range, we increase only the width of the redshift search interval. If this fails to return $N\,{\geq}\,N_{\rm min}$ control galaxies we also iteratively increase the \mstar\ and SFR search intervals as outlined above. A PM (or comparison) galaxy is considered unmatchable and discarded from analysis\footnote{In practice this never occurs during our core analysis, which uses the `best-estimate' \mstar\ and SFR values of Sect. \ref{sect:physquant}, thanks to sufficient overlap in the \mstar\ and SFR distributions of all interacting galaxy samples with the corresponding distributions for xCG galaxies (see Fig. \ref{fig:MSplane} for PMs and Appendix \ref{appsect:compgal_paramspace} for all other interacting galaxy types). For our analysis of systematics, where we test the consistency of sample offsets when adopting alternative \mstar\ and SFR estimates from the literature, it occasionally happens that individual interacting galaxies are excluded due to a lack of control galaxies with matching \mstar\ and/or SFR.} if forming a control set of at least $N_{\rm min}$ xCG galaxies is not possible even after a doubling the initial search intervals in all available dimensions.\medskip

For the control matching itself we implement a variation of the procedure in \citet{ellison11}, which consisted of the following two-step process: (a) calculation of the median property of the control galaxies and then (b) determination of the offset of the galaxy of interest from this median. Our alternative approach here is to instead first pair a given PM (or comparison) galaxy and each of the $N$ control galaxies one-by-one, and calculate the corresponding $N$ individual offset values. These values constitute an $N$-vector of offsets for each PM (or comparison) galaxy. The second step is to draw for all PM (or comparison) galaxies -- in random order but initially without replacement -- a single value from each of the variable-length $N$-vectors and with this spectrum of offset values construct an individual realisation of the distribution of, e.g., gas fraction offsets for the entire PM (or comparison) sample. We repeat this process until we have drawn each of the $N$ offset measurements for a given PM (or comparison) galaxy at least once. The maximal number of repetitions is therefore determined by the object which has the largest number of control-matched galaxies assigned to it. For galaxies with a smaller control sample a given offset value can be drawn more than once.\medskip

Our modification of the \citet{ellison11} control-matching approach allows us to (i) associate an uncertainty to each point in the final offset distribution of a given sample (and hence also to the measurement of, e.g., the median gas fraction offset of that sample), and (ii) take into account CO non-detections (which lead to upper limits on the individual galaxy gas fraction or depletion time estimates) in the control sample using survival analysis. The latter is particularly relevant when matching PMs or the comparison samples with the xCG control galaxies based on redshift and \mstar\ only, as this can involve samples spanning quite distinct SFR-ranges. In the following we utilize two survival analysis algorithms: the \citet{kaplanmeier58} product limit estimator \citep[KM; implemented as described in][]{feigelsonnelson85} for distributions subject to one-sided censoring, and the \citet{schmitt85} estimator \citep[implemented as described in][]{sargent10} when double-sided censoring is at play. Single-sided censoring occurs during our control matching when a PM or comparison galaxy with a CO-detection is paired with an undetected xCG control galaxy; the resulting upper limits on the gas fraction or depletion time of the control galaxy translate to a lower limit (right-sided censoring) for the offset of the PM (or comparison) galaxy with respect to the control galaxy in question. For the comparison samples of mPSBs and DETGs, which include some CO non-detections, the offset distributions can be doubly censored (i.e., lower and upper limits are present).\medskip

In adopting the KM and \citet{schmitt85} estimators we assume that so-called `random censoring' applies to the offset measurements, i.e. the censored offset measurements (known only in the form of upper or lower limits) do not preferentially occur at specific percentiles of the distribution of offset values. The cumulative distribution function of offset values reconstructed via the two estimators changes value at each uncensored data point, by an amount determined by the fraction of objects (counting both objects with censored and uncensored measurements) in the sample that have been placed by the estimators in the interval between two adjacent uncensored measurements. Preferential clustering of censored measurements at specific values would thus lead to visible jumps in the reconstructed cumulative distribution function. The absence of any conspicuous features of this kind in the distribution functions for PMs in Sect. \ref{sect:results} provides reassurance that our assumption of random censoring is in general appropriate. We can intuitively understand this as the consequence of dealing with a higher-order metric, where initially potentially quite similar upper limits on molecular gas mass are scrambled to a much wider range of values when combined with other physical properties (to infer gas fractions or depletion times), with these properties then used to compute offsets relative to control galaxies.

\begin{figure*}
\centering
\includegraphics[width=\textwidth]{f1.ps}
\caption{Comparison between measures of obscured star-formation activity (panels {\it b}/{\it c}) and between obscured and unobscured or dust-corrected star-formation rate (SFR) estimates (panel {\it a}). Large symbols -- post-mergers; small symbols -- xCOLD GASS galaxies. In all panels dark (light) colours are used for star-forming (AGN/composite) spectral types. For galaxies with open symbols no SDSS spectral type is available. Dashed (dotted) lines are drawn at a 2:1 (10:1) offset from the 1$\div$1 locus (solid line).\newline
Panel {\it (a)}: obscured SFR estimate SFR$_{\rm IR(W3/4)}$ vs. GALEX UV SFR-estimate (blue) and dust-corrected SFR derived from SDSS fiber spectra (red). SFR$_{\rm IR(W3/4)}$ is the `best' available star-formation rate from WISE photometry, averaging the estimates from the CE01 and DH02 template libraries (see Sect. \ref{sect:SFRderiv} for details). Panel {\it (b)}: comparison between far-IR (42-122\,$\mu$m) luminosities inferred via bolometric corrections from single-band WISE photometry and directly from IRAS 60+100\,$\mu$m fluxes. Panel {\it (c)}: relation between the IR-luminosities derived from W3 and W4, using the \citet[][DH02 -- blue]{dalehelou02} and \citet[][CE01 -- red]{charyelbaz01} template libraries.
\label{fig:SFR_IRnUV}}
\end{figure*}

\section{Intrinsic galaxy properties: SFR, \mstar\ and $M({\rm H_2})$}
\label{sect:physquant}

In this section we derive galaxy-integrated star-formation rates, stellar masses and molecular gas masses. These are the basis for calculating the molecular gas depletion times ($\tau_{\rm depl.}\,{\equiv}\,M({\rm H_2})$/SFR) and molecular-to-stellar mass ratios $M_{\rm gas}/\mstar$ (``\fgas", in the following), which we aim to compare between `normal' control galaxies drawn from the xCG survey and PMs, as well as all other comparison samples from Sect. \ref{sect:compsampintro}.\medskip

Overall, the galaxies in our samples of interacting/starbursting galaxies and in the xCG reference sample span several orders of magnitude in star-formation rate and stellar mass, as well as several spectral types (e.g., AGN-host/composite vs. star-forming). Galaxy interactions in particular can trigger a variety of processes with emission signatures that may bias stellar mass and/or SFR estimates \citep[e.g.][]{cardoso17, buchner24}. A comparative study like ours can address this by either (i) using SFR and \mstar\ estimates based on diagnostics tailored to be accurate for a specific galaxy type or evolutionary phase, or (ii) adopting a single, one-size-fits-all approach based on the same underlying observational tracers for all galaxies. (i) bears the risk of introducing systematic offsets due to methodological differences, while (ii) guarantees overall methodological consistency, but may not provide robust measurements for galaxies in some specific evolutionary stages. SFR and \mstar\ measurements play a central role in this paper for characterising galaxy physical properties (including CO-to-H$_2$ conversion factor estimates, see Sect. \ref{sect:XCO}), and for the control-matching of interacting with normal galaxies (Sect. \ref{sect:matching}). We therefore repeat the key steps of our analysis using several representative and commonly used SFR and \mstar\ measurements for SDSS galaxies. Appendix \ref{appsect:altSFRnMstell} presents the different literature catalogs we used for this assessment of systematic uncertainties. This section introduces our `best-estimate' SFR and \mstar\ values for all galaxies, and the statistical corrections we apply to compensate for remaining systematic shifts where necessary.

\subsection{Star-formation rate measurements}
\label{sect:SFRderiv}

\subsubsection{SFRs for star-forming spectral types}
\label{sect:SFRderiv_SFG}
To directly probe both unobscured and obscured activity for the full range of different galaxy types included in our analysis, we infer total SFRs via the hybrid UV+IR recipes of \citet{hao11}. \citet{leslie18} have also shown that the hybrid SFR recipes in \citet{hao11} eliminate inclination-related biases for normal disk galaxies -- which are numerous in the xCG sample -- when implemented with IR-luminosities estimated from mid-IR (MIR) photometry. We adopt the hybrid SFR estimate for star-forming PMs and comparison/reference galaxies \citep[\texttt{bptclass}\,=\,1 \& 2 in the MPA-JHU catalog\footnote{Via the \texttt{bptclass} flag the MPA-JHU catalog distinguishes between galaxies dominated by star formation, composite objects and galaxies with a substantial AGN contribution. The dividing lines adopted to separate these three categories in the \citet[BPT]{baldwin81} diagram are from \citet{kauffmann03} and \citet{kewley01}.};][]{brinchmann04}. In Sect. \ref{sect:SFRderiv_others} we describe the SFR measurements for AGN hosts and composite spectral types.\medskip

To derive the unobscured (obscured) SFR contribution we cross-correlated the SDSS positions with GALEX GR6/7 (unWISE) photometry obtained through the MAST/GalexView server (from the \citealp{lang16} forced photometry catalog). For GALEX we selected far- and near-UV (FUV \& NUV) flux measurements from the deepest possible coverage maps where this photometry passed all of our quality checks (ordered by increasing depth, the imaging surveys included in the GR6/7 data release are: AIS, MIS, GII, DIS). All GALEX fluxes were corrected for Galactic extinction \citep{schlegel98} and colour $K$-corrected to the appropriate rest-frame wavelength based on the UV spectral slope determined from the FUV and NUV fluxes. For galaxies detected in only one GALEX filter, we applied a colour $K$-correction based on the median UV spectral slope of the sample. For both GALEX and unWISE we visually screened the position-based cross-identifications for overdeblended/ambiguous matches and manually consolidated these in cases where this was necessary. Adoption of SFR estimates from existing SDSS catalogs (see, e.g., Appendix \ref{appsect:altSFRnMstell}), without this visual consolidation, would have led to less accurate results particularly for interacting galaxies with their at times complex structural features. Fig. \ref{fig:SFR_IRnUV}a compares the dust-obscured star-formation rate SFR$_{\rm IR}$ of PMs to the unobscured contribution SFR$_{\rm UV}$. The median obscured-to-unobscured ratio is $\sim$12, roughly double that for `normal' xCG galaxies with similar SFR, in line with other literature findings that interaction-induced star formation often takes place in heavily dust-obscured environments \citep[e.g.,][]{veilleux02,dacunha10,ellison13a,lanz13,alonso-herrera16}.\medskip

Hybrid SFRs were calculated with a combination of GALEX and unWISE photometry via
\begin{equation}
\left(\frac{\sfr}{\msun/{\rm yr}}\right) = \mathcal{C_{\rm UV}}\Bigg[\left(\frac{L_{\rm UV}}{\rm erg/s}\right) + a_{\rm UV}\left(\frac{L_{\rm TIR}}{\rm erg/s}\right)\Bigg]~.
\label{eq:HaoSFR}
\end{equation}
Here $\mathcal{C_{\rm UV}}$ and $a_{\rm UV}$ are the FUV or NUV coefficients as provided in Tables 2 \& 3 in \citet{hao11} and $L_{\rm TIR}$ is the total infrared (3-1100\,$\mu$m) luminosity inferred from the W3 or W4 flux density. (W3 is the 12\,$\mu$m, W4 is the 22\,$\mu$m WISE filter.) To account for the luminosity-dependence of MIR-to-TIR bolometric corrections, due to the temperature vs. luminosity and PAH-fraction vs. luminosity relations of local IR-emitting galaxies, we used the \citet[CE]{charyelbaz01} and \citet[DH]{dalehelou02} SED libraries\footnote{For the DH02 SED library the dependence on luminosity is imposed manually via the empirically calibrated relation between FIR colour $S_{\rm 60}/S_{\rm 100}$ and $L_{\rm IR}$ from \citet{marcillac06}. As our final TIR-luminosity measurement we adopt the average of the two values calculated using the CE01 and DH02 libraries. The systematic error due to choice of template library in general dominates (by approx. a factor 30) the total error budget of $L_{\rm TIR}$ measurements for PMs.}. Prior to applying the MIR-to-TIR scaling factor, we subtract stellar contributions from the W3/4 photometry. These were estimated from the WISE W1 (3.4\,$\mu$m) flux density following \citet{ciesla14}. Specifically, we adopt the average of the correction factors for late- and early-type galaxies (which at 3.6\,$\mu$m only differ by $\sim$2\%; see Table B.1 in \citealp{ciesla14}), and apply this mean correction to all galaxies in our samples. As a test of the robustness of this extrapolation from the MIR bands, we compare far-IR (FIR; 42-122\,$\mu$m) luminosities directly measured from IRAS 60 \& 100\,$\mu$m fluxes following \citet{sandersmirabel96} with our own FIR luminosity estimates in Fig. \ref{fig:SFR_IRnUV}b. For a subset of 91 bright and purely star-forming PM and xCG galaxies with IRAS coverage, we find that the two alternative luminosity derivations agree at the $\sim$10\% level (median $L_{\rm FIR}$(W3/4)/$L_{\rm FIR}$(IRAS) ratio: 0.89$^{+0.05}_{-0.03}$).\medskip

In order of preference, we adopt the following combinations of UV and IR photometry to measure SFRs: FUV+W3, FUV+W4, NUV+W3, NUV+W4. If a given photometry combination was not available (due to non-detections or poor quality flags) the next lower option was instead adopted\footnote{~For SFGs with SFR$_{\rm IR}\,{>}$\,1\,\msun/yr that lack UV photometry ($<$10\% of the BPT-classified SFGs in the PM and xCG samples), we adopt the dust-obscured SFR estimate as a proxy of the total SFR, as in low-$z$ galaxies with these levels of star-formation activity in general SFR$_{\rm IR}\,\gg$\,SFR$_{\rm UV}$.}. We favour FUV over NUV photometry as the NUV band can have contributions from more evolved stars. We favour W3 over W4 flux densities as the relation between monochromatic WISE luminosity and $L_{\rm IR}$ is tighter at W3, and because the W3 filter is less sensitive to the presence of (residual) AGN-heated hot dust  by virtue of its wide bandpass, which samples a combination of PAHs, nebular lines, and continuum \citep[e.g.][]{donoso12, salim16, cluver17}. Moreover, the higher sensitivity of the W3 band \citep{jarrett13} facilitates the derivation of consistent SFR estimates for a broader range of galaxies in our comparison and reference samples. However, despite the empirical finding in \citet{cluver17} that MIR-to-TIR ratios are subject to a larger scatter when using W4 flux densities, Fig. \ref{fig:SFR_IRnUV}c shows that luminosities inferred from W3 and W4 photometry are nevertheless highly consistent for star-forming galaxies; averaged between the CE01 and DH02 libraries, the median $L_{\rm TIR}$(W3)/$L_{\rm TIR}$(W4) ratio for the 239 PM and xCG galaxies with both W3 and W4 detections and flag values \texttt{bptclass}\,=\,1/2 is 1.03$\pm$0.04.

\subsubsection{SFRs for composite or AGN, and unknown spectral types}
\label{sect:SFRderiv_others}
AGN-related emission in UV and MIR bands (due to the big blue bump and dusty-torus emission, respectively) can falsify SFR measurements that rely on monochromatic UV and/or MIR flux densities. For galaxies classified as AGN-host/LINER (\texttt{bptclass}\,=\,4/5) we adopt SFR-estimates derived through an UV-to-optical SED-fitting analysis with CIGALE in \citet[S16]{salim16}. We also adopt the \citet{salim16} SFRs for composite objects (\texttt{bptclass}\,=\,3), if no hybrid IR+UV SFR measurement involving WISE W3 fluxes is available. A W3-based IR-contribution to the hybrid SFR is deemed acceptable for composites following the assessment in \citet{salim16,cluver17} that SFRs inferred from the 12\,$\mu$m band are less impacted by AGN related continuum emission than is the case for the W4 filter at 22\,$\mu$m.\medskip

To galaxies with an unknown spectral type we assign SFRs from \citet{salim16}, or if these are not available an SFR-estimate from the MPA-JHU catalog \citep{brinchmann04}. We note that galaxies with unknown spectral type often have low star-formation activity, and that mid-IR brightness breaks down as an SFR indicator for galaxies with \ssfr$_{\rm SED}\,{<}\,0.01$\,Gyr$^{-1}$, where dust heating is dominated by old populations. The SED-fits of \citet{salim16} should thus in any case provide a more robust estimate of the SFR than a hybrid UV+IR recipe.

\subsubsection{Compensation for systematic shifts between different SFR estimators}
\label{sect:SFRrenorm}
The different SFR estimators mentioned in the two preceding sections display systematic offsets that can reach up to 0.2\,dex for PMs and xCG reference galaxies (see Appendix \ref{appsect:altSFRnMstellintro}, where we also quantify the magnitude of these systematics for all our comparison samples). These are genuine offsets caused by differing methodology and are not simply due to a different choice of IMF. We use the offsets measured for the xCG sample (500+ galaxies) to shift SFR measurements from the \citet{salim16} and MPA-JHU catalogs to on average match the normalisation of our hybrid UV+IR SFR-estimator. The corresponding offsets are 0.15 and 0.11\,dex (log(SFR$_{\rm UV+IR}$/SFR$_{\rm \{S16,MPA-JHU\}}$)), respectively, and follow from the median offset values displayed in the lower right corners of the panels in the central column of Fig. \ref{appfig:SFRsyscomp} in Appendix \ref{appsect:altSFRnMstellintro}. SFR shifts calculated for the xCG reference sample are robust in terms of their formal statistical uncertainty. However, it is important to realise that they are not free of systematics as, e.g., xCG has a different mix of galaxy spectral types and evolutionary stages than the PMs and other comparison samples. Such systematics are especially significant for mPSBs and DETGs, where large offsets between different SFR measurements are observed (see Appendix \ref{appsect:altSFRnMstell}). We discuss their implications for the conclusions of this paper in Sects \ref{sect:compgal_offsets} and \ref{sect:discussion}.\medskip

The homogenized sets of SFR-measurements -- consisting of hybrid UV+IR SFR-estimates whenever possible, and usually \citet{salim16} SFRs for composite/AGN objects or objects with lacking photometric coverage -- represent our `best-estimate' star-formation rates for the rest of the paper. In Table \ref{tab:prop} we list the according values for PM galaxies with their 1\,$\sigma$ uncertainties that account for both measurement errors (e.g., on MIR and UV fluxes, and UV spectral slope estimates) and systematic errors (SED-dependent MIR-to-TIR extrapolations, and coefficients in the hybrid SFR-recipe in eq. \ref{eq:HaoSFR}).\newpage

\subsection{Stellar mass measurements}
\label{sect:Mstellderiv}

We consider three different types of \mstar\ measurements in our analysis: (i) stellar masses presented in \citet[M14]{mendel14}, and based on bulge+disk and Sérsic profile photometric decompositions of SDSS $ugriz$ photometry, (ii) stellar masses from the UV-to-optical SED-fitting analysis in \citet{salim16}, who use photometry optimised to yield accurate colour information, and (iii) stellar masses from the `legacy' MPA-JHU data base \citep{brinchmann04}.\medskip

As our primary source of stellar mass information we use the \citet{mendel14} catalog. We adopt the \citet{mendel14} masses resulting from a bulge+disk decomposition of SDSS images as our `best-estimate' \mstar\ values for star-forming and composite objects (\texttt{bptclass}\,=\,1--3). These stellar mass estimates from \citet{mendel14} tend to lie near the median of the mass estimates from the other catalogs, making them a good choice for a `best-estimate' mass to underpin the principle conclusions of the paper. For AGN host galaxies (\texttt{bptclass}\,=\,4/5) we adopt stellar masses from the \citet{salim16} catalog as our `best-estimate' \mstar\ values. Where neither of these two measurements is available we assign the stellar mass reported in the MPA-JHU data base.\medskip

As done for SFRs in Sect. \ref{sect:SFRrenorm}, we attempt to place `best-estimate' stellar masses on a consistent scale (defined by the normalisation of masses in the \citealp{mendel14} catalog) by correcting for systematic offsets between the different estimators. The corresponding offsets are -0.1\,dex  (log(\mstar$_{\rm M14}$/\mstar$_{\rm S16}$)) and 0.04\,dex (log(\mstar$_{\rm M14}$/\mstar$_{\rm MPA-JHU}$)), and follow from the median offset values displayed in the central column of Fig. \ref{appfig:Mstellsyscomp} in Appendix \ref{appsect:altSFRnMstellintro}. In Table \ref{tab:prop} we list our `best-estimate' \mstar\ values for PM galaxies together with their 1\,$\sigma$ uncertainties.

\subsection{CO line luminosities and molecular gas masses}
\label{sect:XCO}

CO line luminosities are calculated from the aperture-corrected, velocity-integrated line fluxes $I_{\mathrm{CO}(J{\rightarrow}J-1)}$ (derived as described in Appendices \ref{appsect:fluxmeas}/\ref{appsect:apcorr}) as
\begin{eqnarray}
\left(\frac{L'_{\mathrm{CO}(J{\rightarrow}J-1)}}{\rm K\,km/s\,pc^2}\right) &=& \frac{3.25{\times}10^7}{\left(1+z\right)^{3}} \left(\frac{\nu_{\rm obs}}{\rm GHz}\right)^{-2} \left(\frac{D_L}{\rm Mpc}\right)^2\nonumber\\ 
&&\times \left(\frac{I_{\mathrm{CO}(J{\rightarrow}J-1)}}{\rm Jy\,km/s}\right)~,
\end{eqnarray}
where $\nu_{\rm obs}\,{=}\,J{\times}\frac{\nu_{\rm rest}(\COone)}{1+z}$ is the observer-frame frequency of the $J$-th rotational transition of CO and $\nu_{\rm rest}(\COone)$\,=\,115.271\,GHz. For PMs the line luminosity of the ground-state transition, $L'_{\rm \COone}$, is listed in Table \ref{tab:fluxes}.\medskip

\begin{figure*}
\centering
\includegraphics[width=.6\textwidth]{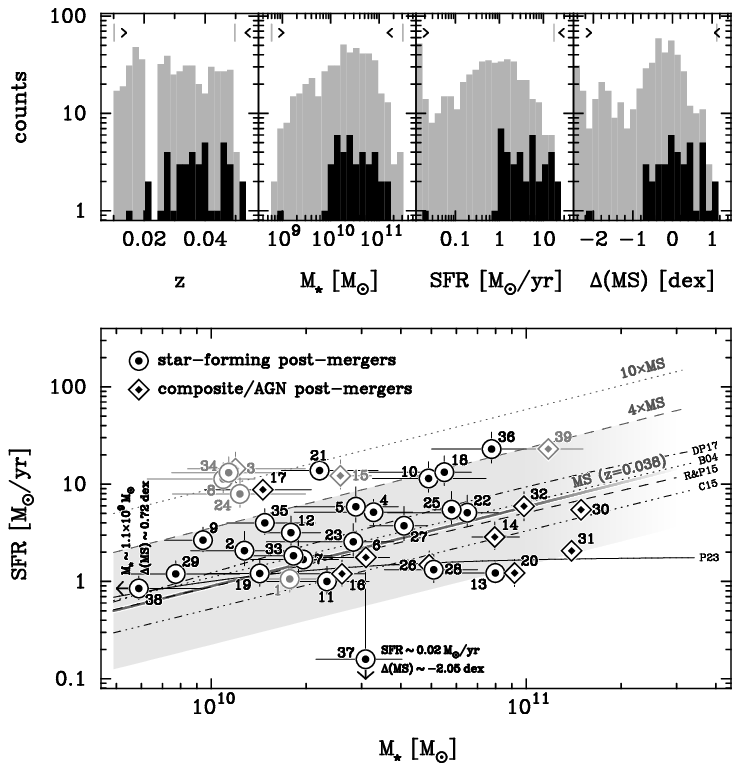}
\caption{{\it Top:} Redshift, stellar mass, SFR and MS-offset (from {\it left} to {\it right}) distributions for the post-merger (black histograms) and xCOLD GASS samples (grey histograms). Black arrowheads (light grey vertical lines) along the upper edge of each panel indicate the range of values covered by post-mergers (xCOLD GASS galaxies).\newline
{\it Bottom:} Location of post-merger galaxies in the SFR vs. \mstar\ plane. All post-mergers are labelled with their sequential ID (see Table \ref{tab:prop}). Objects with warm IRAS colours are plotted with light grey symbols, post-mergers with AGN signatures in the optical spectra are highlighted with diamond-shaped symbols. The grey shaded area spans the $\pm$3\,sigma dispersion of the main sequence of star-forming galaxies at the median redshift of the post-merger sample ($z\,{\sim}\,0.038$). The grey dashed (dotted) line traces the locus of four-fold (ten-fold) sSFR-excess with respect to the MS. The average low-$z$ MS locus (dark grey line) was derived based on the four literature trend lines shown in black (B04 -- \citealp{brinchmann04}, C15 -- \citealp{chang15}, R\&P15 -- \citealp{renzinipeng15}, DP17 -- \citealp{duarte-puertas17}; see Sect. \ref{sect:PMonMSnSK} for further details).
\label{fig:MSplane}}
\end{figure*}

We compute CO-to-H$_2$ conversion factors following the `2 Star-Formation Mode' (2-SFM) framework of \citet{sargent14}. The 2-SFM framework distinguishes between `normal' star-forming galaxies (residing on the galaxy main sequence, MS) and starburst (SB) galaxies, which mostly lie above the MS, but in small numbers also occur within the MS locus when subject to only a small boost of star formation. The threshold above which SBs dominate the star-forming galaxy population by number lies at an MS-offset of $\Delta({\rm MS})\,{\sim}$\,3 in the 2-SFM framework (here $\Delta({\rm MS})\,{\equiv}$\,\ssfr/\ssfrMS\ is the galaxy sSFR relative to the ridge line of the MS, \ssfrMS; see \citealp{sargent12} for details). Below this threshold\footnote{The 2-SFM description of the star-forming galaxy population does not capture well the behaviour of green-valley or quiescent galaxies. To galaxies significantly below the MS we thus simply assign a MW-like conversion factor, \aCO\,=\,4.35\,\msun/(K\,km/s\,pc$^2$).}
we assign galaxies an \aCO\ value which varies with gas-phase metallicity according to the \citet{wolfire10} prescription, which we normalise to a Milky Way-like CO-to-H$_2$ conversion factor of 4.35\,\msun/(K\,km/s\,pc$^2$) at solar metallicity. Galaxy metallicities are estimated statistically from their stellar mass and SFR via the fundamental metallicity relation (FMR) of \citet{mannucci10}. This ensures uniformity across all samples as metallicity measurements based on optical emission lines are not available for all types of interacting and control galaxies, nor even for all galaxies within a given sample (e.g., Table \ref{tab:prop}). The FMR is constrained by data \citep[see Fig. 1 in][]{mannucci10} down to \mstar\,${\sim}\,2{\times}10^9$\,\msun\ and SFR\,$\sim$\,0.05\,\msun/yr, covering the vast majority of the parameter space in which the interacting and post-starburst galaxies in all our samples reside (see Fig. \ref{fig:MSplane} and Appendix \ref{appsect:compgal_paramspace}). Nevertheless,  inferring the gas-phase metallicities of the galaxies with the lowest SFRs in the mPSB and DETG samples via the FMR will inevitably be subject to some systematic uncertainty.\medskip

Above the threshold $\Delta({\rm MS})\,{\sim}$\,3 we assign \aCO\ values that deviate from the conversion factors of MS-galaxies by an amount that scales continuously with the intensity of the starburst, matching the inferred dependence of \aCO\ on interaction stage in ISM simulations of merging galaxies \citep[e.g.][]{renaud19}. Specifically, this recipe for assigning CO-to-H$_2$ conversion factors makes use of the statistical linkage -- described by the ``SFR boost factor" in the 2-SFM framework (see eq. 27 in \citealp{sargent14}) -- between a starburst and its progenitor MS-galaxy state. In addition to being lower, our starburst \aCO\ values thus also reflect the metallicity-dependence of the conversion factors for `normal' progenitor galaxies (eq. 29 in \citealp{sargent14}); at fixed $\Delta({\rm MS})$, starburst \aCO-values are higher for low-\mstar, low-metallicity galaxies than for more metal-enriched, massive galaxies. (See Fig. 16 in \citealp{sargent14} for a map of these \aCO-variations in the $z\,{\sim}$\,0 SFR-\mstar\ plane.)\medskip

It should be emphasised that the interacting and post-starburst galaxies that are the subject of this paper may not in all cases follow the scaling relations underpinning the calculation of CO-to-H$_2$ conversion factors in the 2-SFM framework. For instance, one might well expect some merging pair galaxies or PMs that lie on the MS to have a more starburst-like \aCO, rather than the MS-like conversion factor we have assigned them based on their $<$50\% statistical probability of being a starburst, given their \mstar\ and SFR. In the absence of the ancillary information that would be required to make a more refined case-by-case judgement (e.g., kinematical data, CO excitation or ISM temperature measurements), we can quantify this systematic uncertainty by re-deriving our key results also with purely observational proxies for gas fractions and depletion times, namely \lco/\mstar\ and \lco/SFR. In this way we confirm in Sect. \ref{sect:PMoffsets_best} and Appendix \ref{appsect:obsproxPM} that our results are not the artefact of assuming a specific recipe when calculating \aCO\ values. Finally, we note that \citet{accurso17b} presented CO-to-H$_2$ conversion factors for xCG galaxies based on radiative transfer modelling and an analysis of [CII]/\COone\ luminosity ratios. For star-forming xCG galaxies we find close agreement between the mean conversion factor inferred via their method ($\overline{\alpha}_{\rm CO}$\,=\,4.43\,K\,km/s\,pc$^2$), and via the 2-SFM framework ($\overline{\alpha}_{\rm CO}$\,=\,4.36\,K\,km/s\,pc$^2$). \aCO\ values from \citet{accurso17b} show a somewhat larger dispersion that is primarily due to their best-fit relation between \aCO\ and metallicity being steeper in the relevant stellar mass range than the one in \citet{wolfire10}, on which our own \aCO\ estimates rely.\medskip

Molecular gas masses \mhtwo\ are inferred from the \COone\ line luminosity via
\begin{equation}
\left(\frac{\mhtwo}{\msun}\right) = \left(\frac{\aCO}{\rm \msun(K\,km/s\,pc^2)^{-1}}\right)\,\left(\frac{L'_{\rm \COone}}{\rm K\,km/s\,pc^2}\right)~,
\end{equation}
and are listed for all PM galaxies in Table \ref{tab:fluxes}. The \aCO\ values used to infer the molecular gas mass follow from the $L'_{\rm \COone}$ and \mhtwo\ values recorded in the table. The majority of PM molecular gas masses lie in the range $10^9\,{<}\,\mhtwo/\msun\,{<}\,10^{10}$ (median: $3.5{\times}10^9$\,\msun), with only a few galaxies displaying larger or smaller values.

\begin{figure*}
\centering
\includegraphics[width=.98\textwidth]{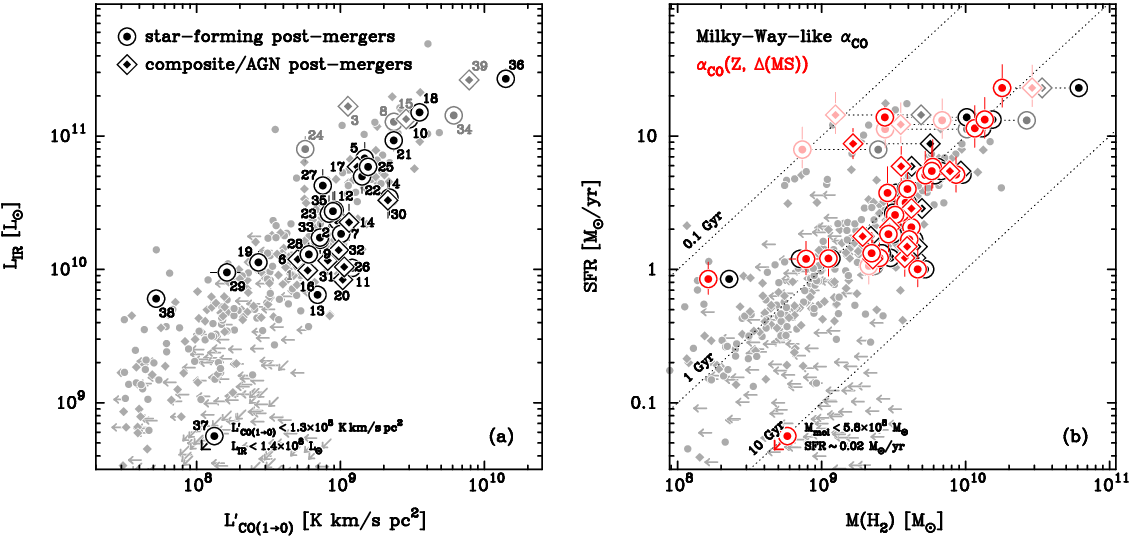}
\caption{Integrated Schmidt-Kennicutt relations for post-merger galaxies (large symbols, labelled with their sequential ID from Table \ref{tab:prop} in panel ({\it a}), with xCOLD GASS data in background. In both samples, diamond-shaped symbols denote AGN/composite objects. Post-mergers with warm IRAS colours are plotted with light grey/red symbols.\newline
{\it (a)}: Observational proxy of the star-formation law, the \lir(8-1000\,$\mu$m) vs. $L'_{\COone}$ relation, with galaxy-integrated (i.e. aperture-corrected) CO line luminosities derived assuming a spatially invariant gas fraction. {\it (b)} Correlation between molecular gas mass \mhtwo\ and galaxy-integrated SFR. For post-mergers, molecular gas masses reported with black/grey points assume a Milky Way-like CO-to-H$_2$ conversion factor \aCO\,=\,4.4\,\msun/[K\,km/s\,pc$^2$] for $H_2$-mass derivations. For the bright/light red data points a conversion factor dependant on metallicity and `starburstiness' was adopted. The xCOLD GASS data is plotted assuming the same variable \aCO.
\label{fig:intSK}}
\end{figure*}

\section{Results}
\label{sect:results}

\subsection{Locus of IRAM post-mergers on galaxy scaling relations}
\label{sect:PMonMSnSK}
The distributions of the 39 PMs with respect to redshift, stellar mass, SFR and offset $\Delta$(MS) from the main sequence\footnote{For the slope and normalisation of the MS, in the following we adopt average parameters from power-law fits to the MS of low redshift galaxies as presented in \citet{brinchmann04}, \citet{chang15}, \citet{renzinipeng15} and \citet{duarte-puertas17}. Prior to averaging fit parameters, all MS relations were placed on the same \citet{chabrier03} IMF scale and shifted to a common reference redshift $\langle{z}\rangle\,{=}$\,0.038, the median redshift of our PM sample. The literature-averaged MS locus derived in this way is very similar in normalisation to the \citet{brinchmann04} and \citet{renzinipeng15} fits, with an only marginally steeper slope $d$log(SFR)/$d$log(\mstar) of 0.82. The shape of the MS locus is selection-dependent \citep[e.g.,][]{salmi12, leslie20}, with selection function-dependent flattening impacting in particular the $\Delta$(MS) values inferred for galaxies with high stellar mass. For illustration we plot in  Fig. \ref{fig:MSplane} the curved MS-relation of \citet[P23]{popesso23}, with which $\Delta$(MS) values for the most massive PMs in our sample shift upward by a factor of 4--5. For $>$70\% of PMs $\Delta$(MS) values vary less than 2-fold when adopting an MS with turn-over at \mstar\,${\simeq}\,10^{10}\,\msun$.} of star-forming galaxies is shown in the foreground of all individual panels in the upper half of Fig. \ref{fig:MSplane} (black histograms). The corresponding histograms for the 488 xCG galaxies constituting our reference sample are plotted in light grey. The xCG sample is well-suited to select control galaxies as it spans the full parameter space occupied by PMs, and counts of xCG galaxies exceeds those of PMs in general by at least a factor 3 but often closer to 10-fold in all bins. Additionally, xCG extends towards substantially lower stellar masses and SFRs. This will allow us to exploit xCG also for control-matching to galaxies in the other comparison samples of interacting galaxies, which in some cases include objects with significantly lower \mstar\ or SFR than are present in the PM sample (see Appendix \ref{appsect:compgal_paramspace}).\medskip

In the lower half of Fig. \ref{fig:MSplane} PM galaxies are plotted in the SFR-\mstar\ plane where they occupy the region SFR\,$\gtrsim$\,1\,\msun/yr and \mstar\,$\gtrsim$\,10$^{10}$\,\msun, with the exception of two objects (PMs \#37 \& 38). These lower bounds reflect the selection criteria in Sect. \ref{sect:PMintro} rather than a fundamental physical property of PMs (e.g. \citealp{bickley22} show that different selection methodologies can impact the physical parameter space from which PMs are drawn). A majority of our PMs reside on the galaxy MS, but $\sim$20\% of the sample can be considered starbursts with a $\geq$4-fold sSFR-offset from the MS ridge line at fixed \mstar. A single PM in our sample is located substantially below the MS in the quiescent galaxy regime. Starbursting PMs above the MS often (5/9 IRAS-detectable objects) have the warm IRAS far-IR colours $S_{\nu}(60\,{\mu}{\rm m})/S_{\nu}(100\,{\mu}{\rm m})\,{>}$\,0.63 typically observed for low-$z$ LIRGs and ULIRGs \citep{marcillac06}. These are frequently associated with merger-induced star-formation activity with short depletion times in the literature \citep[e.g.][]{haan11, larson16}, providing some additional support to our choice of assigning low CO-to-H$_2$ conversation factors \aCO\ to high-sSFR objects (see Sect. \ref{sect:XCO}). One further object with warm far-IR colours (PM \#1) is located close to the core of the MS, a position which is robust to exchanging our `best-estimate' W3+FUV SFR (Sect. \ref{sect:SFRderiv_SFG} for SFR-estimates using W4+FUV or IRAS photometry. The ULIRG-like far-IR colour may thus hint to starburst activity having raised this heavily distorted PM system (see first row of Fig. \ref{appfig:spectra}) up onto the MS from a previously lower activity level. Finally, we note that some of the most massive PMs have spectral classifications as AGN hosts or composite objects. This is not a specific outcome related to our choice of stellar masses from \citet{mendel14}, but a genuine trend for high-mass galaxies to have \texttt{bptclass}\,$\geq$\,3 that is also present in the xCG reference sample (see Appendix \ref{appsect:altSFRnMstell}).\medskip

We plot the position of PMs with respect to the integrated Kennicutt-Schmidt relation in Fig. \ref{fig:intSK} (panel ({\it a}) -- directly observed quantities; panel ({\it b}) -- physical quantities), alongside the xCG reference sample. The majority of CO-detected xCG galaxies cluster around a line of constant depletion time in SFR--\mhtwo\ space with $\tau_{\rm depl.}\,{\sim}$\,1\,Gyr \citep[see also][]{saintonge11}. The median depletion time of our PM sample is 1.06\,Gyr when adopting a variable \aCO\ to convert line luminosities \lco\ to H$_2$ masses. For comparison, the assumption of a constant, Milky-Way-like conversion factor leads to similar gas masses and depletion times for most PMs  (median depletion time in this case: 1.24\,Gyr). CO-detected xCG galaxies are distributed essentially symmetrically around the median depletion time with $\sigma\,{\approx}$\,0.23\,dex (factor 1.7 scatter). In contrast, PMs have a broader and more skewed distribution with a symmetrised 1\,$\sigma$ scatter of $\sim$0.43\,dex (factor 2.7 dispersion). Fig. \ref{fig:intSK}a illustrates that this broader distribution for PMs persists in the parameter space of \lir(8-1000\,$\mu$m) vs. $L'_{\COone}$, the directly observed proxies of SFR and \mhtwo, i.e. this behaviour is not the consequence of applying variable \aCO-values on a galaxy-by-galaxy basis. Both panels of Fig. \ref{fig:intSK} also contain visual evidence of a surplus of PMs with long depletion times compared to xCG galaxies. We will place this evidence on quantitative ground via control-matching to the xCG reference sample in the next section, and investigate the interplay between fuel reservoir (gas fraction) and SFE in determining the SFRs of PMs.

\subsection{Gas fraction and depletion time offsets in post-mergers}
\label{sect:PMoffsets_best}

\begin{figure*}
\centering
\begin{tabular}{ccc}
\includegraphics[width=.45\textwidth]{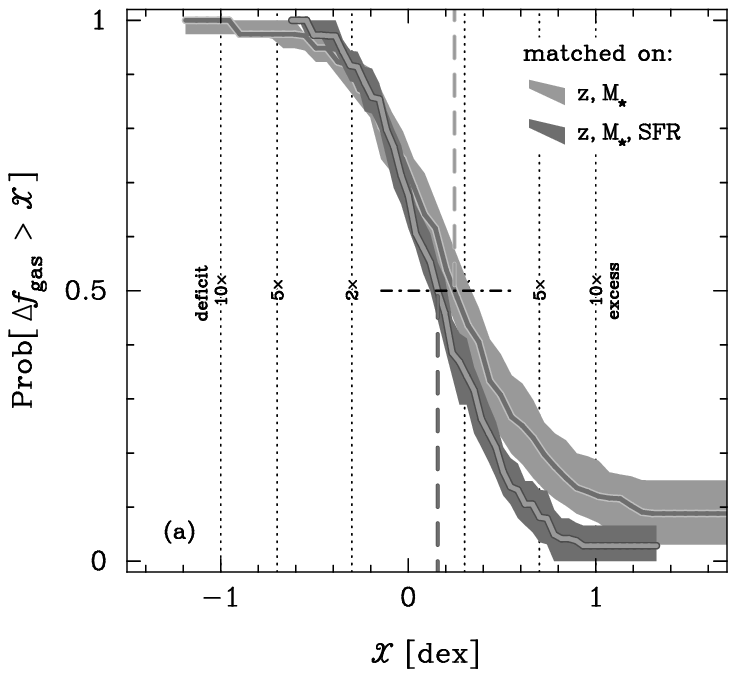} & & \includegraphics[width=0.45\textwidth]{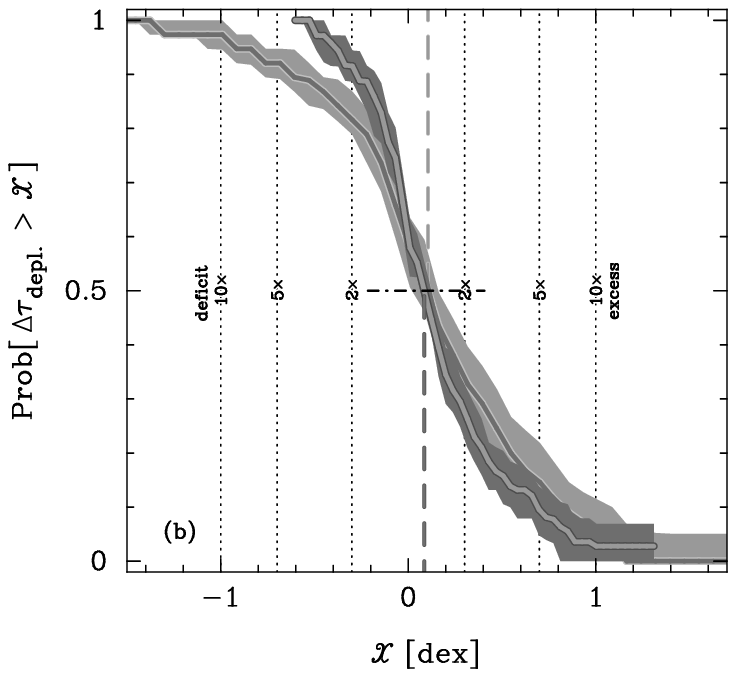}
\end{tabular}
\caption{Offset of post-merger gas fractions $\fgas\,{=}\,M({\rm H}_2)/\mstar$ (panel {\it a}) and molecular gas depletion times $\tau_{\rm depl.}\,{=}\,M({\rm H}_2)/\sfr$ (panel {\it b}) compared to normal galaxies from xCOLD GASS. The offsets measured for the post-merger sample are shown in form of a `survival distribution', i.e. a normalized cumulative distribution function taking into account both CO-detected and undetected xCOLD GASS control galaxies. The cumulative distribution function represents the probability of drawing, from the post-merger sample, a galaxy with \fgas\ (panel {\it a}) or $\tau_{\rm depl.}$ (panel {\it b}) in excess of $\mathcal{X}$ compared to control galaxies. $\mathcal{X}$ is an auxiliary variable spanning the full range of individual galaxy $\Delta\fgas$ or $\Delta\tau_{\rm depl.}$ offset measurements occurring in the PM sample. Distributions traced in light (dark) grey are for control matching on redshift $z$ and \mstar\ ($z$, \mstar\ \& SFR), respectively. Shaded areas span the 68\% confidence region.\newline
The vertical dashed lines project the median of the cumulative distribution (at a $y$-axis value of 0.5) onto the $\mathcal{X}$-axis. Vertical dashed lines extending to the upper (lower) $x$-axis mark the median for control matching on redshift \& \mstar\ (on redshift, \mstar\ \& SFR). Both for control-matching on redshift \& \mstar, and on redshift, \mstar\ \& SFR, post-mergers on average have enhanced gas fractions and depletion times relative to xCOLD GASS control galaxies.
\label{fig:physcontrolmatch_best}}
\end{figure*}

In this section we measure the offset of PM galaxies from normal, non-interacting xCG galaxies with respect to two quantities: the molecular-to-stellar mass ratio \fgas, and the molecular gas depletion time $\tau_{\rm depl.}$\,=\,SFE$^{-1}$. For this we implement control-matching as introduced in Sect. \ref{sect:matching}, and we derive gas fraction and depletion time offsets using our `best-estimate' SFR and \mstar\ derivations (Sects. \ref{sect:SFRderiv} \& \ref{sect:Mstellderiv}).\medskip

As per Sect. \ref{sect:matching} we adopt initial matching tolerances of $\Delta{z}$\,=\,0.01, $\Delta{{\rm log}(\mstar)}$\,=\,0.1\,dex and $\Delta{{\rm log(SFR)}}$\,=\,0.1\,dex for our selection of xCG control galaxies. If necessary, we grow the default matching tolerances iteratively until $N_{\rm min}$\,=\,5 xCG control galaxies can be assigned to a given PM. When matching on redshift and \mstar\ only\footnote{The SFR distribution of PMs (see Fig. \ref{fig:MSplane}), and also of some comparison galaxy samples (see Appendix \ref{appsect:compgal_paramspace}), are not well matched with the SFR distribution of the full xCG sample. For gas-related physical properties that scale with SFR this can lead to unrepresentative offset values when control-matching on redshift and \mstar\ only. To counteract such biases as best possible we only draw mass- and redshift-matched xCG control galaxies from above a suitable SFR threshold determined as follows: (i) the smallest value SFR$_{\min}$ in the interacting galaxy sample, as long as SFR$_{\min}$ is not a strong low-SFR outlier compared to the overall SFR distribution of the sample; (ii) 0.3\,dex below the 10$^{\rm th}$ percentile of the SFR distribution of the interacting galaxy sample, if SFR$_{\min}$ lies more than a factor 10 below this percentile. When additionally matching on SFR as a third parameter imposing these cuts is not necessary.}, a minimum of 5 matching xCG galaxies were always available already within the default matching intervals, with control samples involving up to 36 xCG objects in the most densely populated regions of parameter space. The median number of control galaxies assigned to a given PM was 18 for matching on redshift and \mstar. When matching on SFR as well, the initial search intervals had to be extended for $\sim$50\% of the PMs, by an average growth factor of 1.3. For one PM the requisite $N_{\rm min}$\,=\,5 xCG control galaxies could not be identified even after doubling of the initial search intervals, our cut-off point for discarding such objects when constructing offset distributions. The median (maximum) number of control galaxies assigned to a given PM was 5 (9) for matching on redshift, \mstar\ and SFR.\medskip

Throughout this paper offsets are defined as log-scale quantities, i.e. the offset of the $i$-th PM from the $j$-th control-matched xCG galaxy is calculated as
\begin{equation}
\Delta Y_{{\rm PM}\,\#i} = {\rm log}(Y_{{\rm PM}\,\#i}) - {\rm log}(Y_{{\rm control}\,\#j})~.
\end{equation}
Here $Y$ stands for the galaxy property of interest, e.g. in this section \fgas\ and $\tau_{\rm depl.}$, or subsequently molecular-to-atomic gas mass ratios \mhtwo/\mHI\ (Sect. \ref{sect:PM_H2overHI}) and line excitation ratios $r_{21}$ (Sect. \ref{sect:PM_r21}). An offset of $\Delta Y$\,=\,1.0 thus implies a ten-fold enhancement over that of a control galaxy. For the median offset of all PMs with respect to control-matched galaxies we use the symbol $\langle\Delta Y\rangle$.\medskip

\begin{figure*}
\centering
\begin{tabular}{ccc}
\includegraphics[width=0.45\textwidth]{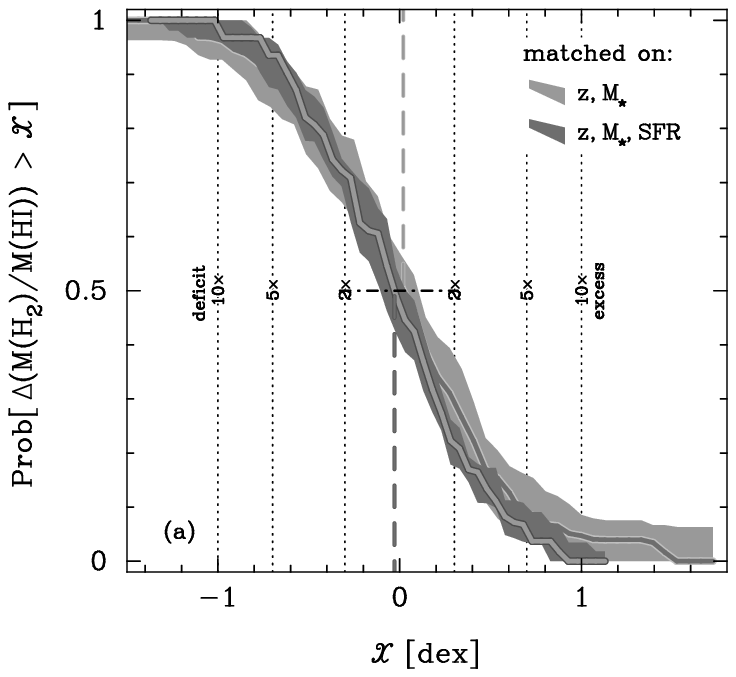} & & \includegraphics[width=0.45\textwidth]{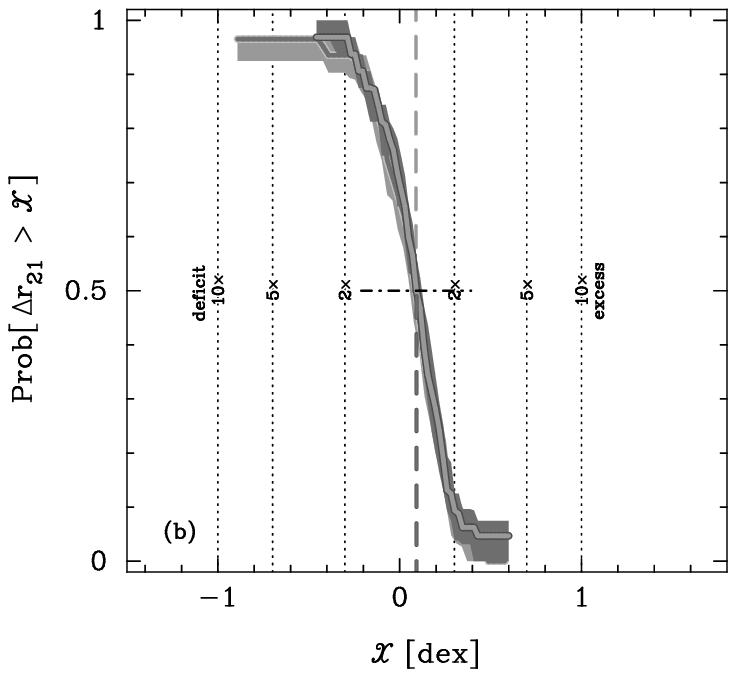}
\end{tabular}
\caption{Cumulative offset distributions for galaxy-integrated post-merger molecular-to-atomic gas mass ratios $\mhtwo/M({\rm HI})$ (panel {\it a}) and \COtwo-to-\COone\ excitation factors $r_{21}$ (panel {\it b}). Post-mergers have very similar $\mhtwo/M({\rm HI})$ mass ratios as xCG control galaxies, but display a 25\% higher excitation of the \COtwo\ transition.
\label{fig:PM_HInr21}}
\end{figure*}

Offset distributions are plotted in the form of cumulative probability distributions in Fig. \ref{fig:physcontrolmatch_best}, for gas fractions (panel {\it a}) and depletion times (panel {\it b}). The probabilistic nature of the distributions owes to the fact that CO-detected PMs can be paired with undetected xCG control galaxies, resulting in a lower limit on the offset measurement of the individual PM. The KM product limit estimator \citep[see further details in Sect. \ref{sect:matching}]{kaplanmeier58} incorporates the information contained in lower limits and uses these in combination with the well-localised measurements for CO-detections to reconstruct the normalised cumulative distribution function of \fgas\ or $\tau_{\rm depl.}$ offsets in PMs relative to the normal galaxy population. (In the field of survival analysis this output of the KM estimator is referred to as the `survival distribution'.) At a fixed position $\mathcal{X}$ on the horizontal axis\footnote{In Figs. \ref{fig:physcontrolmatch_best} and \ref{fig:PM_HInr21} the $x$-axis variable $\mathcal{X}$ is an  auxiliary variable spanning the full range of offsets (e.g., for the quantities \fgas\ or \tdep) measured for individual PMs.} of Fig. \ref{fig:physcontrolmatch_best}, the value of the cumulative offset distribution function gives the probability of finding, among the galaxies in our PM sample, gas fraction (panel {\it a}) or depletion time (panel {\it b}) measurements ${>}\mathcal{X}$. In both panels of Fig. \ref{fig:physcontrolmatch_best} we additionally highlight the 68\% confidence region of the offset distribution functions, based on the scatter measured in our multiple realisations of the KM estimator (see methodological details in Sect. \ref{sect:matching}). Results for control-matching on redshift and \mstar\ (redshift, \mstar\ and SFR) are plotted in light (dark) grey. In this section we report only on offset measurements for galaxy-integrated H$_2$ masses, calculated by upward correction of the CO-flux assuming a spatially invariant gas fraction (aperture correction factor $\mathcal{A}_{\COone}$(const.\,\fgas), see Sect. \ref{appsect:apcorr}). The consistency of these results with offset measurements relying on other aperture correction choices is discussed in Appendix \ref{appsect:aperturesys}.\medskip

PMs display molecular gas fractions higher by $\langle\Delta\fgas\rangle$\,=\,0.25$_{-0.06}^{+0.09}$\,dex (factor 1.8) compared to xCG control galaxies when matched on redshift and \mstar\ (Fig. \ref{fig:physcontrolmatch_best}a). A smaller factor 1.4 shift (0.16$_{-0.04}^{+0.05}$\,dex) is found when additionally control-matching on SFR, due to the median SFR of PMs being higher than for the xCG sample (see Fig. \ref{fig:MSplane}). PM depletion times are also enhanced (Fig. \ref{fig:physcontrolmatch_best}b): by $\langle\Delta\tau_{\rm depl.}\rangle$\,=\,0.11$_{-0.09}^{+0.05}$\,dex (factor 1.3) when employing control-matching on redshift and \mstar, and by a nearly identical amount -- albeit with $\sim$50\% smaller error bars -- when control-matching on redshift, \mstar\ and SFR. The \fgas\ and $\tau_{\rm depl.}$ enhancements found when control galaxies matched on all three properties are selected are consistent within uncertainties, i.e. the gas fraction increase $\Delta\fgas$ directly contributes to longer depletion times at fixed SFR. This reflects the fact that residual gas mass variations are comparatively small for galaxies with a similar SFR. The stellar mass and SFR, that enter the calculation of \fgas\ and $\tau_{\rm depl.}$, therefore mostly contribute a different normalisation to the underlying offset $\Delta$M(H$_2$) measured for individual galaxies. Our finding that the gas fraction enhancement for control-matching on redshift and \mstar\ is higher than for matching on redshift, \mstar\ and SFR is due to low-SFR galaxies having lower gas fractions on average compared to more active systems \citep[e.g.,][and references therein]{saintongecatinella22}. The higher median SFR of PMs compared to xCG then leads to this observed shift. The dependence of $\tau_{\rm depl.}$ on SFR is much shallower than that of \fgas\ on SFR, such that the median depletion time offsets vary by an only negligible amount between the two control matching schemes.\medskip

We close this section by referring the reader to Appendix \ref{appsect:syschecks}, where we discuss the following sources of systematic uncertainty on our findings for PMs: the choice of the SFR and \mstar\ estimator (Appendix \ref{appsect:altSFRnMstell}), different aperture correction strategies to account for incomplete recovery of spatially extended CO emission by the IRAM 30\,m telescope beam (Appendix \ref{appsect:aperturesys}), and the recipe for assigning CO-to-H$_2$ conversion factors \aCO\ to all PMs on a galaxy-by-galaxy basis, depending on statistical metallicity estimates and their `starburstiness', i.e. sSFR-offset from the MS (Appendix \ref{appsect:obsproxPM}). To summarise, these appendices demonstrate that, when varying the SFR and \mstar\ estimators and the control matching approach (i.e. matching on 2 vs. 3 physical parameters), our conclusion that {\it both} the gas fractions and depletion times of PMs are enhanced compared to normal, non-interacting galaxies holds for 90\% of the combinations explored. Aperture correction methodology does not contribute substantially to the systematic error budget. Finally, we show that the gas fraction and depletion time enhancements are not merely the result of adopting our specific recipe for converting CO luminosities to molecular gas masses.

\subsection{Post-merger molecular-to-atomic gas mass ratios}
\label{sect:PM_H2overHI}

In this section we consider the ratio of the mass in the molecular hydrogen phase to that in the atomic (HI) phase, and compare this H$_2$-to-HI mass ratio for PMs and control-matched xCG galaxies. The source catalog of the extended GALEX Arecibo SDSS Survey \citep[xGASS;][]{catinella18} tabulates HI mass measurements for 350 xCG galaxies, based on the velocity-integrated line flux of the 21\,cm line of hydrogen. Upper limits on the HI line flux are available for a further 127 xCG galaxies. In combination with the \COone\ detections and upper limits, the xCG sample provides a pool of 367 control galaxies for which a well-defined constraint (direct measurements or lower/upper limit) on the H$_2$-to-HI mass ratio can be formed. For PMs not drawn from the xCG sample we take HI masses from \citet{ellison15, ellison18}. Overall, we were able to derive H$_2$-to-HI mass ratios for 35 out of 39 PMs, one of which is a lower limit.\medskip

For both matching on redshift and \mstar, and redshift, \mstar\ and SFR, the median offset of PM H$_2$-to-HI mass ratios relative to xCG galaxies is consistent with zero within 1\,$\sigma$ (see Fig. \ref{fig:PM_HInr21}a). Specifically, we measure a median offset of $\langle\Delta(\mhtwo/\mHI)\rangle\,{=}\,-0.03^{+0.06}_{-0.08}$\,dex ($0.02^{+0.08}_{-0.12}$\,dex) for control matching on redshift, \mstar\ and SFR (on redshift and \mstar\ only). For the HI-to-stellar mass ratio we find an offset of $\Delta\fgas$(HI)\,=\,0.21$^{+0.08}_{-0.05}$\,dex when control matching with respect to all three parameters. (Consistent with the median excess of $\Delta\fgas$(HI)\,=\,0.2\,dex reported for a larger sample of HI-detected PMs by \citealp{ellison18}.) Our analysis of PM \mhtwo-to-\mstar\ ratios in Sect. \ref{sect:PMoffsets_best} returned a median excess of $\Delta\fgas$(H$_2$)\,=\,0.16$^{+0.05}_{-0.04}$\,dex for matching on redshift, \mstar\ and SFR. Taken separately, both the atomic and the molecular hydrogen content of PMs is thus enhanced, with the similar levels of enhancement leading to on average normal H$_2$-to-HI mass ratios.

\begin{table*}
\centering
\caption{Median $\Delta\fgas$ \& $\Delta\tau_{\rm depl.}$ offsets measured for all samples relative to xCG control galaxies.}
\label{tab:resultsoverview}
\begin{tabular}{l @{\quad\vline\quad} ccr | lcc}
\hline\hline \\[-2ex]
Sample & \multicolumn{2}{c}{$\langle\Delta\fgas\rangle$} & & & \multicolumn{2}{c}{$\langle\Delta\tau_{\rm depl.}\rangle$}\\
& \multicolumn{2}{c}{[dex]} & & & \multicolumn{2}{c}{[dex]}\\[0.5ex]
(1) & (2) & (3) & & & (4) & (5)\\[1ex]
\hline \hline
\multicolumn{1}{r}{\it \raisebox{-1ex}{Matched on:}} & \raisebox{-1ex}{$z$,\,\mstar} & \raisebox{-1ex}{$z$,\,\mstar,\,{\it SFR}} & & &  \raisebox{-1ex}{$z$,\,\mstar} & \raisebox{-1ex}{$z$,\,\mstar,\,{\it SFR}}\\[2ex]
\hline \\[-2ex]
IPGs & 0.27$_{-0.08\,(-0.31)}^{+0.05\,(-0.02)}$ & 0.10$_{-0.05\,(-0.01)}^{+0.06\,(+0.10)}$ & & & $-0.02_{-0.07\,(-0.11)}^{+0.07\,(+0.07)}$ & 0.06$_{-0.06\,(-0.09)}^{+0.07\,(+0.11)}$\\[2ex]
PMs & 0.25$_{-0.06\,(-0.12)}^{+0.09\,(+0.24)}$ & 0.16$_{-0.04\,(-0.06)}^{+0.05\,(+0.08)}$ & & & 0.11$_{-0.09\,(-0.24)}^{+0.05\,(+0.13)}$ & 0.09$_{-0.02\,(-0.01)}^{+0.04\,(+0.12)}$\\[2ex]
yPSBs & 0.25$_{-0.17\,(-0.07)}^{+0.09\,(+0.24)}$ & 0.18$_{-0.04\,(+0.04)}^{+0.07\,(+0.39)}$ & & & 0.24$_{-0.18\,(-0.18)}^{+0.11\,(+0.36)}$ & 0.13$_{-0.15\,(-0.03)}^{+0.07\,(+0.35)}$\\[2ex]
mPSBs & $-0.39_{-0.17\,(-0.04)}^{+0.17\,(+0.34)}$ & $-0.55_{-0.06\,(-0.19)}^{+0.12\,(+0.65)}$ & & & $-0.49_{-0.31\,(-0.14)}^{+0.22\,(+1.19)}$ & $-0.57_{-0.13\,(-0.41)}^{+0.09\,(+0.69)}$\\[2ex]
DETGs & 0.07$_{-0.24\,(+0.14)}^{+0.40\,(+0.43)}$ & 0.35$_{-0.32\,(-0.78)}^{+0.25\,(+0.60)}$ & & & 0.44$_{-0.29\,(-0.79)}^{+0.20\,(+1.07)}$ & 0.25$_{-0.13\,(-0.51)}^{+0.39\,(+0.76)}$ \\[1ex]
\hline\hline
\end{tabular}
\tablecomments{All offset measurements were derived using the `const. \fgas' aperture correction (see Appendix \ref{appsect:apcorr}), and \mstar\ and SFR values from Sects. \ref{sect:SFRderiv}/\ref{sect:Mstellderiv}. Errors quoted are 68\% confidence intervals. The values in parentheses span the full range of median offsets found in our analysis of systematic uncertainties in Appendix \ref{appsect:altSFRnMstell_compsamps} (see full overview of results in Figs. \ref{appfig:physcontrolmatch_sys} to \ref{appfig:compsamp_tdepl_sysoverview}).}
\end{table*}

\subsection{\COtwo\ line excitation in post-mergers}
\label{sect:PM_r21}

\COtwo\ spectra are available for 334/488 non-interacting xCG galaxies and 32/39 PMs, enabling a comparison of the excitation of the second rotational level of \CO\ between PMs and normal galaxies via the line ratio parameter
\begin{equation}
r_{21} = \frac{L'_{\COtwo}}{L'_{\COone}} = \frac{I_{\COtwo}}{I_{\COone}}\left(\frac{\nu_{\rm obs}(\COone)}{\nu_{\rm obs}(\COtwo)}\right)^2~,
\end{equation}
where the velocity-integrated line fluxes $I_{\rm CO}$ are in units of Jy\,km/s. We could derive well-defined measurements of $r_{21}$ -- i.e., where at least one of the two velocity-integrated line fluxes $I_{\rm CO}$ is directly measured at $S/N\,{>}$\,3 -- for 223 xCG control galaxies and 32 PMs. \COone\ and \COtwo\ line fluxes are both subject to aperture corrections, with median correction factors for the \COone\ (\COtwo) transition of xCG galaxies amounting to 1.5 (2.35) when adopting our preferred aperture correction which is based on the assumption of a spatially invariant \fgas\ within galaxies (see Appendix \ref{appsect:apcorr}). For PMs the median aperture correction factors are very similar at 1.54 (\COone) and 2.32 (\COtwo). To reduce systematic uncertainties related to aperture losses, we give precedence to \COtwo\ velocity-integrated line fluxes from APEX (available for 28 xCG galaxies and one PM) over those measured on spectra from the IRAM 30\,m telescope. For the remaining galaxies we estimate galaxy-integrated line ratios by scaling upward the ``raw" measurement of $r_{21}$ by the ratio of the aperture correction factors $\mathcal{A}_{\COtwo}/\mathcal{A}_{\COone}$. The median value of the individual aperture correction ratios is 1.48 for PMs and 1.57 for galaxies in the xCG reference sample.\medskip

The cumulative distribution of $r_{21}$ offset of PMs relative to xCG control galaxies is shown in Fig. \ref{fig:PM_HInr21}b and characterised by a narrow log-space dispersion of $<$0.2\,dex. The median offset is $\langle\Delta r_{21}\rangle$\,=\,0.09$^{+0.03}_{-0.01}$\,dex for PMs matched on redshift, \mstar\ and SFR, with matching on just the first two quantities producing an identical result. This mild enhancement is in line with the tendency reported by \citet{leech10} for the excitation ratio $r_{31}$ of the next higher transition to increase weakly with merger stage. At a value of 0.63, the overall median $r_{21}$ for xCG galaxies with a measurable \COtwo-to-\COone\ line ratio is somewhat smaller than the characteristic value reported by \citet{saintonge17}, but in excellent agreement with recent findings in \citet{denBrok21, yajima21} and \citet{leroy22} for nearby galaxies mapped in multiple CO transitions, and hence free of systematic uncertainties related to aperture correction methodology. In line with the $\sim$25\% enhancement of PM \COtwo-to-\COone\ line ratios inferred via control matching, the straightforward (i.e. without control matching) overall sample median of the 32 PMs with $r_{21}$ measurements is 0.74. Given the small error bars on the median $\langle \Delta r_{21}\rangle$, and the fact that the average aperture corrections applied to PMs and xCG galaxies do not differ substantially (see above), this suggests a small but statistically significant increase of the excitation of the $J$\,=\,2 level in PMs relative to normal xCG galaxies. We further discuss this result, and sample-internal variations of $r_{21}$ among PMs, in Sect. \ref{sect:PM_ISM}.

\begin{figure*}
\centering
\includegraphics[width=.65\textwidth]{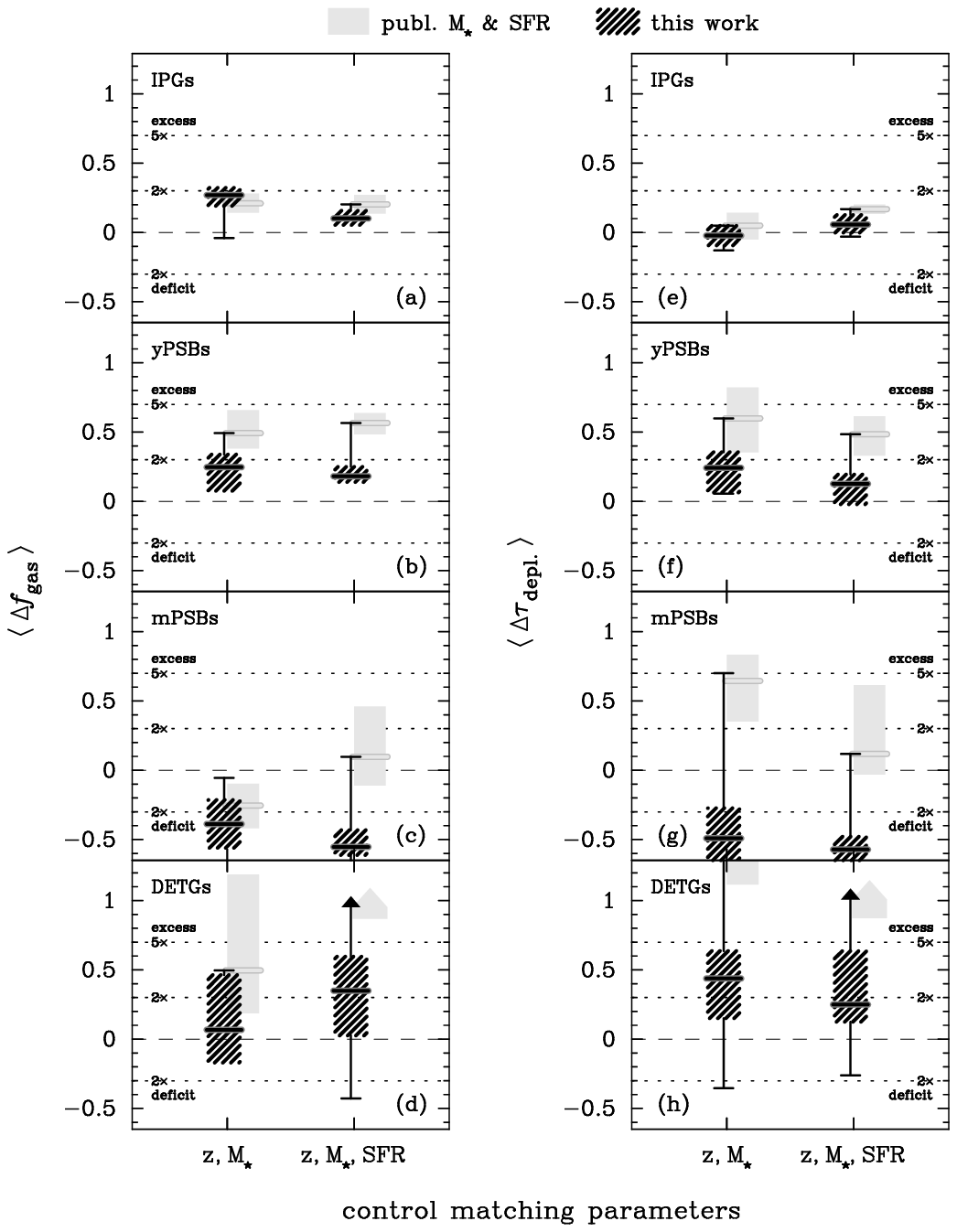}
\caption{Median gas fraction offsets $\langle\Delta\fgas\rangle$ ({\it left}) and depletion time offsets $\langle\Delta\tau_{\rm depl.}\rangle$ ({\it right}) for the comparison samples of (top to bottom): interacting pair galaxies (IPGs), young post-starburst galaxies (yPSBs), mature post-starburst galaxies (mPSBs), and dust lane early-type galaxies (DETGs). In a given column both the results for control matching on the 2-D (3-D) parameter space of $z$ and \mstar\ ($z$, \mstar\ \& SFR) are shown. The hatched rectangles and solid horizontal bars mark, respectively, the 68\% confidence region and median for offset measurements obtained with our `best-estimate' \mstar\ and SFR values. The fine T-shaped error bars extending beyond the 68\% confidence region span the full range of median offsets found in our analysis of systematic uncertainties (see Appendix \ref{appsect:altSFRnMstell_compsamps} for details). For comparison, we also show in light grey the median offsets and associated 1\,$\sigma$ uncertainties derived via control-matching that uses \mstar\ and SFR values published in the original data papers for each sample.
\label{fig:compsamp_overview}}
\end{figure*}

\subsection{Gas fraction and depletion time offsets for comparison samples: pairs, post-starbursts and dust lane early-type galaxies}
\label{sect:compgal_offsets}
Our control-matching approach ensures that we can relate the molecular gas properties of post-mergers in a consistent way to those of the  comparison samples introduced in Sect. \ref{sect:compsampintro}, even if the comparison samples are not matched to PMs in terms of \mstar\ or SFR (cf. Figs. \ref{fig:MSplane} \& \ref{appfig:comparisonMSFRz}). Analogously to the analysis for PMs in Sect. \ref{sect:PMoffsets_best}, this section presents offset measurements for gas fractions and depletion times using our `best-estimate' \mstar\ and SFR values. The results (including a summary accounting of systematic errors) are displayed in Fig. \ref{fig:compsamp_overview} and tabulated in Table \ref{tab:resultsoverview}. A detailed overview of systematic uncertainties -- which are dominated by the choice of the \mstar\ and SFR estimator -- for all comparison samples can be found in Appendix \ref{appsect:altSFRnMstell_compsamps}. The ISM properties of all comparison samples discussed below have been studied in previous works. To assess whether our approach for deriving offsets can lead to conclusions that differ from these papers, we also apply our control-matching technique to the original literature \mstar\ and SFR estimates; the according results are plotted in grey in Fig. \ref{fig:compsamp_overview}.

\subsubsection{Interacting Pair Galaxies -- IPGs}

For IPGs \citet{violino18} and \citet{pan18} report small gas fraction enhancements of $\sim$0.1\,dex when control-matching on redshift, \mstar\ and SFR with an approach similar to ours, with $\tau_{\rm depl.}$ offsets varying around zero. When matching on redshift and \mstar\ only, both studies measure a somewhat larger gas fraction enhancement ($\tau_{\rm depl.}$ reduction) compared to their control galaxies. Here we consider a larger IPG sample combining the 11 objects from \citet{violino18} with a further 26 IPGs from \citet{pan18}.\medskip

The offsets we infer for this new IPG sample are in good agreement with \citet{violino18} and \citet{pan18}, both when using the respective \mstar\ and SFR measurements tabulated in \citet{violino18} and \citet{pan18} and our own `best-estimate' values (measurements reported in grey and with black hatching, resp., in Fig. \ref{fig:compsamp_overview}a/e). The 1\,$\sigma$ errors on the medians are small for IPGs, enabling us to draw the following robust conclusions: IPGs display small and commensurate 15-25\% gas fraction and depletion time enhancements compared to normal galaxies when control-matched on redshift, \mstar\ and SFR; for matching on the two parameters redshift and \mstar, IPG depletion times are indistinguishable from xCG control galaxies, but the gas fraction offset increases to $\sim$70-80\%. The T-shaped error bars in the top row of Fig. \ref{fig:compsamp_overview} show that the systematic uncertainty affecting the median offsets measured is also small. (We quantify the median offsets found for all \mstar\ and SFR estimators tested in Appendix \ref{appsect:altSFRnMstell_compsamps}). The increase (decrease) of the median gas fraction ($\tau_{\rm depl.}$) offset found when matching on redshift and \mstar\ only -- compared to the offsets measured when SFR is fixed as well -- is caused by the IPG sample having larger average SFRs than the full pool of xCG control galaxies, in a similar way as already discussed for PMs in Sect. \ref{sect:PMoffsets_best}.

\subsubsection{Young Post-StarBursts -- yPSBs}

Ours is the first control-matching analysis of the yPSB sample, as \citet{rowlands15} focused on the evolution with starburst age of the H$_2$ content of yPSBs in their work. We can nevertheless assess the agreement of our results with theirs in a qualitative sense, as \citet{rowlands15} referenced their gas fraction and depletion time measurements to the range of values typically spanned by low-$z$ galaxy samples like xCG (see their Figs. 5/7).\medskip

A consistent picture of gas fraction enhancements in yPSBs is conveyed by the 2nd row of Fig. \ref{fig:compsamp_overview}, where sample median offsets are always positive, regardless of the control-matching scheme implemented or the \mstar\ and SFR catalogs adopted. This enhancement is of order a factor 2 in a cross-catalog averaged sense for both matching on redshift and \mstar\, and on redshift, \mstar\ and SFR (see Appendix \ref{appsect:altSFRnMstell_compsamps}, Fig. \ref{appfig:compsamp_fgas_sysoverview}c/d). The median gas fraction enhancements $\langle\Delta\fgas\rangle$ inferred with our `best-estimate' \mstar\ and SFR values are consistent within 1\,$\sigma$ with these cross-catalog averages, but are systematically lower than the median offsets calculated when using \mstar\ measurements from \citet{rowlands15} (grey symbols in Fig. \ref{fig:compsamp_overview}b/f). The fact that stellar masses tabulated by \citet{rowlands15} are on average smaller than alternative \mstar\ derivations (see Fig. \ref{appfig:Mstellsyscomp} in Appendix \ref{appsect:altSFRnMstell}) contributes to this systematic shift.\medskip

Analogously to yPSB gas fraction offsets, when averaged across all \mstar\ and SFR catalogs, the $\tau_{\rm depl.}$ offsets we measure for yPSBs are also comparable when control-matching in the 2-D or 3-D parameter space of redshift and \mstar\ or redshift, \mstar\ and SFR. They provide evidence of an increase of the median $\langle\Delta\tau_{\rm depl.}\rangle$ by 40-80\%. Cross-catalog scatter of median depletion time offsets is larger than for yPSB gas fractions, with median offsets $\langle\Delta\tau_{\rm depl.}\rangle$ varying from almost zero to a factor 3--4 (see Appendix \ref{appsect:altSFRnMstell_compsamps}, Fig. \ref{appfig:compsamp_tdepl_sysoverview}c/d). Drawing a robust conclusion re. the exact depletion time offset is thus hard despite the overall cross-catalog average pointing to a net enhancement (i.e. lower SFE). The finding of a net enhancement of depletion times in yPSBs is qualitatively consistent with the result we obtain when using the original \citet{rowlands15} \mstar\ and SFR values.

\subsubsection{Mature Post-StarBursts -- mPSBs}
\label{sect:mPSB_offsets}

For galaxies in their sample of mPSBs, \citet{french15} find gas fractions spanning the full range from fairly gas-rich star-forming galaxies to gas-poor, low-$z$ early-types, with a distribution skewed towards lower gas fraction values than the COLD GASS sample \citep[Fig. 6 in][]{french15}. They report low SFEs (long $\tau_{\rm depl.}$) for mPSBs, with the exact offset compared to normal SFGs depending on the choice of SFR. Here we carry out the first control-matching analysis for this sample. The challenges of deriving SFR constraints for galaxies in the mPSB sample that represent actual measurements rather than upper limits are discussed in \citet{french15, french18a} and \citet{smercina18}. They directly influence the degree of scatter we observe for our offset measurements for depletion times ($M({\rm H}_2)$/SFR), and/or when control-matching on SFR as a third parameter.\medskip

In keeping with this, the gas fraction offset measurements at fixed redshift and \mstar\ (Fig. \ref{fig:compsamp_overview}c) are least affected by uncertain SFR measurements. Here offsets are consistently shifted into the gas fraction deficit regime $\langle\Delta\fgas\rangle\,{<}\,$0\,dex, leading to a cross-catalog average deficit of approx. -0.3\,dex (factor 2) that recalls the skew towards low gas fractions in \citet{french15}. SFR-related systematics dominate the spectrum of gas fraction offset measurements for matching on redshift, \mstar\ and SFR (Figs. \ref{fig:compsamp_overview}c \& \ref{appfig:compsamp_fgas_sysoverview}f), with individual measurements ranging from a factor 5 deficit to a $\sim$25\% enhancement of gas fractions. SFR estimates based on IR photometry and/or SED fitting (here: the \citealp{chang15} and \citealp{salim16} SFRs, as well as our own hybrid UV+IR measurements) return the highest SFRs for mPSBs. Should these represent overestimates -- as is plausible if the mid-IR emission of mPSBs has contributions from both dust heated by A-type stars and the stellar continuum (\citealp{french15, smercina18}, and references therein) -- control matching on SFR would then pair such mPSBs with xCG control galaxies with too high an SFR (and thus gas fraction), resulting in a spuriously low gas fraction offset in the $\langle\Delta\fgas\rangle\,{<}\,$0\,dex deficit regime. On the other hand, \citet{salim16} showed that the choice of attenuation law strongly impacts the SFRs of E+A galaxies, and argue that SED fits have smaller residuals when adopting an attenuation law that produces higher SFRs for such galaxies. Fig. \ref{appfig:compsamp_fgas_sysoverview}f in Appendix \ref{appsect:altSFRnMstell_compsamps} also highlights that control-matching with systematically lower SFR estimates from the MPA-JHU catalog, based on H${\alpha}$ line emission or the D$_n$4000 index, and which are also adopted by \citet{french15}, leads to the conclusion that gas fractions of mPSBs are marginally enhanced ($\sim$25\%) at fixed redshift, \mstar\ and SFR. We will return to these discrepant findings when discussing the evolution of ISM content during interactions in Sect. \ref{sect:discuISMevo}. Importantly, the time scales on which SFR is probed/averaged for these different tracers may also play a role for PSBs with rapidly decreasing SFR.\medskip

In closing this section on mPSBs we note that the determination of depletion time offsets is subject to the same strong SFR-related systematics (see Figs. \ref{fig:compsamp_overview}g \& \ref{appfig:compsamp_tdepl_sysoverview}e/f). Assigning low H${\alpha}$- or D$_n$4000-based SFR values suggests that  mPSBs have systematically enhanced depletion times, while adopting the higher set of SFRs would lead to the conclusion that their depletion times are systematically shorter (suggesting an enhanced SFE).\newpage

\subsubsection{Dust Lane Early-Type Galaxies -- DETGs}

In their study of DETGs \citet{davis15} reported exceptionally long depletion times (sample median in excess of 6.6\,Gyr). Molecular-to-stellar mass ratios were found to vary substantially, but at a few percent on average matched values typically found for SFGs with a similar stellar mass of around 10$^{11}$\,\msun. When implementing our control-matching approach to determine $\tau_{\rm depl.}$ offsets with the \mstar\ and SFR values given in \citet{davis15}, we confirm a more than 10-fold depletion time enhancement for both our control matching schemes (grey symbols, Fig. \ref{fig:compsamp_overview}h). These depletion time enhancements are lower, but still a factor of a few on average, for four of the other five \mstar\ and SFR catalogs tested (see Figs.  \ref{fig:compsamp_overview}h \& \ref{appfig:compsamp_tdepl_sysoverview}g/h). Even if the measurement involving MPA-JHU \mstar\ and SFR values is an outlier in this respect, we thus still consider the measurement of a $\tau_{\rm depl.}$ enhancement for the DETG sample as robust.\medskip

The gas fractions of DETGs (Fig. \ref{fig:compsamp_overview}d) appear to be enhanced by a smaller amount than the DETG depletion times, with the catalog-averaged median enhancement amounting to $\sim$70\%. Due to large uncertainties, many of the median gas fraction offset measurements $\langle\Delta\fgas\rangle$ are still consistent with zero when control-matching on redshift and \mstar. This is qualitatively in line with our characterisation of the findings in \citet{davis15} at the beginning of this section, even if it should be noted that formally the use of \mstar\ and SFR values published in \citet{davis15} results in a median gas fraction enhancement that is higher than for most other combinations of \mstar\ and SFR (see Fig. \ref{appfig:compsamp_fgas_sysoverview}g/h in Appendix \ref{appsect:altSFRnMstell_compsamps}). Stronger evidence for a positive median gas fraction offset of DETGs -- roughly matching the enhancement of depletion times -- emerges when they are matched to xCG galaxies also on SFR. Such a gas content-driven enhancement of both \fgas\ and $\tau_{\rm depl.}$ in principle conforms to expectations in a scenario where gas reservoirs of initially ISM-poor objects are replenished through wet-dry minor mergers (see discussion in Sect. \ref{sect:discuISMevo}).

\section{Discussion}
\label{sect:discussion}

We have built a data base of 95 galaxies at different interaction stages (39 close kinematic pair galaxies, 39 post-mergers, and 17 dust lane early-type galaxies which have recently experienced a minor merger), plus 42 post-starbursts with a range of post-burst ages. The data for the 39 post-mergers are presented for the first time in this paper, and we have gone to a substantial effort to homogenise the measurements of the physical properties (molecular gas mass, stellar mass, star-formation rate) of the PM sample and of all other comparison samples of interacting or post-starburst galaxies. 120 of the overall total of 137 interacting or post-starburst galaxies are CO-detected with the IRAM 30\,m telescope, enabling us to study the evolution of the molecular gas properties throughout the merger sequence. We do so via a control-matching approach that relies on a reference sample of $\sim$500 galaxies from the xCOLD GASS survey (also observed in \COone\ with the IRAM 30\,m telescope), from which we took care to remove all objects that show evidence of being in an interaction. We draw on the appropriate tools from the field of survival analysis to be able to include also the information contained in the CO non-detections among the interacting galaxies and in the xCG reference sample.\medskip

Though a key consistency check from the point of view of methodological consistency, adopting a fully homogenised set of physical measurements (e.g., via the same tracer of SFR or the same recipe for converting \COone\ line luminosities to H$_2$ mass) for galaxies in substantially different evolutionary stages comes with its own risks. In the preceding section we have laid out in detail which findings we hold to be robust, and which are subject to large systematic uncertainty. In the following we attempt to synthesize this information into a coherent observational picture of the evolution of molecular gas fractions and depletion times with merger stage, and of the interplay between enhanced gas content and star-formation efficiency in shaping the merger events with the highest levels of star-formation activity.

\subsection{The ISM properties of low-$z$ post-merger galaxies}
\label{sect:PM_ISM}

Numerous recent studies have shown that PM galaxies at low redshift have enhanced SFRs (\citealp[e.g.][]{ellison13b, thorp19, bickley22}, but see also \citealp{barrera15}), continuing a trend of rising excess SF activity with advancing merger stage \citep{woodsgeller07, knapenjames09, scudder12, patton13, ellison13b, feng20}. Control-matching as adopted here has quantified these SFR-enhancements to lie in the range $\Delta$SFR\,=\,0.25-0.6\,dex for galaxies matched on redshift and stellar mass, with the strongest increase measured for centrally concentrated SFR \citep{ellison13b} and the enhancements of galaxy-integrated SFRs spanning the lower half of the range quoted above \citep{ellison13b,ellison18}. For the molecular gas phase of PMs, which provides the fuel for the SF process, we measure a gas-to-stellar mass ratio enhancement $\Delta\fgas$ of 0.1-0.5\,dex, depending on the choice of SFR and \mstar\ (see Sect. \ref{sect:PMoffsets_best}), with a ``best-estimate" value of 0.25\,dex. Given the mildly supra-linear galaxy-integrated Kennicutt-Schmidt relation SFR\,${\propto}\,\mhtwo^{\beta}$, with $\beta\,{\simeq}$\,1.2 \citep[e.g.,][]{saintonge12, sargent14}, the similar $\Delta$SFR and $\Delta\fgas$ values for \mstar-matched PMs suggest that -- on average -- increased gas content is the main driver of the excess SFR activity. SFE variations play a secondary role, but are not entirely negligible, as indicated by the fact that the depletion time contrast $\Delta\tau_{\rm depl.}$ between PMs and control galaxies is smaller than the gas fraction enhancement irrespective of the control matching strategy adopted (see discussion of systematics in Sect. \ref{appsect:altSFRnMstell_PMs}). Furthermore, while the spread of gas fraction measurements in the PM and xCG samples is similar (interquartile range 0.5\,dex), depletion time measurements show a larger scatter for PMs than for normal xCG galaxies (0.47 vs. 0.35\,dex), again pointing to an non-negligible interplay between merger activity and SFE variations. We will return to this topic in Sect. \ref{sect:discuSBmerger}.\medskip

In absolute terms, the gas fraction offset between PMs and normal xCG galaxies translates to \mhtwo/\mstar\ ratios of 12\% and 7\% if we qualitatively calculate median gas fractions of these samples. (I.e. without accounting for the scalings of \fgas\ with \mstar\ and SFR, as is explicitly done during control matching.) A key factor in setting absolute gas fraction values is the CO-to-H$_2$ conversion factor \aCO, which is not measured directly but which we assign to galaxies on an object-by-object basis, based on statistical gas phase metallicity estimates and `starburstiness' (see Sect. \ref{sect:XCO}). We use the same prescription to assign \aCO\ values to PMs and xCG control galaxies. The gas fraction enhancement of PMs could shrink or even turn into a deficit were we to posit that the ISM in PMs is starburst-like with correspondingly more turbulent gas and optically thinner CO emission, leading to lower \aCO\ and consequently to lower gas masses. We show in Appendix \ref{appsect:obsproxPM} that gas fractions and depletion times of PMs are enhanced compared to normal galaxies also for the direct observational proxies of these quantities, $\lco/\mstar$ and $\lco/\sfr$, but this scenario cannot test for the impact of such galaxy type-dependent \aCO\ shifts as it again puts PMs and control galaxies on an equal footing as regards the \aCO\ factors (not) applied.\medskip

While line flux ratios between \COone\ and rotational levels higher than $J$\,=\,2 distinguish more clearly between starburst and normal galaxy ISM conditions \citep[e.g.][]{liu21}, their mild median enhancement $\langle\Delta r_{21}\rangle$\,=\,0.09\,dex argues against widespread turbulent starburst-like ISM conditions in PMs and consequently against collectively low, ULIRG-like CO-to-H$_2$ conversion factors. With a median ratio $\langle r_{21}\rangle$\,=\,0.74 the excitation of the $J$\,=\,2 level of PMs on average resembles that of $z\,{\sim}$\,1.5 MS-galaxies \citep{daddi15}, which have roughly Milky Way-like \aCO\ values \citep[e.g.,][]{daddi10a, magdis12}. However, within the PM sample we find a clear link between gas excitation and starburstiness, traced by $r_{21}$ rising with MS-offset $\Delta$(MS) (Pearson's correlation coefficient (PCC) $\rho_{r_{21},\Delta{\rm (MS)}}$\,=\,0.57, significant at the $p\,{<}$\,0.001 level). This qualitatively matches the $r_{21}$ vs. $\Delta$(MS) correlation found by \citet{keenan25} for non-interacting, low-$z$ galaxies. Therefore, there is only a weak tendency for the $r_{21}$-offset $\Delta r_{21}$ of individual PMs to increase with MS-offset (PCC $\rho_{\Delta r_{21},\Delta{\rm (MS)}}$\,=\,0.34, significant at the $p\,{\sim}$\,0.05 level). The PMs with the highest levels of excess star formation ($\Delta$(MS)\,=\,10) on average display the thermal-like line ratios ($r_{21}\,{\simeq}$\,1) that are most readily produced by turbulent, warm and/or dense gas \citep[e.g.,][]{leroy22}. This provides evidence in support of our approach of assigning successively lower \aCO\ values to off-MS starbursts the larger their offset from the MS (see Sect. \ref{sect:XCO}), complementing commonly cited dynamical arguments for low \aCO\ values in low-$z$ starbursts (e.g., \citealp{downessolomon98,hinzrieke06,isbell18} or \citealp{molina20}, but see also \citealp{papadopoulos12} and \citealp{dunne22} for techniques leading to higher CO-to-H$_2$ conversion factors in such systems).\medskip

\begin{figure}
\centering
\includegraphics[width=0.48\textwidth]{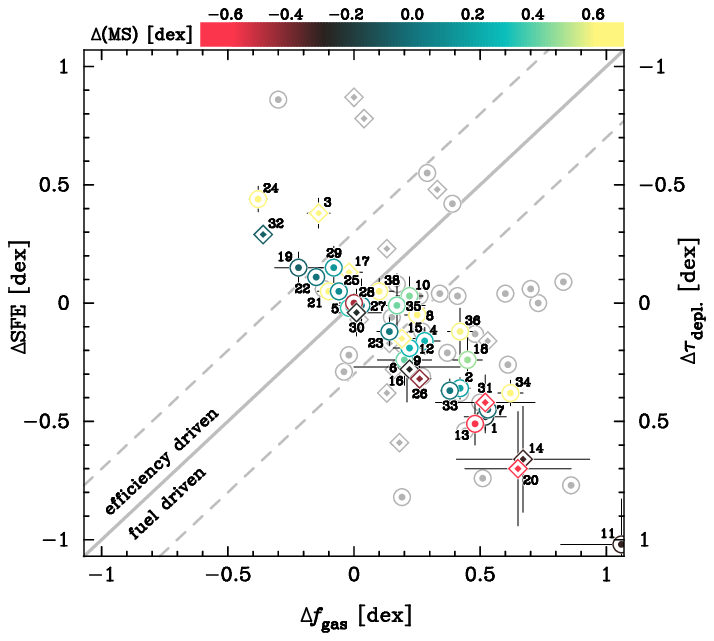}
\caption{Galaxy-integrated $\Delta$SFE vs. $\Delta\fgas$ values for PMs. For offsets calculated in 3D control-matching on redshift, \mstar\ and SFR individual data points are coloured by MS-offset $\Delta$(MS), with galaxies in the off-MS regime appearing with yellow (high-$\Delta$(MS) starbursts) or red (low-$\Delta$(MS) PMs in the green valley) symbols. Diamond-shaped symbols denote AGN/composite objects, error bars the 1\,$\sigma$ error on the measurement. In the background (grey data points) we plot offset measurements derived for control matching on redshift and \mstar. Data points are labelled by their sequential ID from Table \ref{tab:prop}. The solid diagonal line separates fuel-driven ($\Delta\fgas\,{>}\,\Delta\sfe$, lower right wedge of plot) from efficiency-driven objects ($\Delta\sfe\,{>}\,\Delta\fgas$, upper left corner). Dashed lines are drawn at a factor 2 offset from the solid 1:1 line.
\label{fig:dfgasvsdSFE}}
\end{figure}

Despite their lower molecular gas mass per unit \lco, not all high-$\Delta$(MS) PMs are characterised by a high SFE or an SFE that is similar to that of xCG control galaxies with similar SFR levels. In Fig. \ref{fig:dfgasvsdSFE} we report the position of our 39 PMs in the space of gas fraction offset $\Delta\fgas$ vs. depletion time offset $\Delta\tau_{\rm depl.}$ (or, equivalently, $\Delta \sfe$) calculated in Sect. \ref{sect:PMoffsets_best}. Galaxies located in the wedge to the lower right of the figure diagonal are fuel-driven ($\Delta\fgas\,{>}\,\Delta\sfe$), those in the wedge on the upper left are efficiency-driven ($\Delta\sfe\,{>}\,\Delta$\fgas). For PMs matched in the 3D parameter space of redshift, \mstar\ and SFR the offsets $\Delta\fgas$ and $\Delta\sfe$ are closely anti-correlated, reflecting the fact that -- once SFR is fixed -- residual gas mass variations are quite small, with stellar mass and SFR primarily contributing a different normalisation factor for the underlying offsets $\Delta$M(H$_2$). Individual data points are coloured by MS-offset $\Delta$(MS), with galaxies in the off-MS regime appearing with yellow (high-$\Delta$(MS) starbursts) or red (low-$\Delta$(MS) PMs in the green valley) symbols.\medskip

Off-MS PMs with $\Delta$(MS)\,$>$\,0.6\,dex can have both among the largest efficiency enhancements and -- at the opposite end of the PM distribution in $\Delta\sfe$--$\Delta\fgas$ space -- substantial excess gas. While high-$\Delta\fgas$ PMs encompass systems with some of the largest and smallest MS-offsets, the region of efficiency-driven galaxies ($\Delta\sfe\,{>}\,\Delta$\fgas) in Fig. \ref{fig:dfgasvsdSFE} is preferentially occupied by PMs with enhanced SFRs above the ridge line of the MS or even in the off-MS starburst region. We note that this conclusion is independent of the exact shape of the MS, and would hold also were we to adopt a curved MS locus with flattening above \mstar\,${\simeq}\,10^{10}\,\msun$, rather than a single power-law form (see Sect. \ref{sect:PMonMSnSK} and Fig. \ref{fig:MSplane}). In 40\% of PMs\footnote{All fractions quoted in this paragraph are for offsets calculated when control matching on redshift, \mstar\ and SFR. In comparison, matching on redshift and \mstar\ only leads to a $\sim$10\% higher fraction of fuel-driven PMs. The grey data points in Fig. \ref{fig:dfgasvsdSFE} illustrate the more scattered distribution of PMs in $\Delta\sfe$--$\Delta\fgas$ space for this control matching scenario.} the SF activity is driven in roughly equal proportions by gas content and SFE, with $\Delta\fgas$ and $\Delta\sfe$ differing by no more than a factor 2. Clearly efficiency-driven PMs ($\Delta\sfe-\Delta$\fgas$\,{>}$\,0.3\,dex) make up only 10\% of our sample, and 50\% are strongly fuel-driven with $\Delta\fgas-\Delta\sfe\,{>}$\,0.3\,dex. In their spatially resolved ALMA analysis of SF regions in PMs, \citet{thorp22} find SF to be fuel-driven in 52\% of the PM spaxels studied. This is $\sim$20\% smaller than the 71\% we infer in this galaxy-integrated study of PM gas properties, if we count as fuel-driven all PMs for which $\Delta\fgas\,{>}\,\Delta\sfe$. The difference may stem from \citet{thorp22} focusing exclusively on regions of enhanced SFR with respect to the spatially resolved galaxy MS. From our Fig. \ref{fig:dfgasvsdSFE} it is evident that PMs lying substantially below the MS all fall in the wedge of fuel-driven objects, such that had we limited our calculations to PMs with $\Delta$(MS)\,$>$\,0\,dex the fraction of fuel-driven PMs would decrease to $\sim$55\%, closely matching the finding of \citet{thorp22}.\medskip

We infer normal molecular-to-atomic gas mass ratios for PMs on average (see Sect. \ref{sect:PM_H2overHI}). Performing an analogous measurement for the pair galaxies with HI-mass measurements\footnote{HI masses are available for 27/39 IPGs, from the following sources: the xGASS catalog \citep[22 galaxies]{catinella18}, \citet[1 galaxy]{ellison15} and the ALFALFA $\alpha$-100 catalog of \citet[4 galaxies]{haynes18}.} in our IPG sample, we find that these too have H$_2$-to-HI ratios consistent with those of xCG galaxies within 1\,$\sigma$. (The median offset of IPGs is $\langle\Delta(\mhtwo/\mHI)\rangle\,{=}\,-0.09^{+0.13}_{-0.11}$\,dex for matching on redshift and \mstar, and $0.03^{+0.11}_{-0.08}$\,dex when additionally matching on SFR.) \citet{lisenfeld19} studied the H$_2$-to-HI ratios for two sets of low-$z$ close pair galaxies ($r_p\,{=}$\,5-20\,kpc\,$h^{-1}$). They find that morphologically unperturbed close pair galaxies on average have normal H$_2$-to-HI ratios, while pairs consisting of morphologically disturbed or even partially coalesced late-type (S+S) galaxies have molecular-to-atomic gas mass ratios that are increased $\sim$4-fold. (See also the similar conclusion for merging pairs at the pericenter stage in \citealp{yu24}.) As 85\% of the pair galaxies in our IPG sample have projected separations $>$20\,kpc\,$h^{-1}$, it is likely that they are in general representative of an earlier merger stage than the galaxies in \citet{lisenfeld19}.\medskip

Taken together, these results point to a scenario where the balance between molecular and atomic gas mass is initially unaffected by the interaction, but then rises sharply during coalescence. On somewhat longer timescales replenishment of the atomic gas reservoir via gravitational torquing or cooling of relatively pristine halo gas into the galaxy \citep[e.g.,][]{brainecombes93, moster11, moreno19} may then once again produce the normal H$_2$-to-HI mass ratios we measure for PMs. Evidence for this comes from the joint enhancement of both H$_2$ (see Sect. \ref{sect:PMoffsets_best}) and HI gas fractions \citep{ellison18} in PMs, and the observation that the gas phase metallicity of PMs is $\sim$0.1\,dex lower than in control-matched field galaxies \citep{kewley06,ellison13b}. The median \mhtwo/\mHI\ ratio of our PMs is 0.34, implying that the atomic phase by mass on average dominates their gas reservoirs, with only 15\% of PMs having more mass in the molecular phase than in the atomic phase. It has been proposed that interaction-driven starburst galaxies display elevated molecular-to-atomic gas mass ratios \citep[e.g.,][]{mirabelsanders89}. However, the observational evidence for such enhancements during phases of increased SF activity are not clear cut even in the (U)LIRG regime \citep{andreani95} and statistical analyses for the low-$z$ galaxies population indicate a that on average \mhtwo/\mHI\ ratios are constant for normal, non-interacting galaxies from the green valley to the off-MS starburst regime \citep[e.g.,][and references therein]{saintongecatinella22}. Within the PM sample, we find no evidence for systematic variations of molecular-to-atomic gas mass ratios with $\Delta$(MS), SFE or $\Delta$SFE, or CO excitation $r_{21}$, despite the fact that these are all linked to starburstiness to some degree. This suggests that episodes of enhanced \mhtwo/\mHI\ during interactions as reported by, e.g., \citet{larson16}, \citet{kaneko17} or \citet{lisenfeld19} may be short-lived.

\begin{figure*}
\centering
\begin{tabular}{rl}
\includegraphics[width=0.65\textwidth]{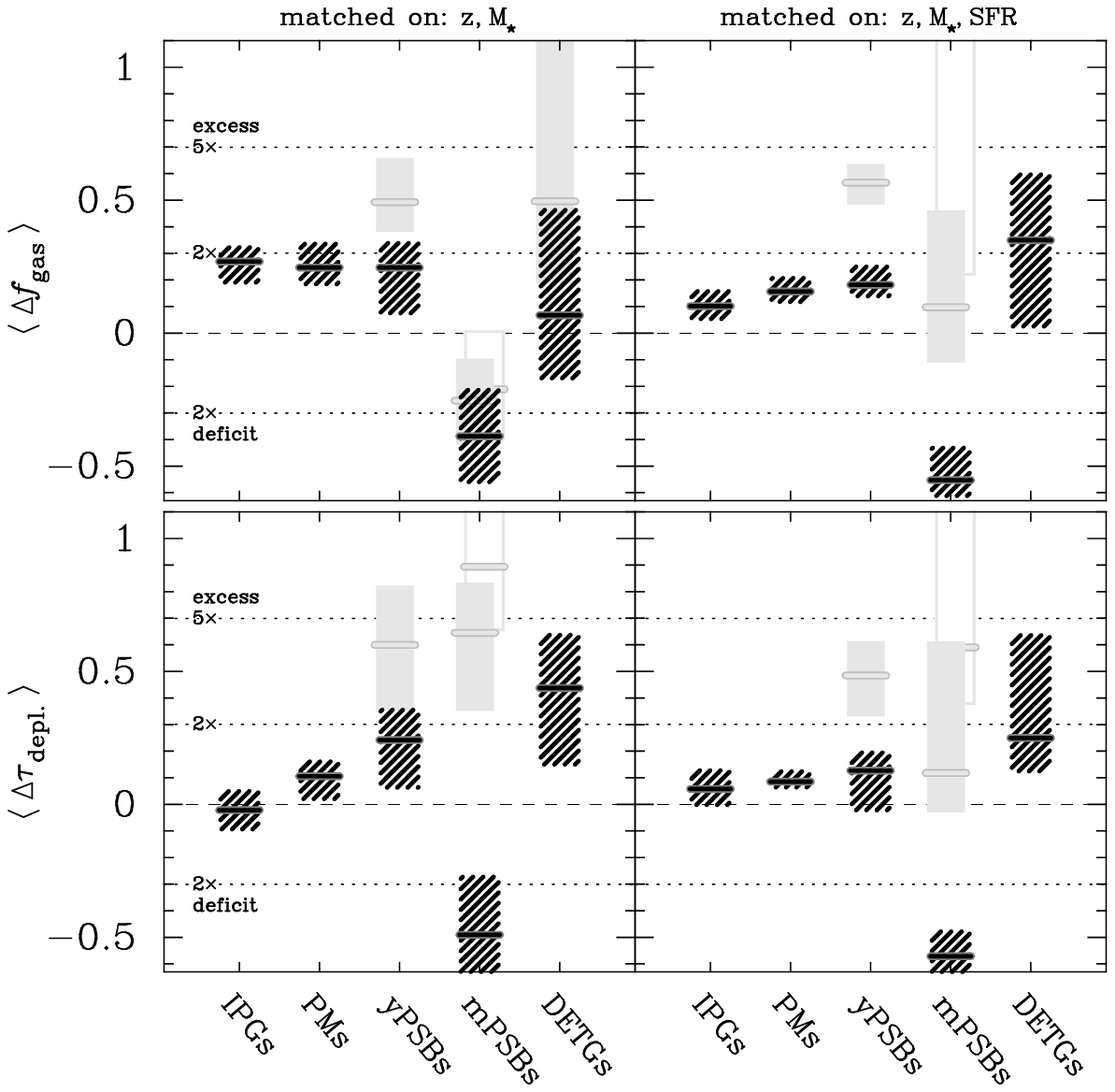} & \includegraphics[width=0.2\textwidth]{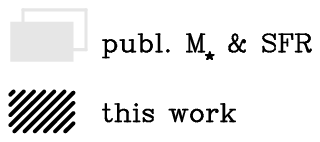}
\end{tabular}
\caption{Overview of median gas fraction ({\it top}) and depletion time offsets ({\it bottom}) for all samples: IPGs -- interacting pair galaxies, PMs -- post-mergers, yPSBs -- young post-starburst galaxies, mPSBs -- mature post-starburst (E+A) galaxies, and DETGs -- dust lane early-type galaxies. The {\it left} ({\it right}) column shows the results for control-matching on redshift \& \mstar\ (on redshift, \mstar\ \& SFR). The hatched black 68\% confidence intervals show the measurements obtained when using our `best' \mstar\ and SFR estimates (see Sect. \ref{sect:physquant}). For yPSBs and DETGs, light grey bars report the offsets measured with the \mstar\ and SFR values from, respectively, \citet{rowlands15} and \citet{davis15}. For mPSBs our `best-estimate' SFRs likely overestimate the true SFR, and measurements based on the H$\alpha$ line (light grey bars) or the D$_n$4000 index (open frames with grey edges) should provide better constraints on gas fraction and $\tau_{\rm depl.}$ offsets.
\label{fig:evosummary}}
\end{figure*}

\subsection{The evolution of the ISM content during galaxy interactions}
\label{sect:discuISMevo}

Fig. \ref{fig:evosummary} provides a summary view of the median gas fraction offsets $\langle\Delta\fgas\rangle$ and depletion time offsets $\langle\Delta\tau_{\rm depl.}\rangle$ of our different samples compared to xCG control galaxies. Where substantial uncertainties are present, we complement the measurements made using our `best-estimate' \mstar\ and SFR values (plotted with black hatched rectangles) with an alternative derivation (grey rectangles) that serves to illustrate the full range of medians found via our exploration of systematics in Sect. \ref{sect:compgal_offsets}.\medskip

The pre-coalescence stage of the merger sequence is covered by our IPG sample. IPGs display molecular gas fraction enhancements compared to normal galaxies, which are consistent with the 0.2-0.3\,dex offsets (factors $\sim$1.5 to 2) reported in previous work \citep[e.g.,][]{casasola04, lisenfeld11, pan18, violino18}. The molecular gas fraction enhancement of PMs is similar to those of IPGs, but we detect some evidence for PM gas fraction offsets to exceed those of IPGs by $\sim$20\% at fixed SFR (i.e., when control-matching on redshift, \mstar\ and SFR). This finding holds for a clear majority of the \mstar\ and SFR combinations we explored, and may be an extension of the trend in \citet{pan18} for tighter pair galaxies to have higher molecular gas fraction enhancements. For both IPGs and PMs, median depletion time offsets are around zero (IPGs) or mildly positive, i.e. track the median gas fraction enhancements of the sample quite closely. In a sample-averaged sense, this is evidence against interactions triggering widespread and/or sustained high-efficiency starbursts. However, as discussed in Sect. \ref{sect:PM_ISM}, and as we will further detail in Sect. \ref{sect:discuSBmerger}, high-SFE IPGs and PMs do exist, and they are preferentially found in the off-MS starburst regime.\medskip

It is in principle conceivable that selection effects could contribute to our measured gas fraction enhancement in PMs as the morphological features on which these PMs are selected are brighter for gas-rich (spiral galaxy) mergers, and \citet{lisenfeld19} have shown that already in the pre-coalescence phase gas fractions are larger in interacting S+S (spiral+spiral) than in E+S (early type+spiral) pairs. However, an increased gas content is in fact expected on theoretical grounds, and also when considering the {\it atomic} (HI) gas content of pair galaxies and PMs, \citet{ellison15, ellison18} inferred an at least mild increase of HI gas fractions through a control-matching analysis similar to ours. The clear observational evidence for enhanced total gas content across both the pre- and post-coalescence phase of the merger sequence points to two mechanisms acting at the same time \citep[see also][]{ellison15, pan18}. Within galaxies, dynamical effects \citep[torques, compressive tides; e.g.,][]{renaud14} facilitate the conversion of atomic gas into a denser molecular phase that is often centrally concentrated or may increasingly become so as interactions progress \citep[e.g.][]{kaneko13, yamashita17}. Compensating the loss to the star-forming H$_2$ phase, the HI reservoir may be replenished through torquing and cooling of ionised halo gas onto the interacting systems \citep[e.g.,][]{brainecombes93, moster11, moreno19}. These processes combined can bring about the higher HI and H$_2$ gas fractions found for IPGs in observations \citep[e.g.,][]{ellison15, violino18} and simulations \citep[e.g.,][]{rafieferantsoa15, moreno19}. At the relative short post-coalescence times of a few 100\,Myr \citet{ellison13b} inferred for our PMs, a purely statistical mechanism may additionally contribute to gas fraction enhancements or to at least initially maintaining them at a reasonably elevated level despite gas consumption through star formation: two interacting galaxies drawn from sub-linear trends $M_{\rm gas}\,{\propto}\,\mstar^{\beta}$ \citep[with $\beta\,{<}$\,1 as found for the HI and H$_2$ masses of low-$z$ galaxies; e.g.,][]{saintonge17} will results in a merger product with overall higher gas fraction \citep{ellison15}.\medskip

Interferometric observations of CO in PMs have revealed a mixture of extended and nuclear disks \citep[e.g.,][]{ueda14}, that may reflect the angular momentum configurations of the pre-merger galaxies. In view of the large number of possible angular momentum configurations and combinations of pre-merger gas fractions we find it interesting that the dispersion of the gas fraction offset distributions for PMs in Fig. \ref{fig:physcontrolmatch_best} remain within 0.3\,dex. We found similarly narrow offset distributions for IPGs, in spite of the fact that these must be sampling not only a variety of angular momentum configurations but also of orbital stages, from first approach to a sampling of various pericentric passages. The narrow $\Delta\fgas$ distributions for IPGs and PMs notably sets these apart from the much broader offset distribution with $\sigma\,{\approx}$\,0.8\,dex measured for dust lane early-type galaxies. This reinforces the conclusion in \citet{davis15} that DETGs are formerly gas-poor massive galaxies, which accreted substantial amounts of new gas through a minor merger (see discussion in Sect. \ref{sect:DETGintro}). Large spreads of gas fraction values in merger remnants are much more readily produced if the high-mass system involved was essentially devoid of gas prior to the interaction. By comparison, gas-rich, high-mass galaxies, to which gas is added via a minor merger, will experience a much smaller change of gas fraction as a result of the accretion of a new gas reservoir. Despite large systematic uncertainties, the range of median gas fraction offsets we find for DETGs implies that their gas fractions are (potentially substantially) enhanced\footnote{If the gas in DETGs is accreted from lower-mass systems, statistical metallicity estimates based on the mass of the DETG likely overestimate the true gas phase metallicity. The resulting underestimate of \aCO\ (and therefore of M(H$_2$) and the gas fraction) would imply an additional boost of the median offset $\langle\Delta\fgas\rangle$ for DETGs compared to estimates plotted in Figs. \ref{fig:compsamp_overview}d and \ref{fig:evosummary}, further reinforcing this conclusion.} and they will plausibly remain so for a long time due to their consistently long depletion times (see Fig. \ref{fig:evosummary}, lower row).\medskip

\begin{figure*}
\centering
\includegraphics[width=.95\textwidth]{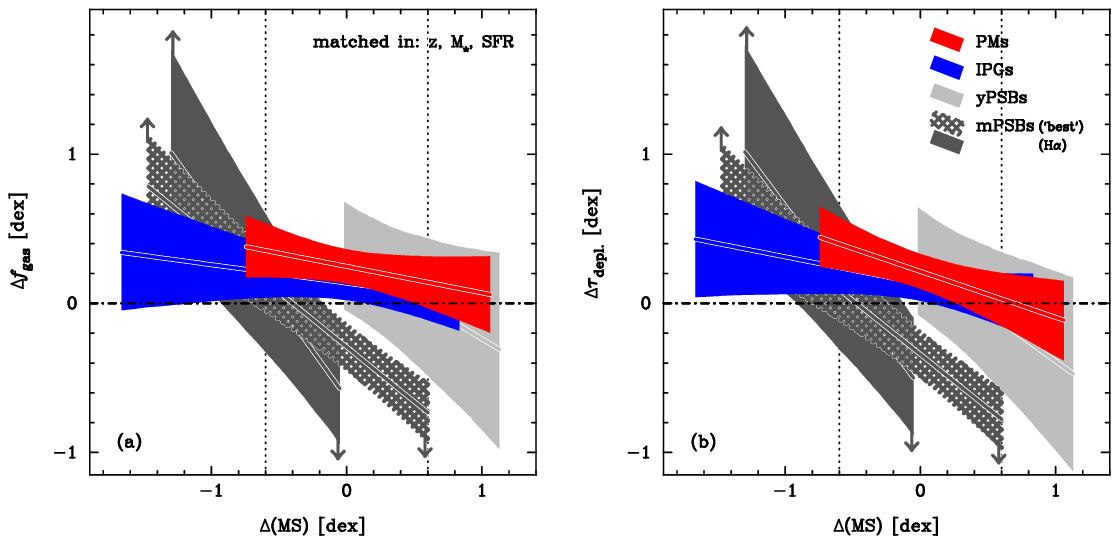}
\caption{Linear fits to the trends of gas fraction offset $\Delta\fgas$ vs. offset from the main sequence $\Delta$(MS) (panel {\it a}) and depletion time offset $\Delta(\tau_{\rm depl.})$ vs. $\Delta$(MS) (panel {\it b}) for post-mergers (PMs; {\it red}), interacting pair galaxies (IPGs; {\it blue}), young post-starbursts (yPSBs; {\it light grey}) and mature post-starbursts (mPSBs; {\it dark grey}). The 3\,$\sigma$ dispersion of the MS is indicated with vertical dotted lines. $\Delta\fgas$ and $\Delta(\tau_{\rm depl.})$ offsets were calculated with control-matching in the 3-D parameter space of redshift, \mstar\ and SFR. Shaded bands mark the uncertainty of the best-fit trend (highlighted itself by a separate line). The $x$-axis range over which the fit is plotted reflects the min./max. $\Delta$(MS) measurements of each sample. For mPSBs the filled (hatched) band show the best fit to offsets calculated when adopting H$\alpha$ (`best-estimate') SFRs. The mPSB fits represent lower limits to the slope; due to the preferential occurrence of lower (upper) limits on the $y$-axis quantities for mPSBs toward lower (higher) $\Delta$(MS), the true trend line slopes will be steeper as indicated by the pair of arrows attached to each fit.
\label{fig:excess_vs_dMS}}
\end{figure*}

Turning to post-starburst galaxies, yPSBs have high gas fractions compared to xCG control galaxies. The positive factor $\sim$1.6-3 gas fraction offset measured for yPSBs (Fig. \ref{fig:evosummary}, upper row) even represents an increase over the enhanced gas fractions of IPGs and PMs, making yPSBs the potentially most gas-rich of all five samples studied here. As discussed in Sect. \ref{sect:PSBintro}, although there is no 1:1 correspondence between PMs and PSBs, the two populations do significantly overlap (e.g., \citealp{wilkinson22} and references therein), and the yPSBs and PMs -- in a sample-averaged sense -- potentially also have somewhat commensurate post-interaction time scales (see next paragraph). Given the high incidence of mergers among PSBs in general \citep{ellison22, li23}, the enhanced gas fractions of yPSBs may simply reflect a particular propensity for starburst activity to occur in the presence of large gas reservoirs, provided external or internal triggers cause a fraction of this reservoir to end up in a dense phase \citep[e.g.,][]{renaud12}. In the previous section we outlined a combination of physical and statistical mechanisms capable of boosting gas fractions in the run-up to and during coalescence. These may play a role in establishing high gas fractions in yPSB progenitors if yPSBs are in many cases connected to mergers, as suggested by the high SFEs for the youngest objects \citep{rowlands15}, resolved PSB kinematics \citep{otter22}, and in some cases structural asymmetries \citep{sazonova21}. With a relatively short amount of time available for gas consumption in young post-starbursts, the yPSBs thus can plausibly still have high gas fractions, esp. given that \citet{rowlands15} find depletion times to rapidly rise as PSB age increases.\medskip

In the older mPSBs evidence for substantial gas consumption is stronger, with molecular gas fractions being consistently suppressed relative to normal galaxies for all \mstar\ and SFR combinations when control-matching on redshift and \mstar. When additionally matching on SFR, most measurements still suggest lower gas fractions for mPSBs than for yPSBs, but strong SFR-related systematics make it harder to determine whether mPSB gas fractions are just decreased compared to yPSBs (while remaining higher than in SFR-matched xCG control galaxies) or suppressed with respect to both yPSBs and normal galaxies. The gas fraction decrease from yPSBs to mPSBs found here follows findings in \citet{french15,french18b} of successively lower fractions as the age of the post-starburst increases (a coherent trend they observe not only for yPSBs and mPSBs, but also Shocked POst-starburst Galaxies (SPOGs) from \citealp{alatalo16}). As PSBs age and gas reservoirs are consumed, \citet{rowlands15} found that within the yPSB sample depletion times increased. We might therefore expect a clear signature for larger depletion time offsets in mPSBs compared to yPSBs, but this is not observed in the bottom row of Fig. \ref{fig:evosummary} for our `best-estimate' \mstar\ and SFR measurements. Adopting SFRs from H$\alpha$ or D$n$4000 (grey rectangles in Fig. \ref{fig:evosummary}) does, however, lead to median depletion time offsets between mPSBs and yPSB that are are broadly compatible with this picture when we control match on redshift and \mstar. Within uncertainties the same also holds for matching on redshift, \mstar\ and SFR. In keeping with the discussion in Sect. \ref{sect:mPSB_offsets}, and based on the independent tests with ionized neon as an SFR tracer in \citet{smercina18}, we consider results for the H$\alpha$ or D$n$4000 SFRs more credible. All measurements combined, these then suggest that decreasing molecular gas fractions in aging PSBs go hand in hand with a decrease of SFE. These decreasing gas fractions may be linked to a reduced gas content already in the atomic phase, as recently found by \citet{ellison25}.\medskip

To summarise, we have assembled a rough interaction-related time sequence flowing from left to right for a given row of panels in Fig. \ref{fig:evosummary}. For pre-coalescence pair galaxies and PMs this sequence is unambiguous, as is the transition from yPSBs to mPSBs, despite there being some overlap in burst age between yPSBs and mPSBs (see Sect. \ref{sect:compsampintro}). PMs typically have post-coalescence times of a few 100\,Myr \citep{ellison13b} and an identical post-burst age was inferred on average for yPSBs in \citet{rowlands15} and \citet{french18b}. Given that the strongest burst of star-formation during mergers occurs around coalescence \citep[e.g.,][]{mihoshernquist94, cox08, renaud14, he23}, we thus expect some temporal overlap between the PM and yPSB population. mPSBs are often spheroid-dominated \citep[e.g.,][]{norton01, yang08} and evolve into gas-poor early-type galaxies, progenitors of DETGs. We have thus placed DETGs at the very right, realising that these are one-off rejuvination events triggered by minor mergers that lead to a very specific kind of post-merger which has little in common with our main PM sample. We will thus primarily focus on the other galaxy samples (IPGs, PMs \& y/mPSBs) in the remainder of the discussion.\medskip

In this simplified time line of events our joint analysis of all samples paints a picture where gas fraction enhancements grow throughout the interacting pair stage \citep[see also][]{pan18} and into the post-merger phase. Low-$z$ interactions trigger a range of SFR enhancements \citep[e.g.,][]{bergvall03, scudder12, brassington15, pearson19}, so the fact that yPSBs have the largest gas fraction excess among all samples (when matched to xCG control galaxies on redshift, \mstar\ and SFR) may simply mean that these are often associated with a specific flavour of post-merger system where star formation -- due to orbital configuration and/or a particularly high gas content \citep[see, e.g.][]{dimatteo07} - experienced an especially strong boost. With increasing time after the burst the gas fractions of PSBs normalise and ultimately decrease until they fall well below those of control-matched normal galaxies. Depletion times follow a similar trajectory as gas fractions, mirroring -- in a sample-averaged sense -- the increased gas reservoirs which take longer to consume in the absence of a widespread and sustained transition to burst-like star-formation in interacting systems. Burst durations and duty cycles will be key in setting the range of observed combinations of gas fraction and depletion time offsets for individual galaxies. We will discuss these variations in Sect. \ref{sect:discuSBmerger}. Combined with literature findings, our analysis suggests that, after a starburst with presumably high SFE (i.e. temporarily short depletion time), the long-term evolution in the post-starburst phase is characterised by increasingly lower SFE as time progresses. Depletion of the gas reservoir and `dynamical suppression' of star formation (e.g., morphological quenching -- \citealp{martig09, gensior20}; or shear effects -- \citealp{davis14}) probably contribute to this slow-down.

\subsection{The starburst-merger connection, or: Is gas content or efficiency the most important driver of low-$z$ starbursts triggered by galaxy interactions?}
\label{sect:discuSBmerger}

\begin{figure}
\centering
\includegraphics[width=0.4\textwidth]{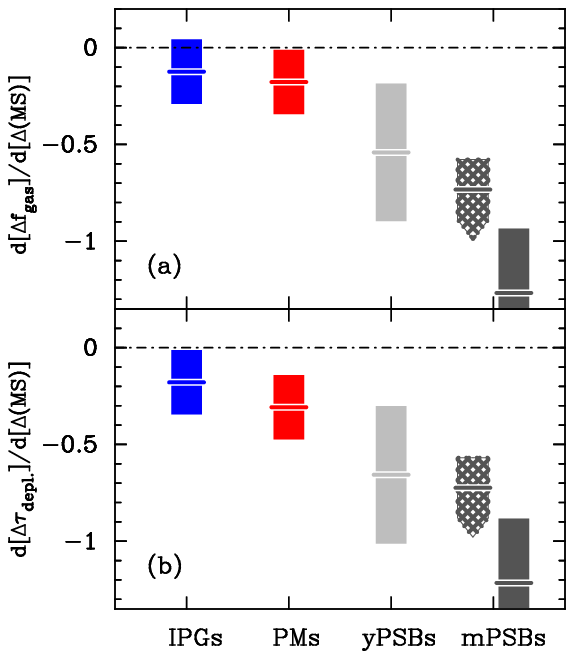}
\caption{Slopes of sample trends for ({\it a}) gas fraction offset $\Delta\fgas$ vs. offset from the main sequence $\Delta$(MS), and ({\it b}) depletion time offset $\Delta\tau_{\rm depl.}$ vs. $\Delta$(MS). The samples plotted are as in Fig. \ref{fig:excess_vs_dMS}: post-mergers (PMs; {\it red}), interacting pair galaxies (IPGs; {\it blue}), young post-starbursts (yPSBs; {\it light grey}) and mature post-starbursts (mPSBs; {\it dark grey}). Slope measurements for mSPBs represent lower limits on the steepness of the trend lines (hatched arrow -- slope constraint for `best-estimate' \mstar\ and SFR values; filled arrow -- SFRs based on the H$\alpha$ line). 
\label{fig:excess_vs_dMS_slopes}}
\end{figure}

In Sect. \ref{sect:PM_ISM} we presented evidence (see Fig. \ref{fig:dfgasvsdSFE}) that PMs with the largest efficiency enhancements $\Delta\sfe$ tend to reside above the MS of star-forming galaxies. We quantify this trend in Fig. \ref{fig:excess_vs_dMS}b, where we plot in red the best-fit trend line of a linear regression to the $\Delta\tau_{\rm depl.}$ vs. $\Delta$(MS) measurements of individual PMs. All offsets $\Delta$ entering the linear regression were derived with control matching on the three properties redshift, \mstar\ and SFR. We find a statistically significant trend towards increasing SFE enhancements with increasing MS-offset $\Delta$(MS), with depletion times for PMs becoming equal or even shorter than those of normal galaxies with matching MS-offset in the strong starburst regime at $\Delta$(MS)\,$\simeq$\,0.7\,dex ($\sim$5-fold MS-offset). This trend is superimposed on an overall baseline of on average longer depletion times for PMs compared to xCG control galaxies (see Sect. \ref{sect:PMoffsets_best}), such that for PMs residing towards the bottom of the MS ($\Delta$(MS)\,$\simeq$\,-0.7\,dex) depletion times are $\sim$2.5 times longer than for non-interacting galaxies. A similar anti-correlation is seen in Fig. \ref{fig:excess_vs_dMS} in $\Delta\fgas$ vs. $\Delta$(MS) space, albeit with a somewhat shallower slope (see Fig. \ref{fig:excess_vs_dMS_slopes}). The shallower slope implies that on average gas fraction offsets of PMs vary by only a factor $\sim$1.5 from the most strongly starbursting PMs with $\Delta$(MS)\,$>$\,1\,dex ($>$10-fold MS-offset) to PMs in the green valley ($\Delta$(MS)\,$\simeq$\,-0.8\,dex), while the efficiency contrast of PMs over this range is twice as large (factor 3 or 0.5\,dex in $\Delta\tau_{\rm depl.}$).\medskip

Galaxies in the pre-merger phase from our IPG sample (linear fits plotted in blue in Fig. \ref{fig:excess_vs_dMS}) show similar slopes as PMs in the space of $\Delta\fgas$ vs. $\Delta$(MS) and $\Delta\tau_{\rm depl.}$ vs. $\Delta$(MS), with trend lines offset in normalisation toward slightly lower $\Delta\fgas$ and $\Delta\tau_{\rm depl.}$ (i.e. higher $\Delta\sfe$) than for PMs, in keeping with the results summarized in Fig. \ref{fig:evosummary} (right-hand column, for control matching on redshift, \mstar\ and SFR). The drivers of interaction-induced star formation -- and esp. of starburst episodes -- are the subject of longstanding debate, with contradictory results suggesting either a key role of enhanced molecular gas content \citep[e.g.,][]{combes94, casasola04, martinez-galarza16, violino18, pan18, lisenfeld19, garay-solis23, yu24} or of a higher SFE \citep[e.g.,][]{sofue93, perea97, cao16, michiyama16, diaz-garcia20, ueda21}. Our results show how these contrasting literature findings can be reconciled: both pre- and post-mergers display gas fraction enhancements on average, but Figs. \ref{fig:excess_vs_dMS} and \ref{fig:excess_vs_dMS_slopes} demonstrate that simultaneously an SFE-gradient with MS-offset (i.e. with `starburstiness') is observed for both samples. Note that these SFE variations are not simply the consequence of assigning low CO-to-H$_2$ conversion factors \aCO\ to pair and PM galaxies in the off-MS regime (see Sect. \ref{sect:XCO}), as we apply the same \aCO\ recipe to both mergers and control galaxies. Instead, these results point to an important role of galaxy interactions in triggering the strongest, high-SFE starburst events (see also \citealp{ellison20} for a spatially resolved assessment of the importance of SFE in driving interaction-induced centrally concentrated starburst activity). Dedicated observations of high-$\Delta$(MS) galaxies lacking interaction signatures could help establish the extent to which other mechanisms like internal gravitational instabilities can produce high-intensity star formation \citep[e.g.,][]{cenci24}.\medskip

There is tentative evidence in Figs. \ref{fig:excess_vs_dMS}b/\ref{fig:excess_vs_dMS_slopes}b that the $\Delta\sfe$-gradient with MS-offset may steepen with interaction stage from IPGs to PMs. The unambiguously steepest slopes are however measured for post-starbursts (grey trend lines in Fig. \ref{fig:excess_vs_dMS}), both in the space of $\Delta\fgas$ vs. $\Delta$(MS) and $\Delta\tau_{\rm depl.}$ vs. $\Delta$(MS). The diversity of the measured gas fraction and SFE offsets of PMs is a key factor behind the shallower slopes $d[\Delta\fgas]/d[\Delta({\rm MS})]$ and $d[\Delta\tau_{\rm depl.}]/d[\Delta({\rm MS})]$ of PMs, compared to PSBs. Exemplary for this is the population of off-MS starburst PMs ($\Delta$(MS)\,$>$\,0.6\,dex), which includes systems spanning the whole range from strong SFE enhancements ($\Delta\sfe\,{\gg}\,\Delta\fgas$) to strong fuel domination (high $\Delta\fgas$; see Fig. \ref{fig:dfgasvsdSFE}). We speculate that off-MS starbursts that are found to be fuel dominated ($\Delta\fgas\,{\gg}\,\Delta\sfe$) may have been recently triggered, leaving insufficient time for substantial gas consumption. Circumstantial evidence for this comes from the fact that among the six high-$\Delta$(MS) starburst PMs showing the highest $\Delta\fgas$ excesses, four objects show signatures of being caught right at final coalescence or shortly afterwards as they display either dual galaxy cores (PMs \#36 \& 39) or evidence of strong colour gradients and asymmetry (PMs \#8, 34). Conversely, of the four off-MS starburst PMs which lie in the SFE driven regime in Fig. \ref{fig:dfgasvsdSFE}, three objects (PMs \#17, 21, \& 24) have quite regular morphologies, suggesting that either (i) the merger and associated onset of triggered star formation may lie further in the past, or (ii) an only fairly minor merger. In the case of (i), gas consumption could lead to a low \fgas\ and high SFE \citep[see, e.g.,][]{gaosolomon99}, and could on the time scale of a few tens of Myrs cause merger remnants to drop from the fairly shallow $\Delta\fgas$ vs. $\Delta$(MS) and $\Delta\tau_{\rm depl.}$ vs. $\Delta$(MS) relations of PMs in Fig. \ref{fig:excess_vs_dMS} to the steeper trend lines followed by yPSBs. Validation of this hypothesis regarding evolutionary pathways in the $\Delta\sfe$ vs. $\Delta\fgas$ diagram would require quantitative age markers for PMs, which is beyond the scope of this paper. We close with the observation that, within uncertainties, the slopes inferred for yPSBs and mPSBs in Fig. \ref{fig:excess_vs_dMS_slopes} are similar and that these two samples differ primarily in terms of the normalisation of their trend lines in $\Delta\fgas$ vs. $\Delta$(MS) and $\Delta\tau_{\rm depl.}$ vs. $\Delta$(MS) space. This is the case regardless of the SFR adopted. The offset in the normalisation of the yPSB and mPSB trend lines reflects progressive gas consumption during the post-starburst phase (see also Fig. \ref{fig:evosummary}), over an indicative timespan of $\sim$1\,Gyr, given the difference in mean post-starburst age of the yPSB and mPSB samples (Sect. \ref{sect:PSBintro}). In view of the substantial overlap between the low-$z$ post-starburst and post-merger populations \citep{sazonova21, ellison22, otter22, wilkinson22, li23}, these findings suggest that the higher the MS-offset during the interaction-induced starbursting phase, the more efficient the starburst and the more rapid the gas consumption (Fig. \ref{fig:excess_vs_dMS}.a). This general picture would be in agreement with the observations-based 2-SFM framework of \citet{sargent14}, or predictions from state-of-the-art numerical simulations \citep[e.g.,][]{hani20}.

\section{Summary and concluding remarks}
\label{sect:summary}

In this paper we take galaxy-integrated CO observations of a sample of SDSS post-merger galaxies as a starting point for a study of the evolution of the neutral gas content during low-redshift galaxy interactions. By comparing the properties of pre- and post-mergers to those of non-interacting control galaxies from xCOLD GASS \citep{saintonge17} with matching redshift, stellar mass and SFR -- and with an identical set of data products -- we are able to quantify to what extent systematic offsets are present for key ISM properties like molecular gas fractions and depletion times. To compute these offsets for both individual interacting galaxies, and for the pre- and post-merger samples on average, we employ survival analysis techniques which allow us to properly account for CO non-detections both among interacting and control galaxies. We complement our analysis of pre- and post-merger ISM properties with analogous measurements for post-starburst from the literature, to shed light on the starburst-merger connection.\medskip

Our main findings are as follows:
\begin{enumerate}
\item SDSS post-mergers have molecular-to-stellar mass ratios that are enhanced by $\sim$50\% ($\langle\Delta\fgas\rangle$\,=\,0.16$_{-0.04}^{+0.05}$\,dex) compared to control galaxies matched on redshift, stellar mass and SFR (see Sect. \ref{sect:PMoffsets_best}/Fig. \ref{fig:physcontrolmatch_best}a). Depletion times are offset to higher values by $\sim$25\% ($\langle\Delta\tau_{\rm depl.}\rangle$\,=\,0.09$_{-0.02}^{+0.04}$\,dex; Fig. \ref{fig:physcontrolmatch_best}b). The fraction of post-merger galaxies for which the star-formation activity is fuel-driven ($\Delta\fgas\,{>}\,\Delta\sfe$), rather than efficiency-driven, is approximately 70\% (see Fig. \ref{fig:dfgasvsdSFE}).
\item The molecular gas fraction enhancement of post-mergers is similar in size to that reported for the atomic gas component \citep{ellison18}. This leads to $H_2$-to-HI ratios that are similar to normal, non-interacting galaxies (see Sect. \ref{sect:PM_H2overHI}/Fig. \ref{fig:PM_HInr21}a). Compared to the latter we find that the excitation of the $J$\,=\,2 rotational level of CO in post-mergers is boosted by a small, but statistically significant amount ($\langle\Delta r_{21}\rangle$\,=\,0.09$^{+0.03}_{-0.01}$\,dex or $\sim$25\%; see Sect. \ref{sect:PM_r21}/Fig. \ref{fig:PM_HInr21}b).
\item We detect evidence for the median $\Delta\fgas$ and $\Delta\tau_{\rm depl.}$ of post-mergers to exceed that of pre-merger interacting pair galaxies by $\sim$20\% at fixed redshift, \mstar\ and SFR (see Fig. \ref{fig:evosummary}). Similarly to post-mergers, we measure the median H$_2$-to-HI ratio of pair galaxies to match that of non-interacting control galaxies. This may imply that H$_2$-to-HI enhancements -- when they do occur \citep[e.g.,][]{larson16,lisenfeld19,yu24} during interactions --  are short-lived.
\item For both pre- and post-mergers, individual galaxy $\Delta\fgas$ and $\Delta\tau_{\rm depl.}$ offsets anti-correlate with their distance from the main sequence of star-forming galaxies, $\Delta$(MS) (see Fig. \ref{fig:excess_vs_dMS}). Starbursting (i.e. high-$\Delta$(MS)) interacting galaxies are on average characterised by an especially high star-formation efficiency. Substantial scatter around this average trend is observed for post-mergers, where the off-MS population contains examples of both strongly fuel-driven post-mergers (high $\Delta\fgas$) and high-$\Delta\sfe$ systems (see Fig. \ref{fig:dfgasvsdSFE}). The most gas-rich starbursting post-mergers may represent an early post-coalescence phase, where fuel consumption due to the starburst has not yet progressed far.
\item Post-starbursts display the steepest dependency of the offsets $\Delta\fgas$ and $\Delta\tau_{\rm depl.}$ on $\Delta({\rm MS})$, with an age-dependent normalisation of the associated trend lines that reflects on-going gas consumption for the remaining gas in mature post-starbursts (see Figs. \ref{fig:excess_vs_dMS} and \ref{fig:excess_vs_dMS_slopes}).
\end{enumerate}

The simultaneous finding of (a) an overall enhanced atomic and molecular gas content in merging systems, and (b) a systematically increasing star formation efficiency going from mergers in the green valley to highly star forming mergers above the main sequence reconciles contrasting literature results advocating that {\it either} (a) {\it or} (b) are the primary driver for interaction-induced star formation. The presence of substantial scatter around these general unifying trends nevertheless highlights that further work is necessary to understand the full range of circumstances and astrophysical processes that control star formation activity along the merger sequence. Progress will come through a combination of larger merger samples, more accurate methods for binning interacting galaxies into different time slices following the merger process, and resolved galaxy studies.\bigskip

\section*{Acknowledgements}
This work is based on observations carried out under project number 196-14 with the IRAM 30\,m telescope. We thank Claudia Marka and the IRAM 30\,m staff for their support in preparing and executing the observations. IRAM is supported by INSU/CNRS (France), MPG (Germany) and IGN (Spain). Our thanks also go to the anonymous referee for their constructive feedback on the preprint version of this article. MTS is indebted to A. Cibinel for her support and helpful discussions throughout this project. MTS is grateful to T. A. Davis, K. D. French and K. Rowlands for sharing additional information on the DETG \& y/mPSB samples, and acknowledges support from a Royal Society Leverhulme Trust Senior Research Fellowship (LT150041) and a Scientific Exchanges visitor fellowship (IZSEZO\_202357) from the Swiss National Science Foundation. SLE acknowledges an NSERC Discovery Grant, JMS support from the National Science Foundation under Grant No 2205551.\\
This research has made use of the VizieR catalogue access tool, CDS, Strasbourg, France. The original description of the VizieR service was published in A\&AS 143, 23. Much of the analysis presented here was carried out in the Perl Data Language (PDL; Glazebrook \& Economou, 1997) which can be obtained from \url{http://pdl.perl.org}.


\appendix

\section{Galaxy-integrated CO spectroscopy}

\subsection{\COone\ and \COtwo\ line flux measurements}
\label{appsect:PMintro_part2}

\subsubsection{Line-extraction in post-merger CO spectra}
\label{appsect:fluxmeas}
We follow \citet{saintonge17} for the measurement of the velocity-integrated \COone\ and \COtwo\ line fluxes in the PM spectra shown in Fig. \ref{tab:fluxes}. For each PM galaxy we manually define an appropriate frequency window for the flux measurement guided by the position of the edges of the line. The velocity-integrated line flux $I_{\rm CO}$ is then obtained by summing the flux in the channels in this window and the spectral noise per 20\,km/s bin, $\sigma_{\rm CO}$, is measured in adjacent line-free channels. The total error on $I_{\rm CO}$ is set to
\begin{equation}
\Delta I_{\rm CO} = \delta v\,\sqrt{N_l}\times\sigma_{\rm CO}~,
\label{eq:dICO}
\end{equation}
where $N_l$ is the number of velocity bins $\delta v$\,=\,20\,km/s required to cover the FWHM (full width at half maximum) velocity width of the line, $W_{50}$. Note that $W_{50}$ is not the result of a Gaussian fit to the line profile, but measured using a custom-made, interactive IDL script for (i) identifying peak fluxes and (ii) approximating the line flanks between 20 and 80\% of the peak flux with a linear fit, leading to an estimate of the respective 50\% amplitudes and thus of the line FWHM \citep{springob05, catinella07, saintonge11}. The central velocity is taken as the mid-point of this line width.\medskip

\begin{table*}
\centering
\caption{Measured \COone\ and \COtwo\ line properties and molecular gas estimates for post-merger galaxies.}
\label{tab:fluxes}
\begin{tabular}{r @{\quad\vline\quad} rccccc @{\quad\vline\quad} rccc}
\hline\hline \\[-2ex]
\# & $I_{\COone}$ & $\sigma$ & $W_{50}$ & $\mathcal{A}$  & $L'_{\COone}$ & \mhtwo & $I_{\COtwo}$ & $\sigma$ & $W_{50}$ & $\mathcal{A}$\\
& [Jy\,km/s] & [mK] & [km/s] & & [$10^9$\,K\,km/s\,pc$^2$] & [$10^{9}$\,\msun] & [Jy\,km/s] & [mK] & [km/s] &\\[0.5ex]
(1) & \multicolumn{1}{c}{(2)} & (3) & (4) & (5) & (6) & (7) & \multicolumn{1}{c}{(8)} & (9) & (10) & (11)\\[1ex]
\hline\hline
 & \multicolumn{6}{c}{\raisebox{-1ex}{\COone}} & \multicolumn{4}{c}{\raisebox{-1ex}{\COtwo}}\\[2ex]
\hline
\multicolumn{11}{c}{\it \raisebox{-1ex}{Newly observed post-merger galaxies (IRAM program 196-14)}}\\[2ex]
\hline \\[-2ex]
 1 &   8.40$\pm$0.53 & 1.2 & 269.7 & 1.15 &  0.58$\pm$0.04 &  2.32$\pm$0.15 &  23.13$\pm$0.78 & 1.4 & 253.8 & 1.44\\
 2 &   6.13$\pm$0.47 & 0.9 & 332.3 & 1.97 &  1.04$\pm$0.08 &  4.56$\pm$0.35 &  13.80$\pm$0.75 & 1.2 & 325.1 & 2.83\\
 3 &  12.52$\pm$0.66 & 1.6 & 235.7 & 1.30 &  1.23$\pm$0.06 &  1.36$\pm$0.07 &  42.04$\pm$1.37 & 2.4 & 263.6 & 1.66\\
 4 &  20.37$\pm$0.66 & 1.5 & 255.4 & 1.81 &  2.38$\pm$0.08 &  9.35$\pm$0.30 &  34.68$\pm$1.30 & 2.1 & 315.7 & 3.20\\
 5 &   7.31$\pm$0.45 & 1.0 & 298.2 & 1.60 &  1.60$\pm$0.10 &  6.43$\pm$0.40 &  20.53$\pm$0.62 & 1.0 & 290.6 & 2.33\\
 6 &  14.44$\pm$0.74 & 1.8 & 213.4 & 1.71 &  0.55$\pm$0.03 &  2.10$\pm$0.11 &  26.04$\pm$1.18 & 2.7 & 146.5 & 2.66\\
 7 &   6.66$\pm$0.44 & 0.9 & 344.7 & 1.49 &  1.10$\pm$0.07 &  4.42$\pm$0.29 &  15.89$\pm$0.76 & 1.2 & 318.1 & 2.18\\
 8 &  29.36$\pm$1.26 & 2.1 & 459.1 & 1.25 &  2.55$\pm$0.11 &  3.01$\pm$0.13 &  82.08$\pm$1.94 & 2.6 & 457.8 & 1.57\\
 9 &   7.21$\pm$0.56 & 1.2 & 267.6 & 1.85 &  0.79$\pm$0.06 &  3.84$\pm$0.30 &   7.17$\pm$1.10 & 2.1 & 228.6 & 4.41\\
10 &  14.58$\pm$0.58 & 0.9 & 515.9 & 2.11 &  3.26$\pm$0.13 & 12.62$\pm$0.50 &  25.05$\pm$1.07 & 1.3 & 507.0 & 3.44\\
11 &  22.61$\pm$1.11 & 2.0 & 418.9 & 2.55 &  1.31$\pm$0.06 &  5.06$\pm$0.25 &  39.45$\pm$1.82 & 2.5 & 415.3 & 4.92\\
12 &  10.01$\pm$0.71 & 1.8 & 201.6 & 1.39 &  0.99$\pm$0.07 &  4.17$\pm$0.30 &  18.31$\pm$1.24 & 2.4 & 215.9 & 2.20\\
13 &   4.88$\pm$0.47 & 0.9 & 362.3 & 1.92 &  0.75$\pm$0.07 &  2.75$\pm$0.26 & $<$2.57 & 1.5 &   --- & 3.36\\
14 &   4.77$\pm$0.50 & 0.9 & 442.1 & 1.86 &  1.25$\pm$0.13 &  4.57$\pm$0.48 &  11.69$\pm$0.81 & 1.1 & 479.3 & 2.79\\
15 &  36.52$\pm$0.60 & 2.0 & 115.9 & 1.32 &  3.11$\pm$0.05 &  3.85$\pm$0.06 &  67.76$\pm$1.65 & 4.4 & 112.7 & 1.92\\
16 &  11.21$\pm$0.73 & 1.8 & 205.0 & 1.66 &  0.64$\pm$0.04 &  2.47$\pm$0.17 &  16.58$\pm$1.51 & 2.8 & 236.8 & 2.58\\
17 &  27.13$\pm$0.91 & 2.3 & 200.5 & 1.40 &  1.42$\pm$0.05 &  1.80$\pm$0.06 &  66.67$\pm$1.45 & 2.8 & 209.2 & 1.82\\
18 &  29.29$\pm$1.26 & 2.5 & 331.1 & 1.43 &  3.85$\pm$0.17 & 14.78$\pm$0.64 &  64.73$\pm$2.51 & 3.9 & 341.8 & 2.00\\
19 &   4.37$\pm$0.60 & 1.5 & 213.8 & 1.42 &  0.29$\pm$0.04 &  1.21$\pm$0.17 &   5.37$\pm$1.03 & 2.1 & 187.9 & 2.05\\
20 &   6.55$\pm$0.44 & 1.0 & 236.5 & 1.65 &  1.13$\pm$0.08 &  4.11$\pm$0.27 &   8.71$\pm$0.70 & 1.3 & 229.3 & 2.38\\
21 &  29.31$\pm$1.28 & 2.9 & 248.8 & 1.34 &  2.55$\pm$0.11 &  2.99$\pm$0.13 &  63.12$\pm$2.89 & 5.0 & 268.2 & 1.84\\
22 &  10.41$\pm$1.30 & 3.2 & 223.0 & 1.32 &  1.53$\pm$0.19 &  5.71$\pm$0.71 &  23.00$\pm$2.99 & 5.8 & 214.2 & 1.77\\
23 &  15.50$\pm$1.01 & 2.4 & 239.3 & 1.45 &  0.90$\pm$0.06 &  3.52$\pm$0.23 &  34.81$\pm$1.90 & 3.4 & 250.7 & 2.30\\
24 &   9.30$\pm$0.44 & 0.9 & 322.1 & 1.32 &  0.62$\pm$0.03 &  0.80$\pm$0.04 &  20.27$\pm$0.72 & 1.1 & 320.7 & 1.73\\[1ex]
\hline
\multicolumn{11}{c}{\it \raisebox{-1ex}{Post-merger galaxies from xCOLD GASS}}\\[2ex]
\hline \\[-2ex]
25 &  32.74$\pm$0.46 & 2.1 &  62.1 & 1.50 &  1.69$\pm$0.02 &  6.34$\pm$0.09 & \multicolumn{4}{c}{---}\\
26 &   5.29$\pm$0.48 & 1.9 &  84.9 & 2.05 &  1.15$\pm$0.10 &  4.28$\pm$0.39 & \multicolumn{4}{c}{---}\\
27 &  16.75$\pm$1.14 & 2.2 & 333.6 & 1.37 &  0.82$\pm$0.06 &  3.12$\pm$0.21 & \multicolumn{4}{c}{---}\\
28 &   9.30$\pm$0.85 & 1.3 & 563.1 & 1.97 &  0.66$\pm$0.06 &  2.42$\pm$0.22 & \multicolumn{4}{c}{---}\\
29 &   2.32$\pm$0.54 & 1.4 & 185.4 & 1.35 &  0.18$\pm$0.04 &  0.85$\pm$0.20 &   3.62$\pm$0.65 & 1.9 &  93.7 & 2.12\\
30 &  12.33$\pm$1.15 & 2.3 & 324.9 & 2.46 &  2.32$\pm$0.22 &  8.47$\pm$0.79 &  12.68$\pm$2.93 & 4.6 & 312.3 & 5.11\\
31 &   5.78$\pm$0.84 & 1.5 & 404.9 & 1.45 &  0.89$\pm$0.13 &  3.23$\pm$0.49 & \multicolumn{4}{c}{---}\\
32 &   8.40$\pm$0.57 & 1.0 & 407.9 & 1.71 &  1.06$\pm$0.07 &  3.88$\pm$0.26 &   6.97$\pm$1.48 & 2.1 & 208.6 & 2.85\\
33 &   8.46$\pm$0.67 & 1.6 & 223.9 & 1.58 &  0.78$\pm$0.06 &  3.16$\pm$0.25 &  10.20$\pm$0.86 & 2.0 & 151.7 & 2.60\\
34 & 106.08$\pm$1.02 & 2.1 & 307.7 & 1.29 &  6.66$\pm$0.06 &  7.48$\pm$0.07 & 352.91$\pm$2.20 & 3.6 & 306.3 & 1.85\\
35 &   6.91$\pm$0.45 & 1.8 &  76.9 & 1.76 &  0.96$\pm$0.06 &  4.29$\pm$0.28 &  12.98$\pm$0.72 & 2.7 &  58.0 & 3.40\\
36 &  56.65$\pm$1.48 & 2.3 & 527.1 & 2.64 & 15.30$\pm$0.40 & 19.49$\pm$0.51 & \multicolumn{4}{c}{---}\\
37 &         $<$1.55 & 1.1 &   --- & 1.23 &  $<$0.14 &  $<$0.63 & \multicolumn{4}{c}{---}\\
38 &   4.44$\pm$0.45 & 1.7 &  96.9 & 1.31 &  0.06$\pm$0.01 &  0.18$\pm$0.02 &  11.92$\pm$0.68 & 1.9 & 100.9 & 1.74\\
39 &  38.07$\pm$1.25 & 2.1 & 478.2 & 1.86 &  8.50$\pm$0.28 &  31.49$\pm$1.04 & \multicolumn{4}{c}{---}\\[1ex]
\hline
\multicolumn{11}{c}{\it \raisebox{-1ex}{Interacting pair galaxy observed during IRAM program 196-14}}\\[2ex]
\hline \\[-2ex]
40 &  51.31$\pm$0.70 & 1.6 & 254.2 & 1.53 &  3.73$\pm$0.05 & 14.29$\pm$0.19 & 168.48$\pm$1.14 & 2.0 & 267.0 & 2.54\\[1ex]
\hline\hline
\end{tabular}
\tablecomments{Cols. 2--7 (cols. 8--11) list quantities derived from \COone (\COtwo) spectra. $\sigma$ (cols. 3/9) is the rms noise on the $T_a^*$ scale in a 20\,km/s channel. Brightness temperatures were converted to flux units with Kelvin-to-Jansky conversion factors inferred by linear interpolation of the data tabulated for the upper and lower end of EMIR bands E0 and E2 at \url{http://www.iram.es/IRAMES/mainWiki/Iram30mEfficiencies}. $W_{\rm 50}$ (cols. 4/10) is the FWHM velocity width. Aperture corrections $\mathcal{A}$ tabulated in cols. 5/11 were derived using the constant \fgas\ assumption (see Sect. \ref{appsect:apcorr}). Velocity-integrated line fluxes $I_{\mathrm{CO}(J{\rightarrow}J-1)}$ (cols. 2/8) are quoted {\it prior} to aperture correction, line luminosities $L'_{\COone}$ in col. 6 are galaxy-integrated values {\it after} application of the aperture corrections in col. 5. Line flux errors are 1\,$\sigma$ uncertainties. Upper limits are placed at the 3\,$\sigma$ level. The spectra of all targets observed through IRAM program 196-14 are shown in Fig. \ref{appfig:spectra} (in Fig. \ref{appfig:pairspectrum} for pair galaxy \#40).}
\end{table*}

\begin{figure*}
\centering
\includegraphics[width=.9\textwidth]{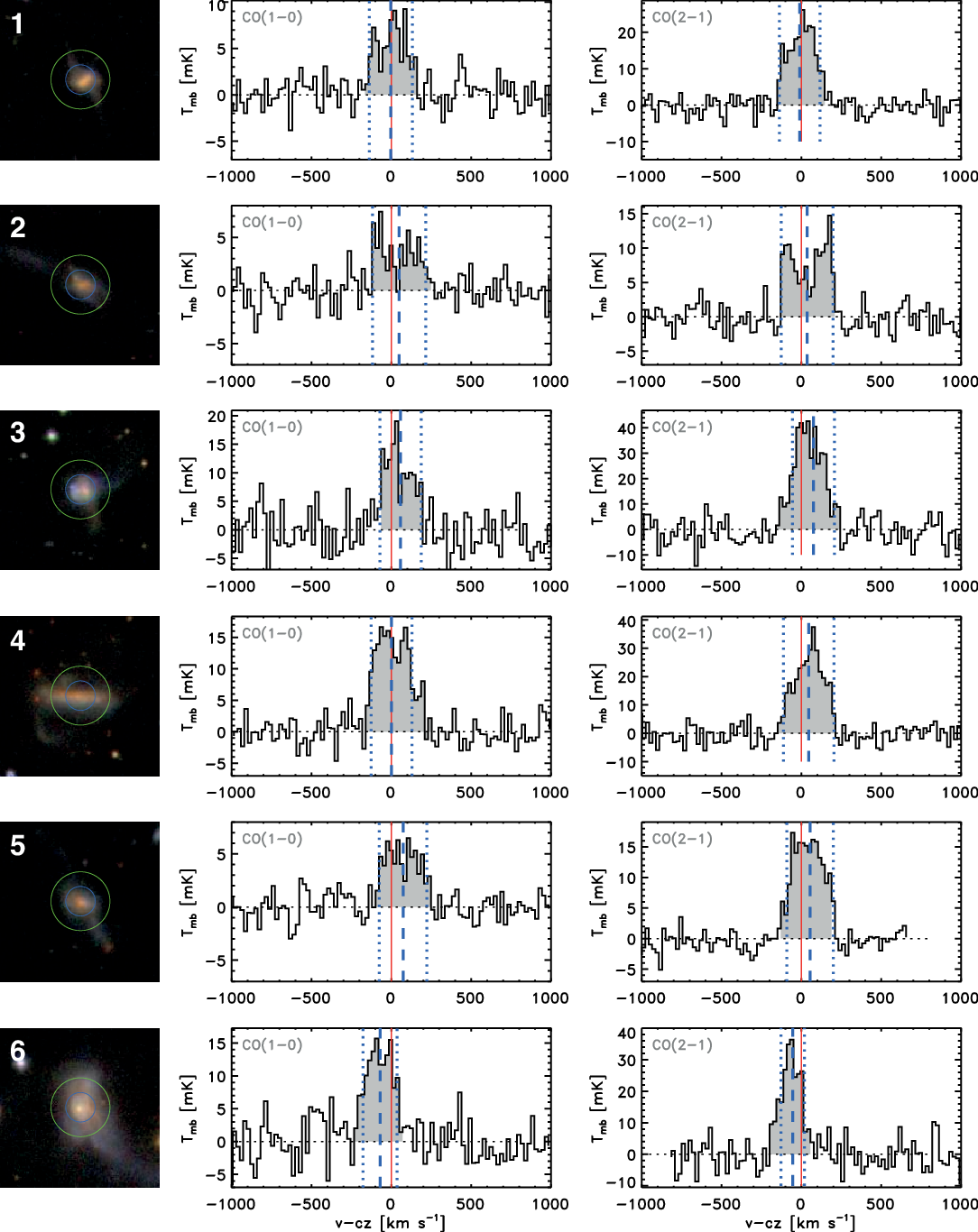}
\caption{SDSS images (1.0$\times$1.0\,arcmin$^2$) and \COone\ and \COtwo\ spectra (20\,km/s velocity resolution) observed with the IRAM-30\,m telescope for the 24 post-merger galaxies targeted by IRAM program 196-14. The position on the images of the $\sim$22$''$ (11$''$) beam of the \COone\ (\COtwo) observations is shown as a green (blue) circle. The region of the spectra integrated to compute the CO line fluxes is highlighted in grey. The central position and width of the CO line is indicated with the blue dashed and dotted lines. The central velocity according to the SDSS optical redshift is shown as the red line. All measured CO fluxes and line widths can be found in Table \ref{tab:fluxes}.
\label{appfig:spectra}}
\end{figure*}

\addtocounter{figure}{-1}
\begin{figure*}
\centering
\includegraphics[width=.9\textwidth]{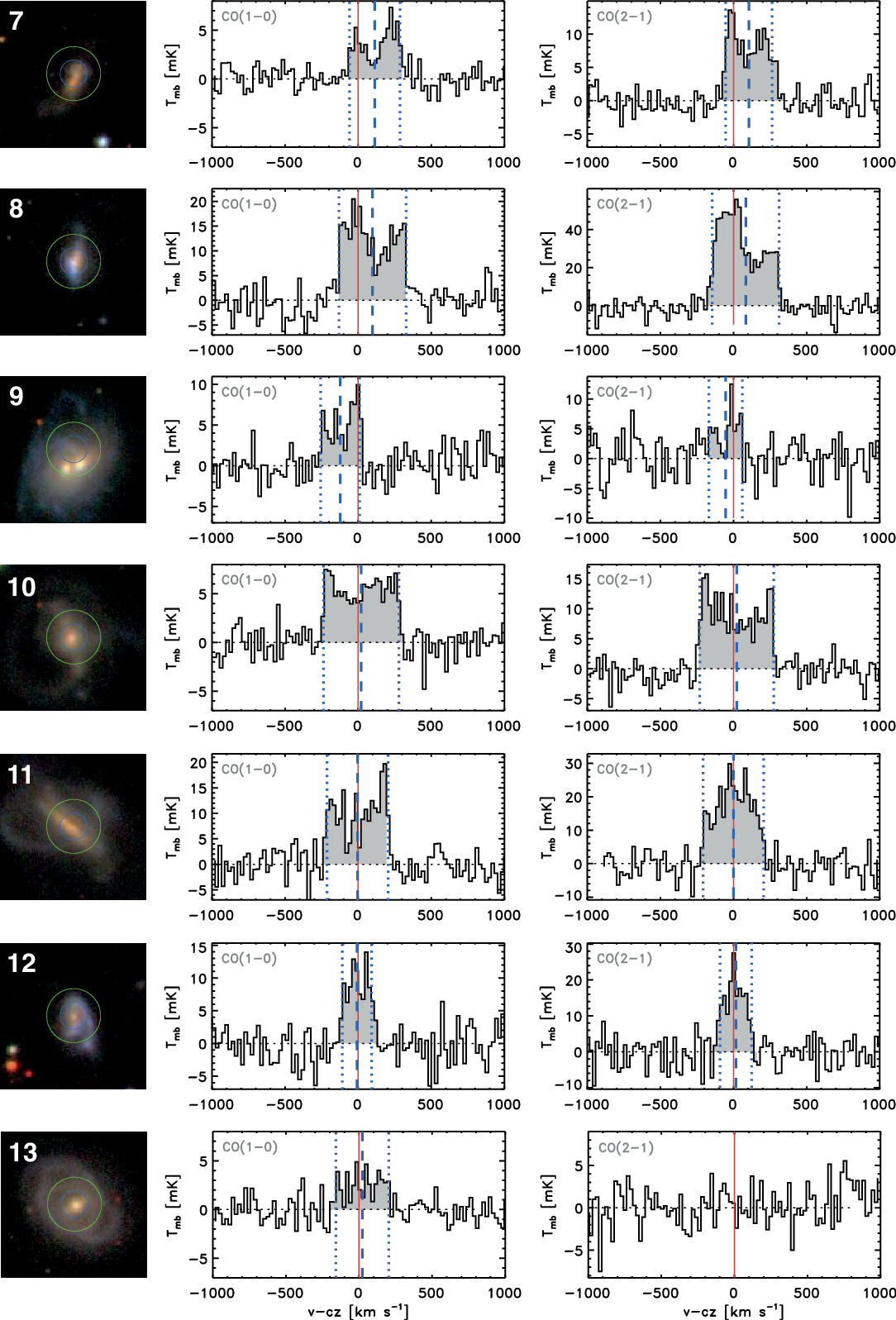}
\caption{continued}
\end{figure*}

\addtocounter{figure}{-1}
\begin{figure*}
\centering
\includegraphics[width=.9\textwidth]{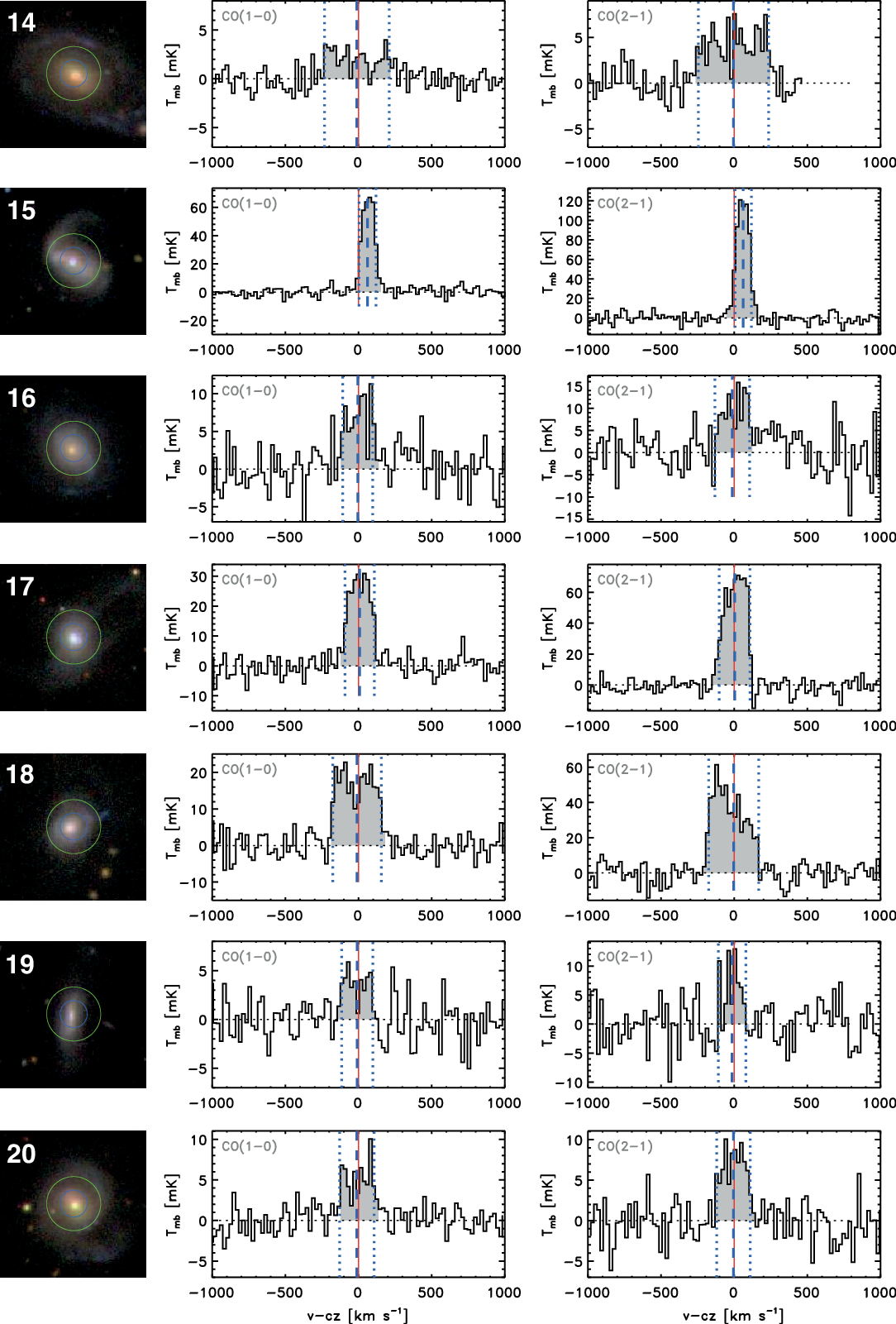}
\caption{continued}
\end{figure*}

\addtocounter{figure}{-1}
\begin{figure*}
\centering
\includegraphics[width=.9\textwidth]{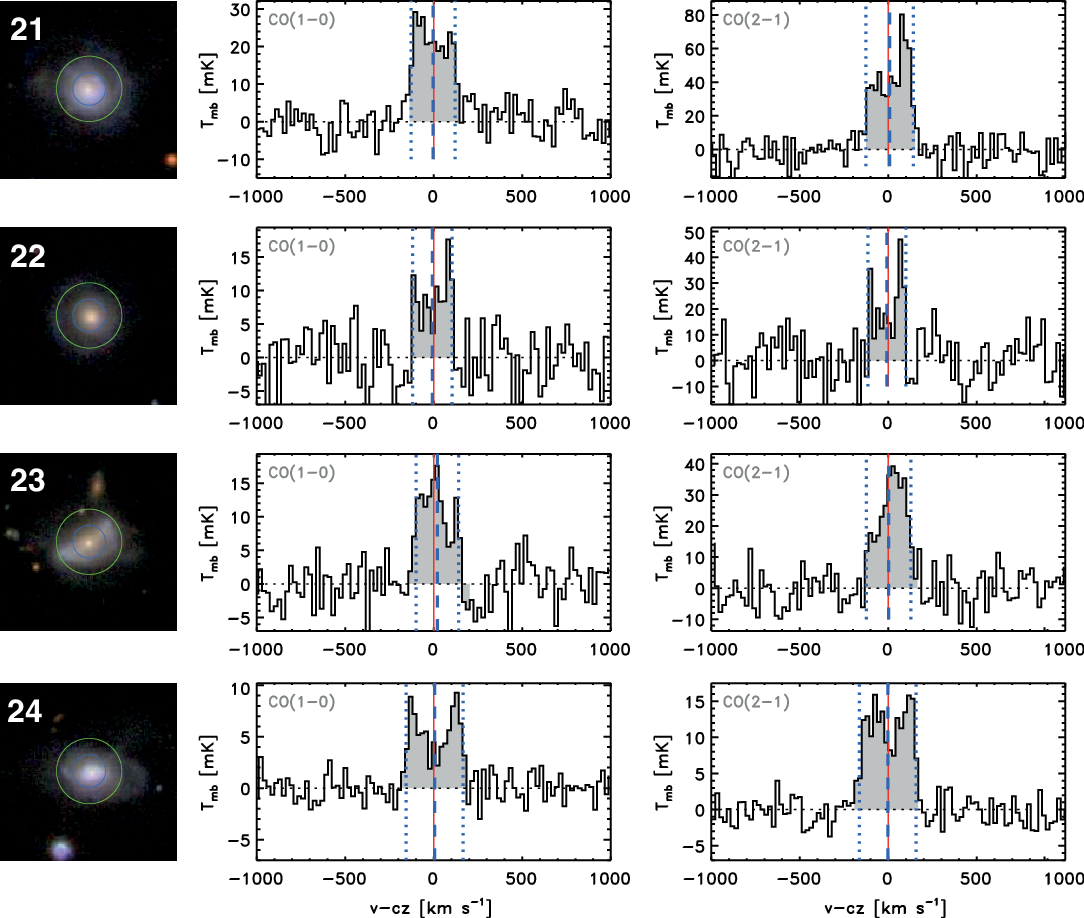}
\caption{continued}
\end{figure*}

\begin{figure*}
\centering
\includegraphics[width=.8\textwidth]{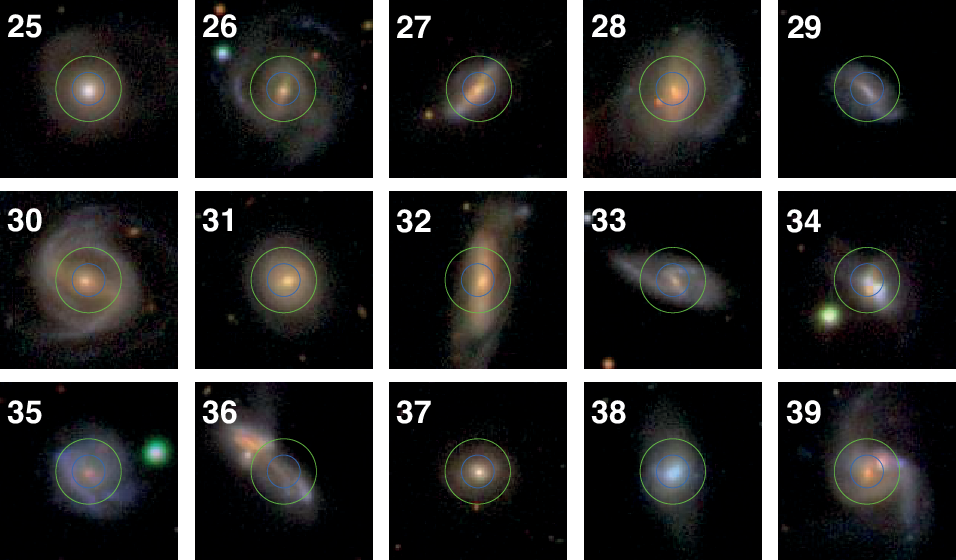}
\caption{SDSS image cut-outs (1.0$\times$1.0\,arcmin$^2$) of the 15 post-merger galaxies identified in the xCOLD GASS sample (see text of Sect. \ref{sect:PMsel} for details).
\label{appfig:addedPMstamps}}
\end{figure*}

Velocity-integrated line fluxes, line widths and channel rms noise values for the 39 PM galaxies are listed in Table \ref{tab:fluxes}. All tabulated line fluxes are `raw' values prior to the application of aperture corrections, which are also given in Table \ref{tab:fluxes} and which we discuss in Sect. \ref{appsect:apcorr}. The 20\,km/s channel sensitivities of our \COone\ spectra range from $\sigma_{\COone}$\,=\,0.9 to 2.3\,mK ($T_{\rm A}^*$ scale), leading to line detections with $S/N\,{\geq}$\,7.3 in all cases but reaching up to $S/N\,{\sim}$\,60 for 587729158966345769 (PM \#15), the PM with the brightest \COone\ line emission in our IRAM sample. 587736543096799321 (PM \#3) was also observed in the COLD GASS project and detected with $S/N$\,=\,13.3 (very similar to the detection significance achieved by our own follow-up, $S/N$\,=\,13.6). For this paper, we adopt the $S/N$-weighted average of the two available flux estimates (COLD GASS -- 11.46\,Jy\,km/s; our observations -- 13.64\,Jy\,km/s) for this object, i.e. 12.52$\pm$0.66\,Jy\,km/s. The 20\,km/s channel sensitivities of our \COtwo\ spectra vary from $\sigma_{\COtwo}$\,=\,1.0 to 5.8\,mK. Line detections of the \COtwo\ transition reach $S/N$-values between 5.2 and 46. For the one \COtwo-undetected PM \#13 (587726101750546619) we tabulate a 3\,$\sigma$ upper limit in Table \ref{tab:fluxes}.\medskip

Finally, in Table \ref{tab:fluxes} we also report the line flux measurements for the galaxy with SDSS ObjID 587724232641937419, initially classified as a PM by \citet{ellison13b}, but later recognized to be undergoing a merger with a nearby galaxy. For completeness, an SDSS cut-out stamp and the CO spectra of this galaxy are shown in Fig. \ref{appfig:pairspectrum}.

\subsubsection{Aperture corrections}
\label{appsect:apcorr}
Due to the Gaussian beam response of the IRAM 30\,m telescope, line emission from galaxy regions offset from the pointing centre is weighted down in relative terms in the unresolved spectra recorded by the single-pixel EMIR receiver. Attempts to compensate for this flux loss often use an aperture correction that assumes an inclined disk model, with parameters reflecting the source structure in optical images \citep[e.g.,][]{saintonge17, gao19}. As our focus here is on interacting galaxies, we instead adopt a non-parametric approach, also based on optical data, which accounts for the presence of asymmetric/irregular morphological features.\medskip

\begin{figure*}
\centering
\includegraphics[width=.78\textwidth]{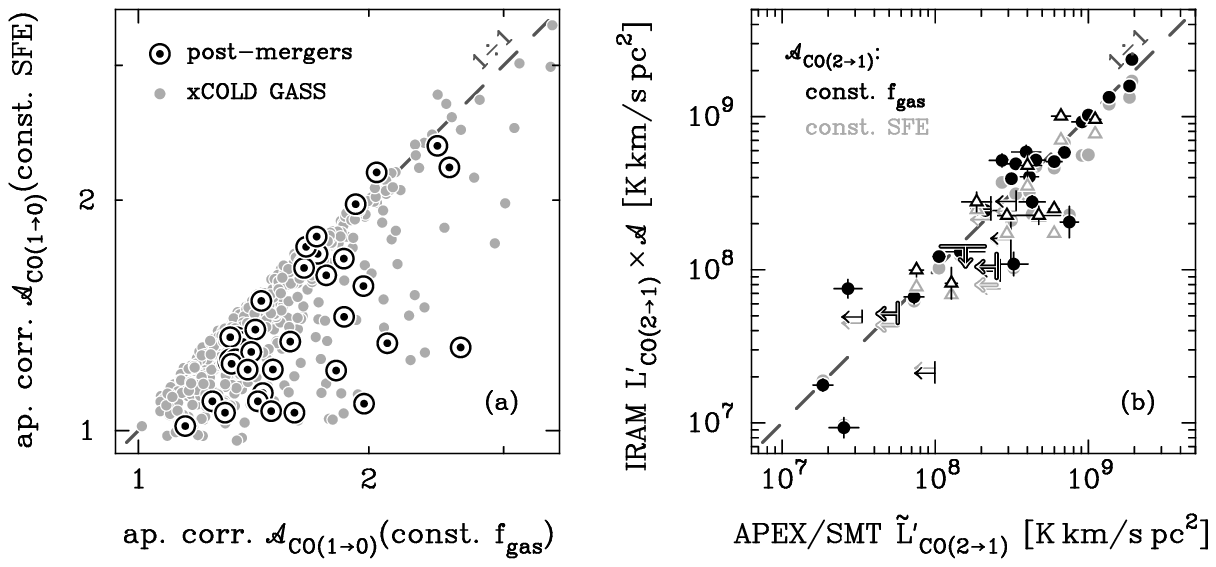}
\caption{Properties of aperture correction factors $\mathcal{A}_{\COone}$ used in our analysis to infer galaxy-integrated CO line luminosities from observed fluxes.\newline
{\it (a)} Comparison of \COone aperture corrections assuming a spatially invariant gas fraction and star-formation efficiency ($\mathcal{A}_{\COone}$(const.\,\fgas), and $\mathcal{A}_{\COone}$(const.\,SFE), respectively). Post-merger (xCOLD GASS) galaxies are plotted with large (small) symbols. {\it (b)} Comparison of aperture-corrected IRAM \COtwo\ line luminosities $L'_{\COtwo}{\times}\mathcal{A}_{\COtwo}$ with an estimate of the total, galaxy-integrated luminosity $\widetilde{L}'_{\COtwo}$ from the smaller APEX/SMT single dish telescopes. Filled dots/arrows -- xCOLD GASS galaxies; open triangles/arrows -- mature post-starburst galaxies. Grey (black) symbols show the offset from the 1$\div$1 relation when adopting the constant SFE (constant \fgas) correction approach. Error bars (1\,$\sigma$) are only shown for data points to which the constant \fgas\ aperture correction was applied. The constant \fgas\ assumption provides a better consistency with the total luminosity proxy $\widetilde{L}'_{\COtwo}$ (see text of Sect. \ref{appsect:apcorr} for details).
\label{appfig:apercorrs}}
\end{figure*}

Specifically, we introduce two different aperture correction measures loosely based on, respectively, proxies of (1) the stellar mass (\mstar) distribution, and (2) the SFR distribution of SDSS galaxies. Adopting aperture correction measure (1) to correct IRAM CO line flux measurements is equivalent to assuming that the gas mass-to-stellar mass ratio, $M_{\rm gas}/\mstar$ (``\fgas"), is constant across galaxies. The approach in (2) instead implies a gas distribution mimicking that of the SFR, and therefore a spatially invariant star-formation efficiency, SFE\,$\equiv$\,SFR/$M_{\rm gas}$. We therefore, in a qualitative way, refer to these two alternatives as the {\it constant $f_{\rm gas}$} and {\it constant SFE} correction from now on. They are calculated as follows:
\begin{enumerate}
\item[(1)] {\it const. \fgas\ correction}: considering a Gaussian beam centred at the IRAM pointing position, we re-compute the object flux weighted by this beam. The ratio of the beam-weighted and total object fluxes are then used to define the relevant aperture correction. Total object fluxes are measured on SDSS $r$-band images, within the SExtractor segmentation maps from \citet{simard11} that mask nearby neighbours if present\footnote{For the morphologically complex PMs \#36 \& 39, which are split in two in the segmentation maps, we construct a merged segmentation map that encompasses the entire extent of the coalescing objects.}. As the emission in red optical and near-IR bands is a good tracer of stellar mass this aperture correction reflects the fraction of the galaxy \mstar-distribution probed by the IRAM telescope beam.
\item[(2)] {\it const. SFE correction}: motivated by the fact that the star-formation rates in the MPA-JHU database are a combination of a fibre SFR (computed following the spectral fitting procedure of \citealp{brinchmann04}) and an ``outer" SFR (computed based on fits to the broad-band photometry), we separately compute the total and beam weighted $r$-band fluxes both within the 3$''$ fibre aperture and in an outer region encompassing the full SExtractor segmentation map, but excluding the fibre aperture\footnote{Since we rely on the $r$-band galaxy flux distribution to calculate the `const. SFE correction', the latter does not directly follow the SFR. However -- by separately considering the fibre and outer SFR estimates -- this correction does encapsulate in an azimuthally averaged sense how the SFR is distributed in the radial dimension within each galaxy.}. The ratios of the beam-weighted and total fluxes within the fibre aperture and in the outer region are then used as weighting factors that are applied to the SDSS fibre and ``outer" SFR measurements. The ratio of this beam-weighted SFR sum and the total SDSS SFR are then used to define the aperture correction factor.
\end{enumerate}
We were able to derive aperture corrections $\mathcal{A}_{\COone}$ for all PMs (see Table \ref{tab:fluxes}), 482/488 of the xCG control galaxies, and overall 98\% of all galaxies which constitute the four additional comparison samples introduced in Sect. \ref{sect:compsampintro}. 
The remaining cases where no aperture corrections could be derived were in general due to incomplete information in the SDSS data base, or the lack of SExtractor segmentation maps from \citet{simard11}. Constant \fgas\ aperture corrections tend to be larger than the constant SFE corrections (see Fig. \ref{appfig:apercorrs}a). For PMs -- plotted with large symbols in Fig. \ref{appfig:apercorrs}a --  the median aperture correction factors are 1.49 (const. \fgas) and 1.28 (const. SFE), for xCG control galaxies 1.42 (const. \fgas) and 1.37 (const. SFE).\medskip

By using the same method to derive aperture corrections for both the interacting/starbursting galaxies, and the pool of control galaxies from xCG, we ensure the {\it relative} consistency between the inferred galaxy-integrated gas properties of these different populations, which is key to our control matching strategy (see Sect. \ref{sect:matching}). Thanks to the availability of \COtwo\ spectra from smaller single dish telescopes it is also possible to test the accuracy of our aperture corrections in {\it absolute} terms\footnote{Additionally, for three xCG galaxies (one of which is PM \#28) galaxy-integrated \COone\ line fluxes are available from the MASCOT survey \citep{wylezalek22} on the 12\,m millimeter single dish telescope at the Arizona Radio Observatory (beam FWHM 55$''$). Adopting our constant \fgas\ correction for these three galaxies, the average ratio of the MASCOT \COone\ line flux to the aperture corrected measurement with the IRAM 30\,m beam response is consistent with unity within 1\,$\sigma$. In particular, for the object with xCG ID 31775 (MASCOT source 9491-6101) -- which is representative of the whole xCG sample in that it is covered by the IRAM 30\,m beam out to approx. twice the $r$-band effective radius -- the constant \fgas\ correction recovers the MASCOT \COone\ line flux to within 20\%.}. \medskip

The FWHM of the IRAM 30\,m telescope beam at the frequency of the redshifted \COtwo\ transition is $\sim$11$''$. As illustrated by the blue circles in the image cut-outs in Fig. \ref{appfig:spectra}, at this angular scale a significant fraction of the \COtwo\ flux gets missed for targets in the redshift range 0.01\,${<}\,z\,{\lesssim}$\,0.05. For a subset of 29 xCG and 13 mature post-starburst galaxies (mPSBs; see Sect. \ref{sect:PSBintro}) \COtwo\ spectra were taken with both the 30\,m telescope and, respectively, either the Atacama Pathfinder EXperiment Telescope (APEX) or the Submillimeter Telescope (SMT) of the Arizona Radio Observatory (see \citealp{saintonge17} and \citealp{french15}). With dish diameters of 12\,m (APEX) and 10\,m (SMT) their FWHM beam widths (APEX -- 27$''$; SMT -- 33$''$) are a much better match to the angular sizes of SDSS galaxies, and the corresponding \COtwo\ line fluxes therefore a good approximation of the total, galaxy-integrated emission\footnote{In keeping with the homogenisation of flux measurements in Sect. \ref{appsect:meashomog} we have remeasured \COtwo\ fluxes for mPSBs using the following steps to ensure that the comparison of aperture-corrected and galaxy integrated fluxes is carried out consistently between xCG and mPSB galaxies. We first produced a noise-weighted, stacked \COtwo\ spectrum for each mPSB galaxy by combining the IRAM 30\,m and the SMT spectra in \citet{french15}. Prior to co-addition of the two spectra, the amplitude of the 30\,m spectrum was scaled by our aperture correction estimate, and the velocity resolution of the SMT spectrum downgraded via linear spline interpolation from 13\,km/s to 20\,km/s to match the resolution of the IRAM \COtwo\ spectrum. We then used the stacked spectrum to set the velocity window for the line flux extraction (see Sect. \ref{appsect:fluxmeas}). This approach improved the reliability of the choice of the velocity window when the $S/N$ of both the IRAM and the SMT spectrum was low. Line channel selection was not strongly susceptible to whether the constant \fgas\ or constant SFE aperture correction was used to scale the IRAM spectrum. Finally, we measured the line integrated \COtwo\ fluxes directly in the individual IRAM and SMT spectra, and retained fluxes reaching $S/N\,{\geq}$\,3 as detections. To convert flux densities in temperature units to Janskys we applied Kelvin-to-Jansky conversion factors with typical values $\sim$7.6\,Jy/K to IRAM 30\,m (conversion from $T_a^*$ scale), and $\sim$44.6\,Jy/K to SMT (conversion from $T_{mb}$ scale) spectra. For xCG galaxies the Kelvin-to-Jansky conversion applied to the APEX \COtwo\ spectra was $\sim$39\,Jy/K (conversion from $T_a^*$ scale).}. In Fig. \ref{appfig:apercorrs}b we compare aperture-corrected IRAM \COtwo\ line luminosities with the APEX and SMT measurements for xCG galaxies (filled symbols) and mPSBs (open symbols). The median ratio between the corrected IRAM 30\,m luminosity, $L'_{\COtwo}{\times}\mathcal{A}_{\COtwo}$, and the proxy of the total luminosity from APEX/SMT, $\widetilde{L}'_{\COtwo}$, is consistent with unity for both `normal' xCG galaxies (median ratio: 1.02$_{-0.12}^{+0.20}$) and mPSBs (median ratio: 0.86$_{-0.46}^{+0.52}$) when using the constant \fgas\ aperture correction. Due to the presence of censored $\lco{\times}\mathcal{A}/\widetilde{L}'_{\rm CO}$ values in the data, these medians and their associated 2$\sigma$ errors were estimated via single-sided survival analysis with the \citet{kaplanmeier58} estimator for the xCG galaxies, and double-sided survival analysis following \citet{schmitt85} for the mPSBs. One of the 13 mPSBs with simultaneous IRAM and SMT \COtwo\ coverage was discarded for this analysis as it lacked an aperture correction. When using the constant SFE aperture corrections the median $\lco{\times}\mathcal{A}/\widetilde{L}'_{\rm CO}$ ratios are 0.88$_{-0.16}^{+0.11}$ (0.70$_{-0.35}^{+0.27}$) for xCG (mPSB) galaxies, indicating that this correction recipe fares less well at recovering the entire lost flux.\medskip

We have relied on \COtwo\ data to test of the validity of our aperture corrections, although the subsequent analysis of this paper hinges on galaxy-integrated \COone\ line fluxes. However, given that the mismatch in angular scale between total galaxy extent and effective area sampled by the IRAM spectra is more severe for the \COtwo\ than the \COone\ transition, this should in fact provide a more stringent test of the methodology. The consistency of the aperture-corrected \COtwo\ line luminosities and the total line luminosity estimates $\widetilde{L}'_{\COtwo}$ thus gives confidence that our approach will retain its accuracy when applied to the \COone\ spectra, which suffer from smaller flux losses. The excellent agreement we find in Sect. \ref{sect:PM_r21} between galaxy-integrated \COtwo-to-\COone\ line ratios inferred with our aperture corrections for normal xCG galaxies on the one hand, and literature measurements on the other hand, provides further evidence for the validity of our aperture correction approach. Finally, we note that the better performance of the constant \fgas\ relative to the constant SFE correction in our data is consistent with the interferometric EDGE-CALIFA survey of low-$z$ galaxies, where molecular and stellar half-mass sizes follow a 1:1 relation on average \citep{bolatto17}. For the results section and final conclusions of this paper we thus only show measurements using the const. \fgas\ aperture correction. When testing the robustness of our analysis in Appendix \ref{appsect:aperturesys}, we quantify by how much results change when adopting different aperture corrections.

\subsection{Cross-sample homogenisation of \COone\ line fluxes and $S/N$ estimates}
\label{appsect:meashomog}

Our control matching scheme relies on measurements being carried out in a consistent way across all samples, as do the intercomparisons of ISM properties between different galaxy samples. Here we discuss the extent to which all CO line flux measurements are already, or have been made, self-consistent.\medskip

For all samples we set the threshold for distinguishing between CO line detections and non-detections at $S/N$\,=\,3 (the errors entering the $S/N$ calculation are discussed further below). 3\,$\sigma$ line flux upper limits for undetected galaxies assume a line width of 300\,km/s. For normal, dynamically unperturbed, non-interacting galaxies a more refined line width estimate based on, e.g., mass (via the CO Tully-Fisher relation) and inclination information might provide more accurate line flux constraints. However, this approach is unlikely to be suitable for many of the interacting galaxies. We therefore adopt a constant 300\,km/s line width for the calculation of line flux upper limits for all types of non-detections and, where necessary, recalculated the upper limits originally tabulated in the literature to reflect this choice.\medskip

Line fluxes for PMs, IPGs and the xCG control galaxies were derived following \citet[][see also Sect. \ref{appsect:fluxmeas}]{saintonge11, saintonge17}, i.e. with direct summation of channel fluxes within a spectral window that effectively extends to zero channel amplitude to either side of the line centre. The associated line flux uncertainties account for statistical noise fluctuations within the velocity interval $W_{50}$ (eq. \ref{eq:dICO}). A very similar procedure is used for yPSBs and DETGs in \citet{rowlands15} and \citet{davis15}, such that line flux measurements are largely consistent with those of the PMs, IPGs and control galaxies, except for the following differences: (i) eq. (\ref{eq:dICO}) for the line flux error is evaluated with $N_l$, the number of line channels, reflecting the full line width down to zero channel amplitude rather than the in principle narrower $W_{50}$ interval, and (ii) compared to eq. (\ref{eq:dICO}), the error calculation involves an additional term $\sqrt{1+N_l/N_b}$ for the uncertainty in the baseline level (here $N_b$ is the number of channels used to fit the baseline). Both (i) and (ii) will lower the $S/N$ relative to our measurements for PMs, IPGs and the xCG control galaxies, but in practice the generally steep flanks of the CO lines in spectra with a velocity resolution of $\sim$20\,km/s (see, e.g., Fig. \ref{appfig:spectra} or Fig. A1 in \citealp{saintonge11}) lead to the ``full" and $W_{50}$ widths often being nearly equivalent, such that the baseline uncertainty term (ii) has a larger impact on the classification of galaxies as detected or undetected at the $S/N$\,=\,3 threshold we adopt. We note that \citet{rowlands15} use a cut at $S/N$\,=\,4 in their analysis, and even with their more conservative flux error galaxy PSB1 in their yPSB sample is detected at $\sim$3\,$\sigma$. Without the baseline uncertainty term its $S/N$-ratio increases to 4.0, such that we treat it as a detection with $I_{\COone}$\,=\,1.23$\pm$0.31\,Jy\,km/s in our analysis. For DETGs the wide frequency coverage afforded by the FTS backend results in $S/N$ shifts due to the baseline uncertainty term that on average are of order 5\% (T. Davis, priv. comm.). Neither of the two undetected DETGs in \citet{davis15} have line emission features that would cause them to rise above the $S/N$\,=\,3 threshold after compensating for this shift.\medskip

Of all samples included in our analysis, the approach of \citet{french15} for mPSBs differs most from ours in that they use all channels out to $\pm$3\,$\sigma$ of a Gaussian fit to the line profile when determining line fluxes and flux errors. While this should lead to no systematic flux offsets compared to our windowing approach, provided their broader frequency range probes only statistical noise fluctuations, the larger number of channels entering the noise calculation in \citet{french15} acts to lower their $S/N$ ratio in relative terms. Consistent with this expectation we find a median ratio of 1.00 between their fluxes and ours if we remeasure them on the spectra in \citet{french15} for their 17 detections, by mimicking the windowing approach used for PMs, IPGs and xCG control galaxies. The $S/N$ ratio of the line detections increases by 37\% on average using our approach, with a further shift (6\% on average; K. D. French, priv. comm.) owing to the inclusion of a $\sqrt{1+N_l/N_b}$ baseline uncertainty term in \citet{french15}. In order to account for this, we reexamined the spectra of mPSBs classified as non-detections in \citet{french15} and identified one further object (EAH16 with $I_{\COone}$\,=\,2.43$\pm$0.64\,Jy\,km/s) which exceeds the 3\,$\sigma$ detection limit using our approach, and which we therefore consider a detection for our analysis.\medskip

Finally, we apply the same frequency-dependent Kelvin-to-Jansky conversion factor (see Sect. \ref{sect:redu}) to all galaxies when converting their brightness temperatures on the $T_a^*$ scale to Janskys. This resulted in minor changes to literature flux values at the level of $\sim$1\% where the same recipe for calculating frequency-dependent Kelvin-to-Jansky conversion factors had not already been implemented.

\section{Exploration of systematic uncertainties}
\label{appsect:syschecks}

Deriving the `best-estimate' galaxy stellar masses, star-formation rates and molecular gas masses that underpin our findings in Sects. \ref{sect:results} and \ref{sect:discussion} involved specific assumptions or methodological choices. Here we quantify to what extent these lead to systematic uncertainty in our results.

\begin{figure*}
\centering
\begin{tabular}{ccccc}
\includegraphics[width=0.3\textwidth]{f14a.ps} & & \includegraphics[width=0.3\textwidth]{f14b.ps} & & \includegraphics[width=0.3\textwidth]{f14c.ps}
\end{tabular}
\caption{Comparison of SFR estimates and associated 1$\sigma$ errors for post-merger galaxies (PMs, {\it left}), control galaxies from xCOLD GASS DR1 (xCG-DR1, {\it centre}), and interacting pair galaxies (IPGs, {\it right}). SFR$_{\rm C15}$ and SFR$_{\rm S16}$ are based on SED fitting and taken from \citet{chang15} and \citet{salim16}. SFR$_{\rm MPA-JHU}$ is the fibre-corrected star-formation rate from the MPA-JHU SDSS catalog \citep{brinchmann04}. SFR$_{\rm UV+IR(W3/4)}$ is a hybrid SFR measured via GALEX UV and WISE photometry (see Sect. \ref{sect:SFRderiv_SFG} for details). SFR$_{\rm published}$ is the SFR estimate reported in the papers (cf. legend in uppermost panel) first introducing each sample.\newline
Dashed (dotted) lines are drawn at a 2:1 (10:1) offset from the 1$\div$1 locus (solid line). Data points are coloured by SDSS spectral type: blue -- star-forming, green -- composite, red -- AGN, black -- unknown. Percentages displayed in the lower-right corner of each panel list the fraction of the galaxies in the respective sample for which a match in the catalogs providing the different SFR-estimates was found. The $\langle\Delta\rangle$ value quoted in the lower right corner of each panel is the median logarithmic ratio SFR$_y$/SFR$_x$ of SFR values plotted on the $y$- and $x$-axis, respectively.
\label{appfig:SFRsyscomp}}
\end{figure*}

\addtocounter{figure}{-1}
\begin{figure*}
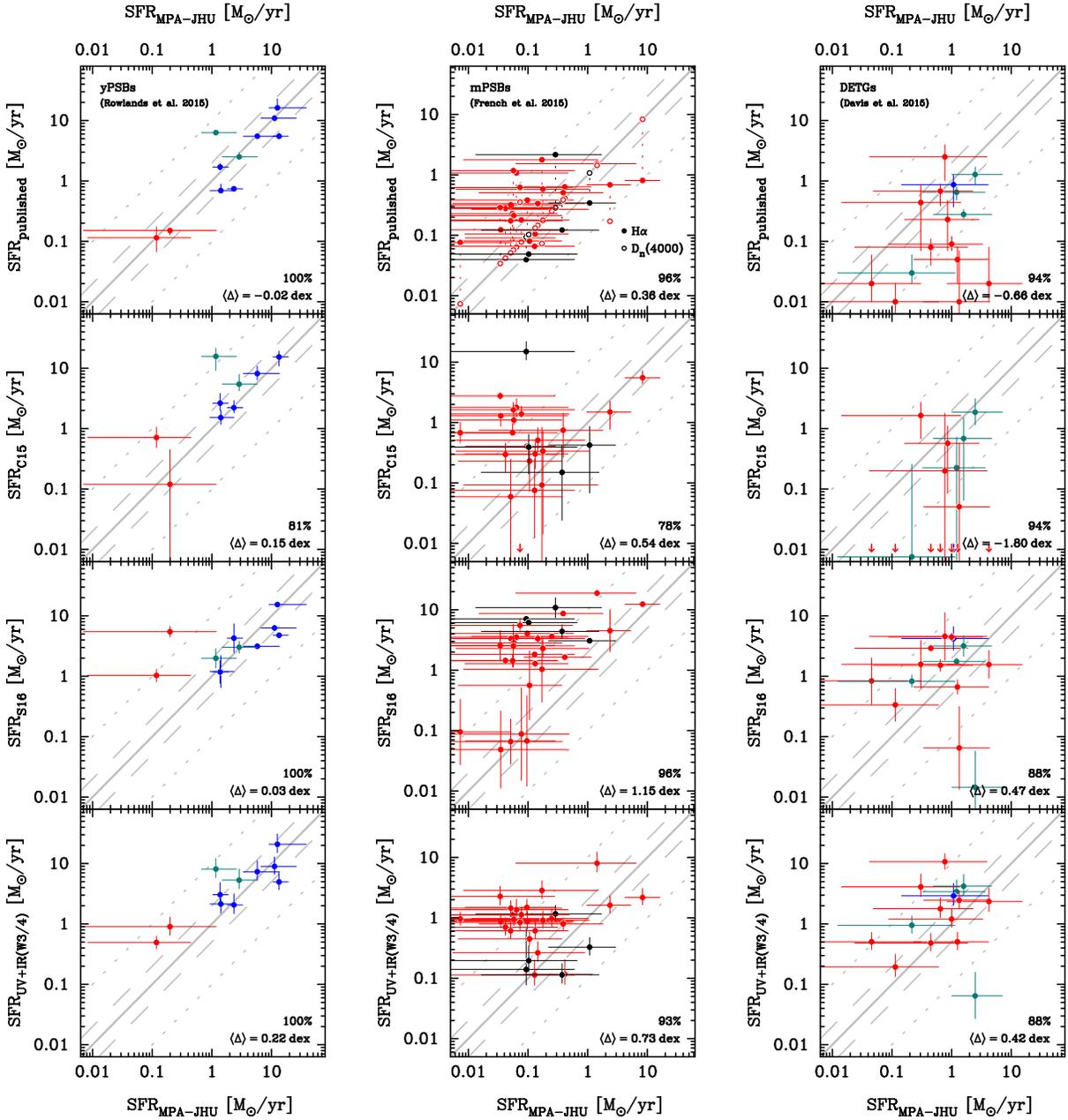

\centering
\begin{tabular}{ccccc}
 \includegraphics[width=0.3\textwidth]{f14d.ps} & & \includegraphics[width=0.3\textwidth]{f14e.ps} & & \includegraphics[width=0.3\textwidth]{f14f.ps}
\end{tabular}
\caption{As for the first part of Fig. \ref{appfig:SFRsyscomp}, but for young post-starburst galaxies (yPSBs, {\it left}), mature post-starburst (E+A) galaxies (mPSBs, {\it centre}), and dust lane early-type galaxies (DETGs, {\it right}). All symbols and annotations as in the first part of Fig. \ref{appfig:SFRsyscomp}, except for mPSBs, where filled (open) dots for SFR$_{\rm published}$ denote SFR-estimates based on the H$\alpha$ line (4000\,{\AA} break). For mPSBs the offset $\langle\Delta\rangle$ quoted for SFR$_{\rm published}$, relative to the MPA-JHU catalog, refers to H$\alpha$ SFRs.}
\end{figure*}

\begin{figure*}
\centering
\begin{tabular}{ccccc}
\includegraphics[width=0.3\textwidth]{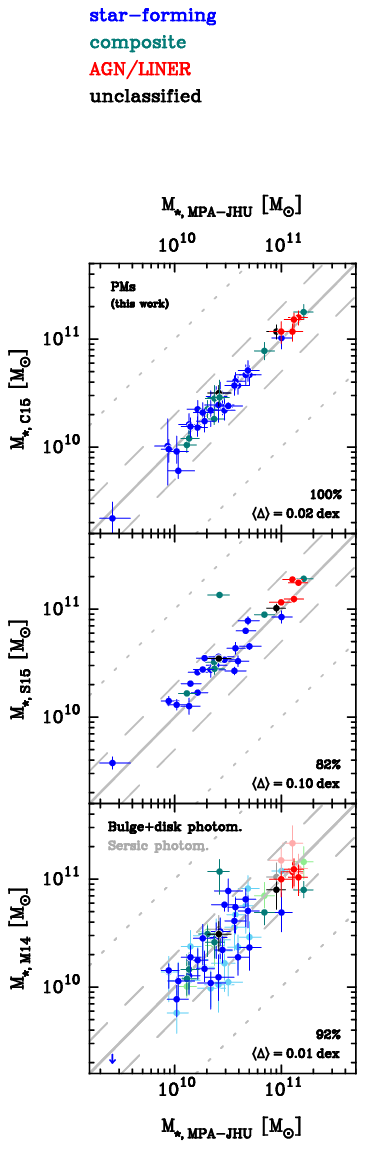} & & \includegraphics[width=0.3\textwidth]{f15b.ps} & & \includegraphics[width=0.3\textwidth]{f15c.ps}
\end{tabular}
\caption{Comparison of stellar mass estimates and associated 1$\sigma$ uncertainties for post-merger galaxies (PMs, {\it left}), control galaxies from xCOLD GASS DR1 (xCG-DR1, {\it centre}), and interacting pair galaxies (IPGs, {\it right}). \mstar$_{\rm ,\,published}$ is the SFR estimate reported in the papers (cf. legend in uppermost panel) first introducing each sample. Further data sources are (from top to bottom): \citet[C15]{chang15}, \citet[S16]{salim16} and \citet[M14]{mendel14}. Two estimates from M14 are shown: masses obtained using single-S\'ersic profile fits (light colours), and two-component bulge+disk decompositions (dark colours) of galaxy surface brightness distributions. All symbols and panel annotations as in Fig. \ref{appfig:SFRsyscomp}.
\label{appfig:Mstellsyscomp}}
\end{figure*}

\addtocounter{figure}{-1}
\begin{figure*}
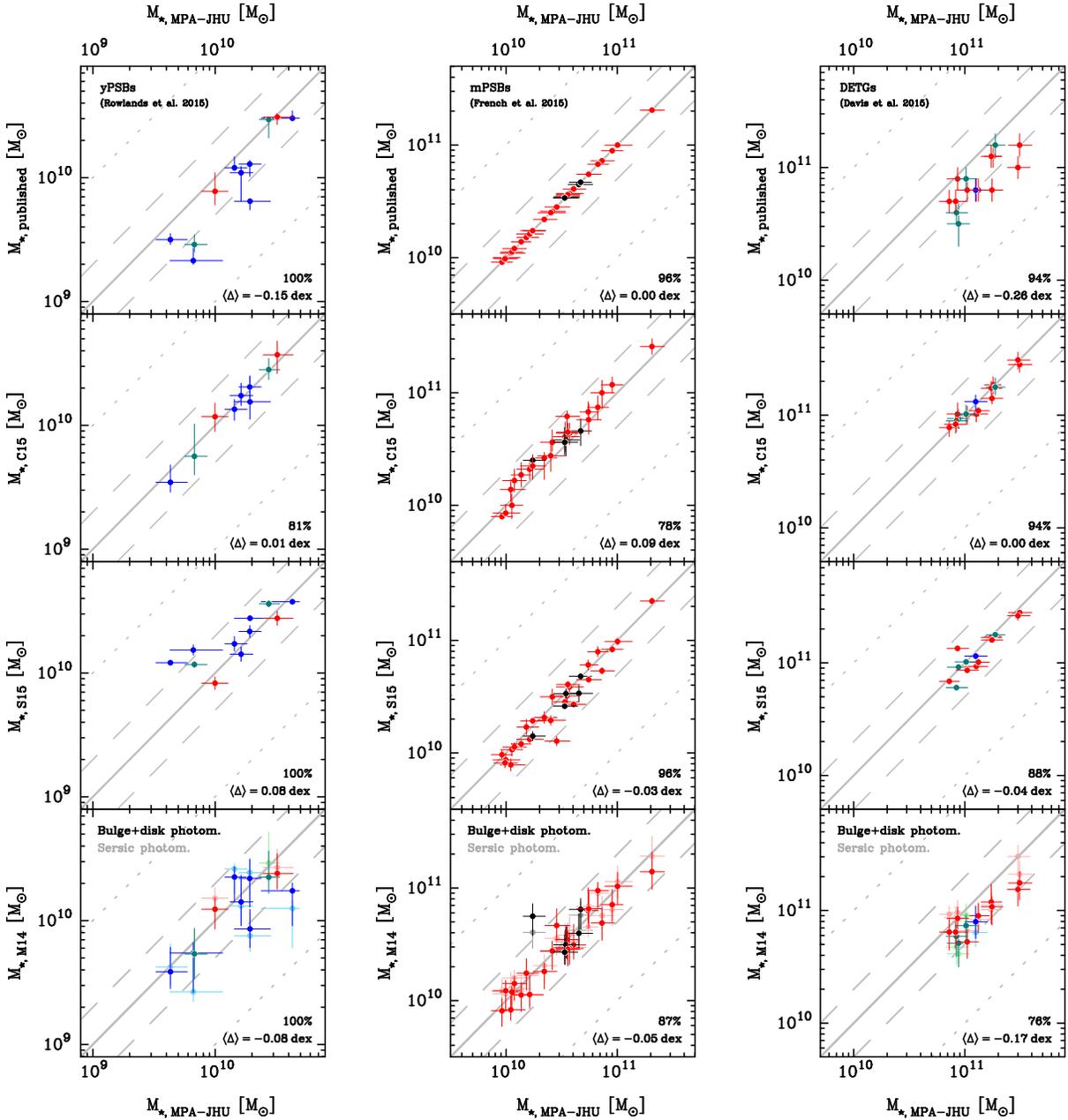

\centering
\begin{tabular}{ccccc}
\includegraphics[width=0.3\textwidth]{f15d.ps} & & \includegraphics[width=0.3\textwidth]{f15e.ps} & & \includegraphics[width=0.3\textwidth]{f15f.ps}
\end{tabular}
\caption{As for the first part of Fig. \ref{appfig:Mstellsyscomp}, but for young post-starburst galaxies (yPSBs, {\it left}), mature post-starburst (E+A) galaxies (mPSBs, {\it centre}), and dust lane early-type galaxies (DETGs, {\it right}). All symbols and annotations as in the first part of Fig. \ref{appfig:Mstellsyscomp}.}
\end{figure*}

\subsection{Impact of the choice of SFR \& \mstar}
\label{appsect:altSFRnMstell}

\subsubsection{Different SFR \& \mstar\ estimates used for consistency checks}
\label{appsect:altSFRnMstellintro}
Galaxy interactions can trigger several processes that may bias either stellar mass or SFR estimates. Different methods for determining stellar masses and SFRs have different and often complementary (dis-)advantages, and/or are subject to systematic offsets (see also \citealp{salim16} for a discussion of such systematics for some of the SFR and \mstar\ derivations mentioned below). Our approach for dealing with this is to repeat the key aspects of our analysis for a representative set of different SFR and \mstar\ estimators, and in this way bracket the systematic uncertainties our main results are subject to, the latter being based on our `best-estimate' SFR and \mstar\ values (see Sects. \ref{sect:SFRderiv} and \ref{sect:Mstellderiv}).\medskip

The SFR and \mstar\ measurements used for our exploration of systematics are taken from:
\begin{enumerate}
\item SED fitting in \citet{salim16}, based on UV-to-optical photometry and using the code CIGALE \citep{noll09, boquien19} that imposes an energy balance constraint during the fitting process (i.e., the dust emission in the mid- and far-infrared is required to be fully consistent with the energy absorbed in the UV and optical),
\item SED fitting in \citet{chang15}, also based on energy balance modelling but using the MAGPHYS code \citep{dacunha08} and applied to photometry spanning the optical-to-MIR regime,
\item the MPA-JHU catalog\footnote{\url{https://wwwmpa.mpa-garching.mpg.de/SDSS/DR7/}}, which is built on SDSS optical photometry and spectroscopy. Stellar masses are based on fits to galaxy-integrated broad-band $ugriz$ photometry and SFRs inferred from either nebular emission line information or the 4000\AA\ break, closely following \citet{brinchmann04}. Aperture corrections are applied to convert SDSS fibre-based SFR measurements to total galaxy SFRs. In addition to representing a different approach than the SED-fitting methods in (1) and (2), these have in general been widely used in the literature and thus provide a familiar reference point.
\item The papers introducing, respectively, the xCG reference sample and all comparison samples. Including the originally published SFR and \mstar\ estimates in our investigation of systematics, allows us to place in direct context our results with those reported in these literature studies.
\end{enumerate}

Figs. \ref{appfig:SFRsyscomp} and \ref{appfig:Mstellsyscomp} provide an intercomparison of SFR and \mstar\ derivations for PMs, galaxies from the xCG reference sample, and galaxies in our four comparison samples.\medskip

Note that, in addition to directly entering the calculation of depletion times and gas fractions, as well as the selection of suitable control galaxies, the SFR and \mstar\ values also impact the choice of CO-to-H$_2$ conversion factors \aCO\ that are necessary to calculate molecular gas masses. Specifically, inferring an \aCO\ via the 2-SFM framework requires a measurement of the MS-offset (see Sect. \ref{sect:XCO}). To ensure that the input MS-offsets are internally consistent for each combination of SFR and \mstar\ values we adjust the normalisation of the main sequence of star-forming galaxies, based on the systematic shifts between different SFR and \mstar\ catalogs quantified in Figs. \ref{appfig:SFRsyscomp} and \ref{appfig:Mstellsyscomp}.

\subsubsection{SFR- and \mstar-related systematics for post-mergers}
\label{appsect:altSFRnMstell_PMs}
This section investigates the extent to which our finding of systematically longer depletion times and higher gas fractions for PMs depends of the choice of \mstar\ and SFR measurements adopted.\medskip

\begin{figure*}
\centering
\includegraphics[width=.65\textwidth]{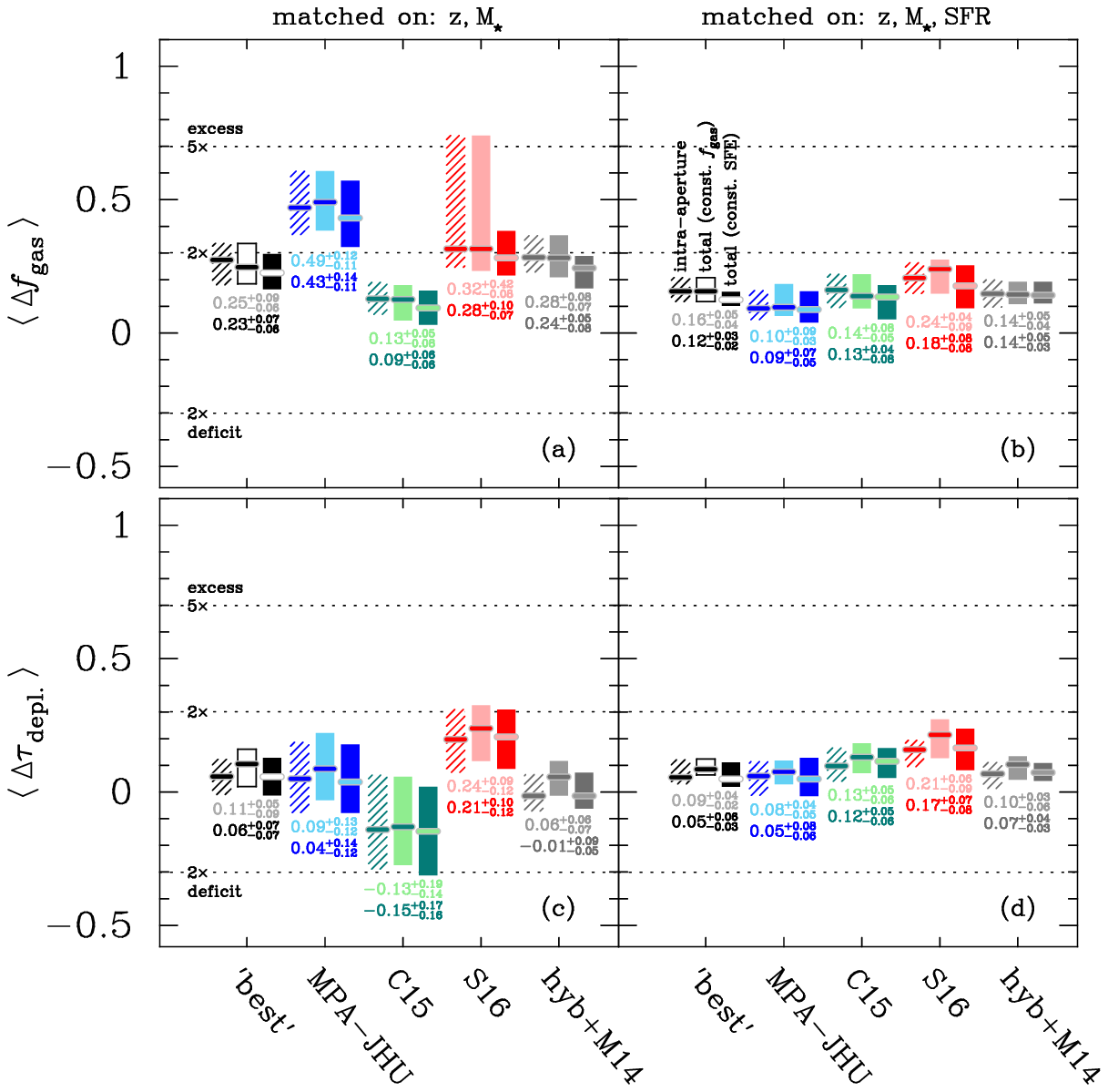}
\caption{Median gas fraction ({\it top}) and depletion time offsets ({\it bottom}) of post-merger galaxies compared to xCOLD GASS when different sets of \mstar\ and SFR estimates are used. (`best' -- spectral type-dependent SFR from Sect. \ref{sect:SFRderiv}, stellar masses from \citealp{mendel14}; MPA-JHU -- \citealp{brinchmann04}; C15 -- \citealp{chang15}; S16 -- \citealp{salim16}; hyb+M14 -- `hybrid' UV+IR SFRs \& stellar masses from \citealp{mendel14}.) The {\it left} ({\it right}) column shows the results for control-matching on redshift \& \mstar\ (on redshift, \mstar\ \& SFR).\newline
The 68\% confidence intervals associated with each median are visualized as rectangles. Results for three different aperture correction approaches are shown. From left to right, for each set of \mstar\ and SFR measurements: offset calculated within the \COone\ beam by downward correction of \mstar\ and SFR ({\it `intra-aperture'} -- hatched bar); galaxy-integrated offset calculated by upward correction (aperture correction factor $\mathcal{A}_{\COone}$(const.\,\fgas)) of the CO-flux assuming a spatially invariant gas fraction ({\it `total (const. \fgas)'} -- light-colour bar); galaxy-integrated offset calculated by upward correction (aperture correction factor $\mathcal{A}_{\COone}$(const.\,SFE)) of the CO-flux assuming a spatially invariant star-formation efficiency ({\it `total (const. SFE)'} -- dark-colour bar). The median offset values and associated 1\,$\sigma$ uncertainties for the const. \fgas\ (const. SFE) aperture correction are reported for each set of \mstar\ \& SFR estimates with light (dark) font.
\label{appfig:physcontrolmatch_sys}}
\end{figure*}

Fig. \ref{appfig:physcontrolmatch_sys} summarises the outcomes of repeating the control-matching analysis applied to our `best-estimate' \mstar\ and SFR values in Sect. \ref{sect:PMoffsets_best} with the catalogs of \citet{brinchmann04}, \citet{chang15}, \citet{mendel14} and \citet{salim16}, as well as our own purpose-extracted UV and MIR photometry underpinning the calculation of hybrid SFRs described in Sect. \ref{sect:SFRderiv_SFG}. The results for control-matching on the 2 (3) properties of redshift and \mstar\ (redshift, \mstar\ and SFR) are shown in the left-hand (right-hand) figure column.\\
When control-matching on the three properties redshift, \mstar\ and SFR we find very good consistency between the median gas fraction and depletion time offsets, $\langle\Delta\fgas\rangle$ and $\langle\Delta \tau_{\rm depl.}\rangle$, derived for the four literature \mstar\ and SFR catalogs. Furthermore, the respective medians differs by at most 0.1\,dex (25\%) from the median gas fraction and $\tau_{\rm depl.}$ enhancements calculated in Sect. \ref{sect:PMoffsets_best} using our `best-estimate' \mstar\ and SFR values, regardless of the aperture correction approach adopted. Our primary measurement presented in Sect. \ref{sect:PMoffsets_best} is typically close to the average of the other median offsets derived in this quantitative assessment of systematics. The modest gas fraction and $\tau_{\rm depl.}$ enhancements reported in Sect. \ref{sect:PMoffsets_best} for PMs when control matching on redshift, \mstar\ and SFR are reproduced for all alternative \mstar\ and SFR catalogs tested here.\medskip

When control-matching on redshift and \mstar\ only, the dispersion of median offsets inferred for the different catalogs becomes larger (see left-hand row of Fig. \ref{appfig:physcontrolmatch_sys}). Nevertheless, they nearly always overlap with our primary `best-estimate' result within 1\,$\sigma$ (the one exception being the gas fraction offset measured with MPA-JHU \mstar\ and SFRs), and as before our `best-estimate' measurement typically defines the 50$^{\rm th}$ percentile of the overall five alternatives explored. The finding that PM gas fractions and depletion times are enhanced is preserved in all cases bar one: when matching on redshift and \mstar\ only, adoption of the \citet[C15 in Fig. \ref{appfig:physcontrolmatch_sys}]{chang15} catalog results in a shorter median depletion time for PMs compared to xCG control galaxies, rather than the increase found with the `best-estimate' \mstar\ and SFRs (Sect. \ref{sect:PMoffsets_best}), and with all other alternatives tested. Finally, we note that the population scatter of the PM offset measurements (and the errors on the associated median offsets) in general do not differ strongly for the different catalogs under all control-matching scenarios.

\setlength{\tabcolsep}{0pt}
\begin{figure*}
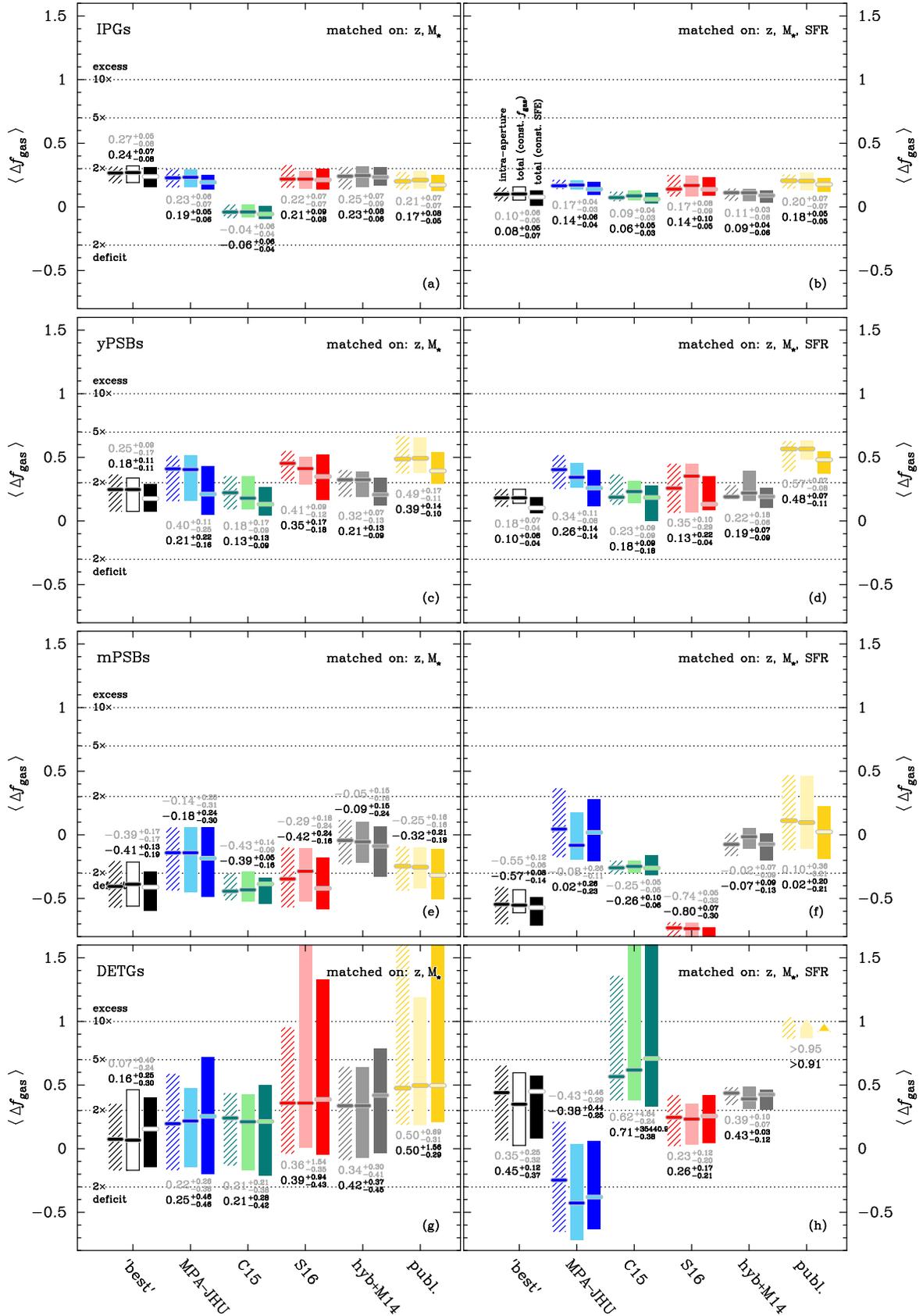

\centering
\begin{tabular}{rl}
\includegraphics[width=0.43\textwidth]{f17a.ps} & \includegraphics[width=0.43\textwidth]{f17b.ps}\\
\includegraphics[width=0.43\textwidth]{f17c.ps} & \includegraphics[width=0.43\textwidth]{f17d.ps}\\
\includegraphics[width=0.43\textwidth]{f17e.ps} & \includegraphics[width=0.43\textwidth]{f17f.ps}\\
\includegraphics[width=0.43\textwidth]{f17g.ps} & \includegraphics[width=0.43\textwidth]{f17h.ps}\\
\end{tabular}
\caption{Gas fraction offsets $\langle\Delta\fgas\rangle$ (median and associated 68\% confidence regions) for the comparison samples of (top to bottom): interacting pair galaxies (IPGs), young post-starburst galaxies (yPSBs), mature post-starburst galaxies (mPSBs), and dust lane early-type galaxies (DETGs). The left-hand (right-hand) column shows the results for control-matching on redshift \& \mstar\ (on redshift, \mstar\ \& SFR). Yellow bars: offsets derived with \mstar\ and SFR estimates from the publications originally introducing these samples (for mPSBs the measurements shown assume H$\alpha$-based SFRs). All other colours and annotations as in Fig. \ref{appfig:physcontrolmatch_sys}.
\label{appfig:compsamp_fgas_sysoverview}}
\end{figure*}

\begin{figure*}
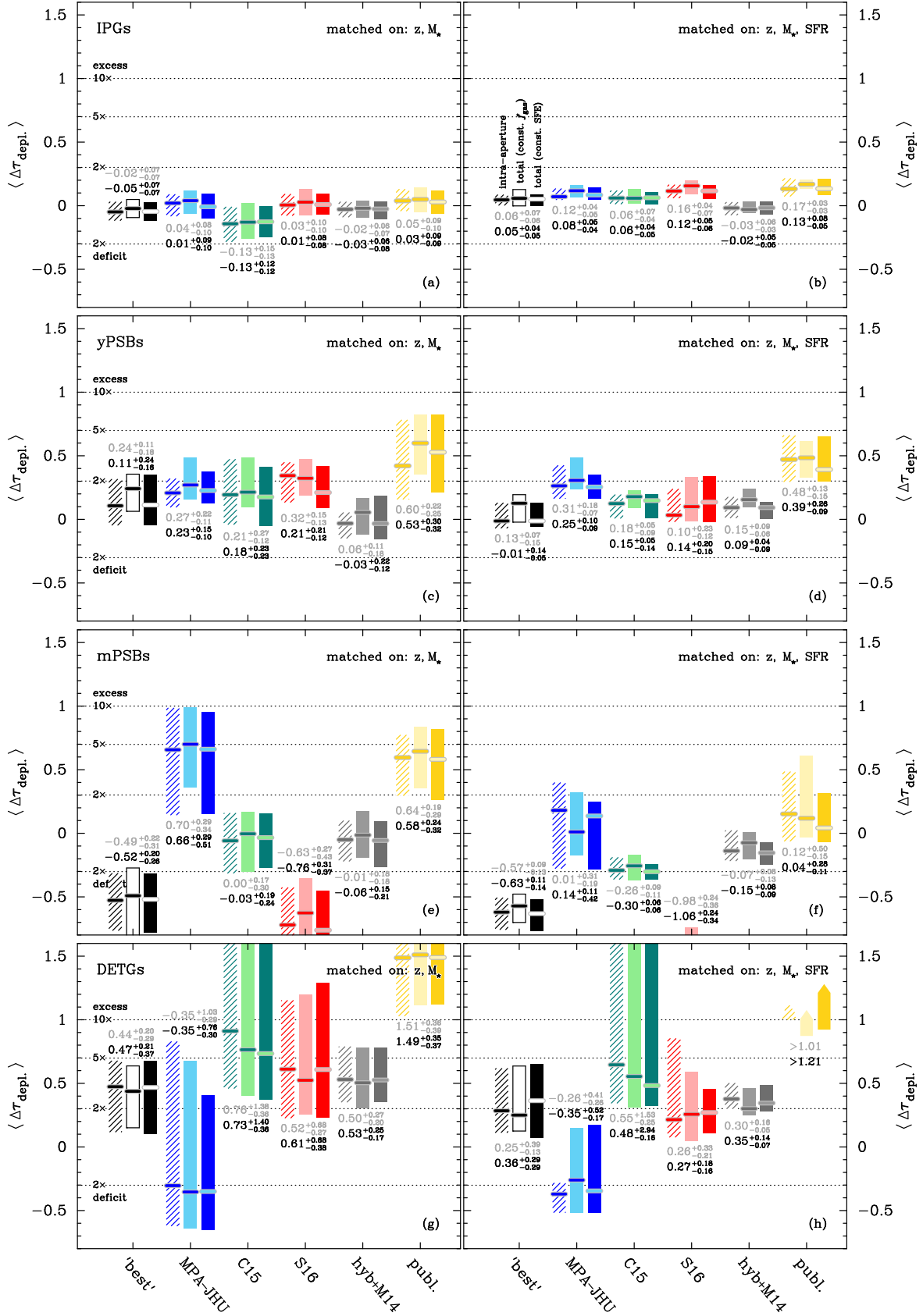

\centering
\begin{tabular}{rl}
\includegraphics[width=0.43\textwidth]{f18a.ps} & \includegraphics[width=0.43\textwidth]{f18b.ps}\\
\includegraphics[width=0.43\textwidth]{f18c.ps} & \includegraphics[width=0.43\textwidth]{f18d.ps}\\
\includegraphics[width=0.43\textwidth]{f18e.ps} & \includegraphics[width=0.43\textwidth]{f18f.ps}\\
\includegraphics[width=0.43\textwidth]{f18g.ps} & \includegraphics[width=0.43\textwidth]{f18h.ps}\\
\end{tabular}
\caption{As in Fig. \ref{appfig:compsamp_fgas_sysoverview}, but for the median depletion time offsets $\langle\Delta\tau_{\rm depl.}\rangle$.
\label{appfig:compsamp_tdepl_sysoverview}}
\end{figure*}
\setlength{\tabcolsep}{6pt}

\subsubsection{SFR- and \mstar-related systematics for comparison samples}
\label{appsect:altSFRnMstell_compsamps}
This section expands on the results of Sect. \ref{sect:compgal_offsets}, by presenting the median gas fraction and depletion time offsets obtained for our four comparison samples when \mstar\ and SFR catalogs other than our `best-estimate' values are employed (i) in the control-matching process, and (ii) to calculate gas fractions and depletion times. Figs. \ref{appfig:compsamp_fgas_sysoverview} and \ref{appfig:compsamp_tdepl_sysoverview} quantify the related systematic uncertainty in analogous form as Fig. \ref{appfig:physcontrolmatch_sys} for PMs. Additionally, for the comparison samples, we apply the control-matching technique to the original literature \mstar\ and SFR estimates. The corresponding offset measurements are shown in yellow in all panels of Fig. \ref{appfig:compsamp_fgas_sysoverview} and \ref{appfig:compsamp_tdepl_sysoverview}. The numerical values of the offset measurements are plotted beside each data point in the figures. The spread in the median values across all different \mstar\ and SFR estimators determines the length of the T-shaped error bars for systematic uncertainties in Fig. \ref{fig:compsamp_overview} in the main body of the paper (Sect. \ref{sect:compgal_offsets}). As discussed in Appendix \ref{appsect:aperturesys}, and shown for PMs in Sects. \ref{sect:PMoffsets_best} and \ref{appsect:altSFRnMstell_PMs}, systematics related to aperture correction methodology are a secondary effect also for all comparison samples (see the high degree of consistency in Figs. \ref{appfig:compsamp_fgas_sysoverview} and \ref{appfig:compsamp_tdepl_sysoverview} between the three adjacent medians/error bars for each set of \mstar\ and SFR measurements).

\begin{figure*}
\centering
\begin{tabular}{ccc}
\includegraphics[width=0.45\textwidth]{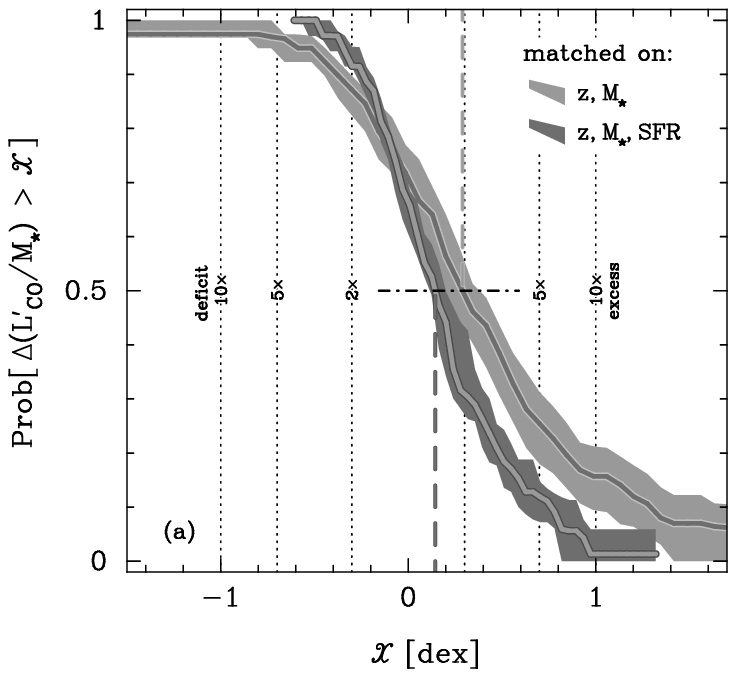} & & \includegraphics[width=0.45\textwidth]{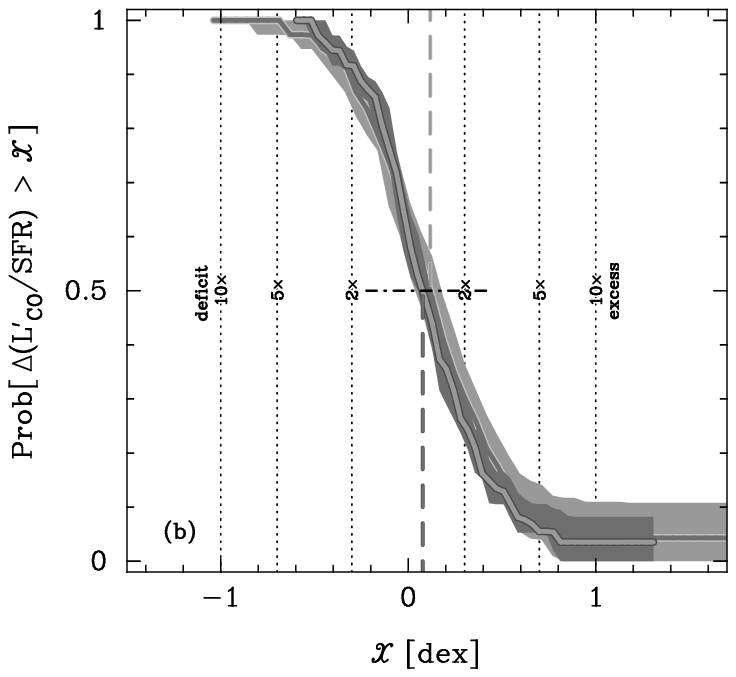}
\end{tabular}
\caption{As in Fig. \ref{fig:physcontrolmatch_best}, but for the observational proxies of gas fraction and depletion time, \lco/\mstar\ and \lco/SFR. The gas fraction and depletion time enhancements measured for our post-merger sample are not an artefact of specific assumptions in the calculation of CO-to-H$_2$ conversion factors; the results shown in this figure agree with those for the {\it physical} gas fraction and depletion times in Fig. \ref{fig:physcontrolmatch_best}.
\label{appfig:proxycontrolmatch_best}}
\end{figure*}

\subsection{Impact of different aperture correction approaches}
\label{appsect:aperturesys}

The `constant \fgas' correction (see Sect. \ref{appsect:apcorr}), which enters all galaxy-integrated molecular gas masses used throughout Sects. \ref{sect:results} and \ref{sect:discussion}, is one of four scenarios we investigated for dealing with telescope aperture effects. These are: {\it outward} correction of the CO luminosity to derive a `total', galaxy-integrated gas mass with the (i) $\mathcal{A}_{\COone}$(const.\,\fgas) or (ii) $\mathcal{A}_{\COone}$(const.\,SFE) aperture correction factors from Sect. \ref{appsect:apcorr}, or {\it inward} correction of (iii) \mstar\ with $\mathcal{A}_{\COone}$(const.\,\fgas) and of (iv) SFR with $\mathcal{A}_{\COone}$(const.\,SFE). Approaches (i) \& (iii) should in principle be equivalent when computing gas fraction offsets, and (ii) \& (iv) for the $\tau_{\rm depl.}$ offset analysis. This expectation is borne out to a good degree by the data in Figs. \ref{appfig:physcontrolmatch_sys} and \ref{appfig:compsamp_fgas_sysoverview}/\ref{appfig:compsamp_tdepl_sysoverview}, though small deviations occur due to statistical fluctuations introduced by our resampling of the KM estimator. In these three figures we report -- for each set of SFR and \mstar\ values -- the results for aperture corrections (i), (ii) and (iii) when plotting sample median offset $\langle\Delta\fgas\rangle$, and the results for corrections (i), (ii) and (iv) when plotting $\langle\Delta\tau_{\rm depl.}\rangle$ measurements. Shifts in these medians offsets relative to xCG caused by different aperture correction approaches remain confined to within the 1\,$\sigma$ errors on these medians. The main systematic uncertainty on the magnitude of \fgas\ and $\tau_{\rm depl.}$ offsets is thus contributed by the method used to infer \mstar\ and SFR values. In comparison, the choice of aperture correction has a smaller impact.

\subsection{Impact of the prescription for assigning \aCO\ values}
\label{appsect:obsproxPM}

Our molecular gas masses for all galaxies assume a CO-to-H$_2$ conversion factor that varies with metallicity and `starburstiness' (parametrised by a galaxy's offset from the MS, see Sect. \ref{sect:XCO}). For PMs specifically, in this section we re-derive gas fraction and depletion time offsets using the ratios \lco/\mstar\ \& \lco/SFR, the direct observational proxies of \fgas\ \& $\tau_{\rm depl.}$. This is also equivalent to assigning a constant \aCO\ value to all galaxies.\medskip

The median enhancement of \lco/SFR ratio of PMs differs from the median $\Delta\tau_{\rm depl.}$ measured in Sect. \ref{sect:PMoffsets_best} by only 0.01\,dex for both control matching approaches (see Fig. \ref{appfig:proxycontrolmatch_best}). When matching on redshift, \mstar\ and SFR, the median offset of the gas fraction proxy \lco/\mstar\ is smaller by 0.02\,dex than the median $\Delta\fgas$. We find a somewhat larger shift from median $\Delta\fgas$\,=\,0.25\,dex to median $\Delta(\lco/\mstar)$\,=\,0.29$\pm$0.09\,dex when control-matching only on redshift and \mstar. While these two values are consistent within 1\,$\sigma$ errors, we attribute the increase to the fact that in relative terms more PM galaxies reside in the off-MS starburst regime than xCG control galaxies. The down-weighting of such systems via a low \aCO\ in the parameter space of physical gas fractions leads to the somewhat smaller average gas fraction enhancement visible in Fig. \ref{fig:physcontrolmatch_best}. The shallow dependence of $\tau_{\rm depl.}\,{\propto}\,{\rm SFR}^{-0.2}$ (compared to SFR$^{0.8}$ for gas fractions; e.g., \citealp{saintonge12, sargent14}) may explain why there is no commensurate shift of the median $\Delta(\lco/{\rm SFR})$ when matching on redshift and \mstar\ only.\medskip

In summary, this test highlights that the conclusion that gas fractions and depletion times of PMs are enhanced compared to normal, non-interacting galaxies is not simply the consequence of adopting our specific recipe for converting CO luminosities to molecular gas masses.

\section{Distributions of physical properties for comparison samples}
\label{appsect:compgal_paramspace}

\begin{figure*}
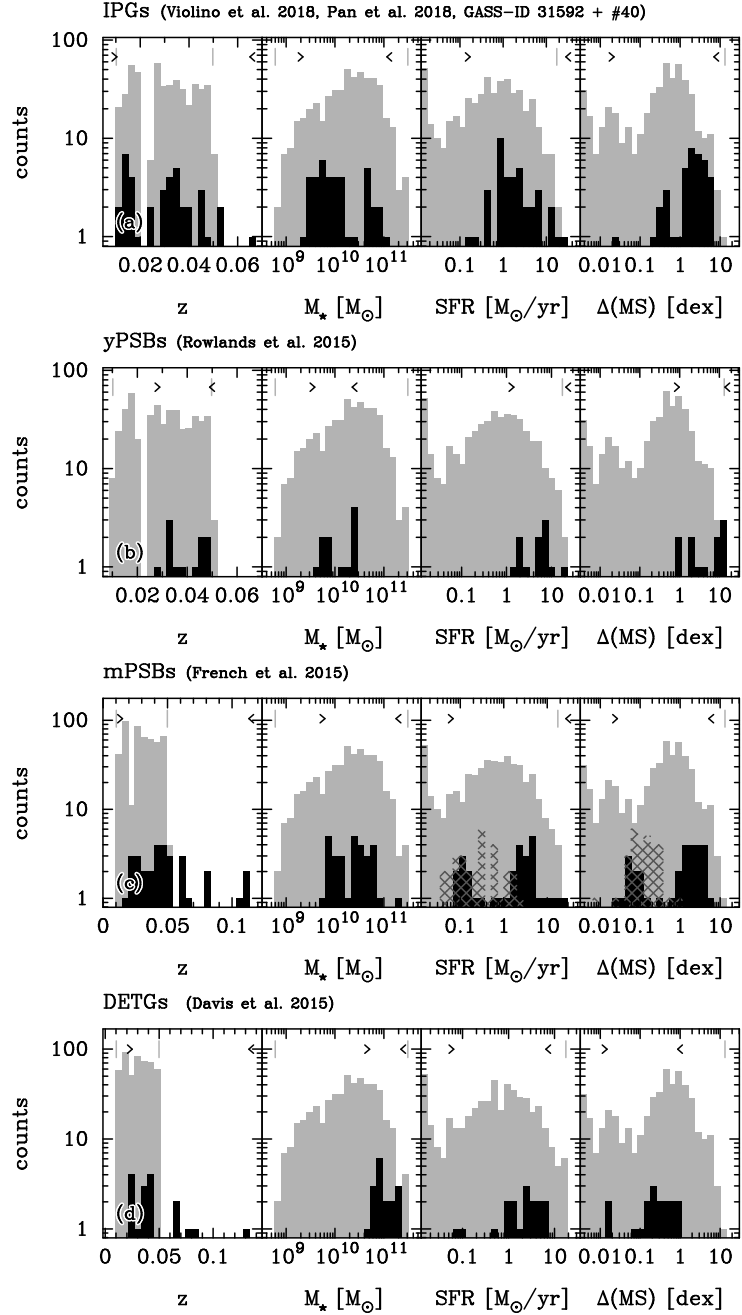

\centering
\begin{tabular}{c}
\includegraphics[width=0.6\textwidth]{f20a.ps}\\ \includegraphics[width=0.6\textwidth]{f20b.ps}\\ \includegraphics[width=0.6\textwidth]{f20c.ps}\\ \includegraphics[width=0.6\textwidth]{f20d.ps}
\end{tabular}
\caption{Redshift, stellar mass, SFR and MS-offset (from {\it left} to {\it right}) distributions of galaxies in our comparison samples (from top to bottom: IPGs -- interacting pair galaxies, yPSBs -- young post-starburst galaxies, mPSBs -- mature post-starburst (E+A) galaxies and DETGs -- dust lane early-type galaxies). The distribution of galaxies in the xCOLD GASS reference sample is shown in grey in the background. For mPSBs the cross-hatched histograms show the distribution of H$\alpha$-based SFRs and MS-offsets. Black arrowheads (light grey vertical lines) along the upper edge of each panel indicate the range of values covered by galaxies in the four comparison samples (by xCOLD GASS galaxies).
\label{appfig:comparisonMSFRz}}
\end{figure*}

Fig. \ref{appfig:comparisonMSFRz} illustrates the distribution of galaxies in our four comparison samples (introduced in Sect. \ref{sect:compsampintro}) with respect to the three quantities used for control-matching (redshift, stellar mass, SFR -- see Sect. \ref{sect:matching}), as well as MS-offset.

\section{CO data for a pair galaxy observed in IRAM project 196-14}
\label{appsect:ourpair}

Fig. \ref{appfig:pairspectrum} shows the \COone\ and \COtwo\ spectrum for a pair galaxy included in the target list of our IRAM project 196-14. Physical properties and CO measurements for this galaxy are listed under entry \#40 in Tables \ref{tab:prop} and \ref{tab:fluxes}, respectively. For the analysis of this paper, this galaxy was included in the comparison sample of interacting pair galaxies (see Sect. \ref{sect:IPGintro} for details).



\end{document}